\documentclass[aps,prd,showpacs,amssymb,floatfix,nofootinbib]{revtex4}
\usepackage{graphicx,amsmath,subfigure,ulem}
\pdfoutput=1
\def\reff@jnl#1{{\rm#1\/}}
\def\apjl{\reff@jnl{Astrophys. J. Lett.}}              
\def\mnras{\reff@jnl{MNRAS}}           

\begin{document}
\title{LISA extreme-mass-ratio inspiral events as probes of the black hole mass function}
\author{Jonathan R. Gair}
\affiliation{Institute of Astronomy, Madingley Road, Cambridge, CB3 0HA, UK}
\email{jgair@ast.cam.ac.uk}
\author{Christopher Tang}
\affiliation{St Catharine's College, Cambridge, CB2 1RL, UK}
\author{Marta Volonteri}
\affiliation{Department of Astronomy, University of Michigan, Ann Arbor, MI, USA}

\date{\today}

\begin{abstract}
One of the sources of gravitational waves for the proposed space-based gravitational wave detector, the Laser Interferometer Space Antenna (LISA), are the inspirals of compact objects into supermassive black holes in the centres of galaxies --- extreme-mass-ratio inspirals (EMRIs). Using LISA observations, we will be able to measure the parameters of each EMRI system detected to very high precision. However, the statistics of the set of EMRI events observed by LISA will be more important in constraining astrophysical models than extremely precise measurements for individual systems. The black holes to which LISA is most sensitive are in a mass range that is difficult to probe using other techniques, so LISA provides an almost unique window onto these objects. In this paper we explore, using Bayesian techniques, the constraints that LISA EMRI observations can place on the mass function of black holes at low redshift. We describe a general framework for approaching inference of this type --- using multiple observations in combination to constrain a parameterised source population. Assuming that the scaling of EMRI rate with black hole mass is known and taking a black hole distribution given by a simple power law, ${\rm d}n/{\rm d}\ln M = A_0 (M/M_*)^{\alpha_0}$, we find that LISA could measure the parameters to a precision of $\Delta(\ln A_0)\sim 0.08$, and $\Delta(\alpha_0)\sim 0.03$ for a reference model that predicts $\sim1000$ events. Even with as few as $10$ events, LISA should constrain the slope to a precision $\sim0.3$, which is the current level of observational uncertainty in the low-mass slope of the black hole mass function. We also consider a model in which $A_0$ and $\alpha_0$ evolve with redshift, but find that EMRI observations alone do not have much power to probe such an evolution.
\end{abstract}

\maketitle

\section{Introduction}
The proposed Laser Interferometer Space Antenna (LISA)~\cite{lisaCQG} will be sensitive to gravitational waves from systems containing supermassive black holes in the $10^4M_{\odot}$--$10^7M_{\odot}$ range. This mass range is very hard to probe electromagnetically, and only five such systems have been robustly identified from dynamical measurements \citep{gultekin2009}, including the black hole in the centre of our own galaxy. Placing constraints on the mass function of low-mass black holes has, however, key astrophysical implications. 

It has been suggested \citep{volonteri2008} that the shape of the mass function of massive black holes at the low-mass end is a key diagnostic to derive constraints on the mechanism that formed black hole seeds. Seed black holes are predicted to form in the mass range $\sim 100-10^5 M_\odot$, depending on the specific physical process involved \citep{volonteri2008}.  As black holes grow from low-mass seeds, it is natural to expect that at least some black holes have not grown efficiently and still trace the initial conditions.  Clearly, black holes at the high-mass end of the mass function have increased their mass by accretion, or they have experienced mergers and dynamical interactions. Any dependence of the black hole mass on the initial seed mass is erased. However, the distribution of low-mass black holes still retains some ``memory'' of the original seed mass distribution. The expectation is that ungrown seeds produce a peak in the mass function that corresponds to the peak of the seed mass function.  The higher the efficiency of seed formation, the more pronounced is the peak. 

Additionally, the mass function of low-mass black holes can provide insights on the co-evolution of black holes and their hosts. Observations show that the masses of black holes correlate with the mass, luminosity and the stellar velocity dispersion of the host \citep[and references therein]{gultekin2009}.  These correlations imply that black holes evolve along with their hosts throughout cosmic time. Lauer et al.~\cite{Lauer2007} suggest that at least some of these correlations break down at the largest galaxy and black hole masses \citep[but see][]{Bernardi2007}. One unanswered question is whether this symbiosis extends down to the lowest galaxy and black hole masses \citep{Greene2008}, due to changes in the accretion properties \citep{Mathur2005}, dynamical effects \citep{Volonteri2007}, or cosmic bias \cite{Volonteri2009}.  

Since current measurements of black hole masses extend barely down to $M_{\rm bh}\sim 10^6 M_\odot$, these features cannot be observationally tested with present data. LISA observations will significantly improve our understanding of the astrophysics of black holes in this mass range. 
LISA will detect mergers between supermassive black holes with these masses out to very high redshift, which will probe the early assembly of these systems and their host galaxies. LISA will also detect gravitational waves generated when compact objects (white dwarfs, neutron stars or black holes) are captured by and inspiral into supermassive black holes in the centres of galaxies~\cite{astrogr}, which are referred to as extreme-mass-ratio inspirals (EMRIs). The EMRI events will mostly be at low redshift, $z \lesssim 1$, and can therefore be used to probe the quiescent population of $\sim10^4M_{\odot}$--$10^7M_{\odot}$ black holes that remain today.  In this paper, we focus on this second type of source and explore what constraints LISA might be able to place on the low redshift population of black holes in this mass range. 

A typical EMRI event will have frequency $\sim 1$mHz and will be observed by LISA for $\sim 1$ year, and so we will detect $\sim 10^5$ cycles of the waveform. This allows LISA to make very precise measurements of the parameters of the host system~\cite{AK,HG09}, and it is hoped that EMRI observations can be used to carry out high precision tests of the spacetime geometry of the central object~\citep[and references therein]{astrogr}. For a typical EMRI event with signal-to-noise ratio (SNR) of $\sim 30$, we would expect to recover the redshifted mass of the central black hole to a precision $\Delta\ln((1+z) M) \sim 10^{-4}$ and the distance to the source to a precision of $\sim 3\%$. The spin of the central black hole, the redshifted mass of the inspiralling object and the orbital parameters (initial radius and eccentricity) should also be measured to very high accuracy, $\Delta\ln X\sim10^{-4}$--$10^{-3}$. While such precise measurements for individual systems are interesting and very important if the data is to be used for high precision tests of relativity, astrophysically it is the statistics of the set of EMRI events that LISA observes that will be of greatest value in constraining models. It is this application of LISA EMRI observations that we focus on in the current paper.

The distribution of events that LISA detects will depend on three factors --- the number density of possible source systems; the rate at which EMRIs occur in systems with particular parameters; and the sensitivity, in terms of distance reach, of the LISA detector to particular types of EMRI event. The last effect can be estimated theoretically in advance of the LISA mission, so we concentrate on what LISA can tell us about the first two effects. A particular model for the black hole distribution and the rate of inspirals per black hole does not precisely predict the number of events that LISA will observe, since inspirals start stochastically in any given galaxy. However, a particular model does predict the rate at which observable EMRIs of particular type will occur, and hence the probability distribution of the observed events. The LISA observation is a sample from this distribution, and we wish to infer the parameters of the underlying model from this sample. This can be done with a simple application of Bayes Theorem. Bayesian techniques have been employed widely in the context of gravitational wave data analysis~\cite[see, for instance,][]{cornishMCMC,christensenMCMC}, but this has been primarily to make statements about the parameters of individual sources. We can also use Bayesian methods to make statements about the underlying population from which the sources we detect are drawn and it is that which we will do here. Using LISA supermassive black hole merger events to constrain astrophysical models was considered in~\cite{plowmanSMBH}. The focus of that work was to derive an ``error kernel'' which would map the intrinsic distribution of source parameters onto the observed distribution of source parameters, and for model selection they used a variant of the Kolmogorov-Smirnov test. Our work differs from that approach not only in the fact that we consider EMRIs, but in the use of a Bayesian framework for the analysis and a parameterised model for the underlying distribution that we wish to constrain.

LISA will be able to measure the product of the EMRI rate per black hole with the number density of black holes, but it is not clear if it will be able to decouple the two effects. The scaling of the EMRI rate per black hole with the black hole parameters can, in principle at least, be constrained in advance through numerical simulations. One such analysis was carried out in~\cite{hopman09}. Various assumptions were made in that analysis which may not be valid for black holes in the LISA mass range, and we will discuss these issues further in Section~\ref{sec:probdist}. However, for the purpose of this work we will assume that the black hole rate scaling is known, and is given by the results in~\cite{hopman09}. This allows us to interpret our results in terms of the underlying distribution of black hole masses. An alternative interpretation that does not rely on this assumption is that we are constraining the convolution of the black hole number density with the EMRI rate per black hole.

The parameters of individual sources will not be measured perfectly by LISA, due to noise in the detector and confusion from other sources in the data stream. It is possible to include parameter errors consistently when making statements about the population, and we will describe how this is done in practice in Section~\ref{parerr}. However, this requires having properly sampled posterior distributions for all of the sources in the data set. An alternative approach is to bin the sources according to their maximum likelihood parameters. Although sources may end up in the wrong bins due to the parameter errors, the majority of sources will be correctly classified. We use the binning approach in this paper, since it allows us to assess LISA's ability to constrain the black hole population without requiring the computationally expensive simulation of LISA noise and recovery of posteriors for hundreds of EMRI sources. We will demonstrate that our conclusions are not affected by the inclusion of reasonable parameter estimation errors, which indicates that our results are an accurate reflection of what will be achievable in practice.

In this paper we take a single power law model for the black hole mass function and consider both a redshift-independent case of the form ${\rm d}n/{\rm d}\ln M = A_0 (M/M_*)^{\alpha_0}$, and a redshift-dependent case of the form ${\rm d}n/{\rm d}\ln M = A_0 (1+z)^{A_1} (M/M_*)^{\alpha_0-\alpha_1 z}$. In the former case, we find that LISA will be able to measure the parameters to precisions $\Delta(\ln A_0) \sim 0.08$ and $\Delta(\alpha_0) \sim 0.03$, and in the latter case to precisions $\Delta(\ln A_0) \sim 0.2$, $\Delta(\alpha_0) \sim 0.06$, $\Delta(A_1) \sim 0.7$ and $\Delta(\alpha_1) \sim 0.2$. These precisions scale with the number of observed events like $N_{\rm obs}^{-1/2}$, and have been normalised to a reference case that predicts $\sim1000$ events. We find that changing our assumptions about the performance of the LISA mission affects these conclusions only through the change in the number of events predicted. We also find that the precisions are somewhat improved if the black holes in the EMRI systems tend to have large spins.

The paper is organised as follows. In Section~\ref{theory} we describe the theoretical framework that we employ in this analysis. This includes a discussion of Bayes Theorem, a description of the probability distribution for EMRI events that LISA will detect, the proper treatment of parameter measurement errors and a summary of existing constraints on the shape of the black hole mass function. In Section~\ref{sec:res}, we present our results for both the redshift-independent and redshift-dependent models. This section includes a discussion of the effect of measurement errors and variation in the bin size used in the analysis. We also demonstrate how the results change as we vary the true black hole population in the Universe and as we change our assumptions about the performance of the LISA detector and the spin of the black holes. Finally, in Section~\ref{sec:discuss} we summarise our results and discuss directions for future research.

\section{Theoretical framework}
\label{theory}
\subsection{Bayesian inference}
Bayes' Theorem provides a way to infer the posterior probability distribution of the parameters, $\vec{\lambda}$, of a model based on a hypothesis $H$, given some observed data $D$
\begin{equation}
p(\vec{\lambda}| D, H) = \frac{ p(D|\vec{\lambda},H) \, p(\vec{\lambda}|H)}{p(D|H)},
\label{bayes}
\end{equation}
where $p(\vec{\lambda}| D, H)$ is the posterior probability distribution, $p(D|\vec{\lambda},H)$ is the likelihood, $p(\vec{\lambda}|H) = \pi(\vec{\lambda})$ is the prior probability on the model parameters and $p(D|H)$ is the evidence for the hypothesis $H$ given the data. The evidence plays a key role in model selection, but it enters Bayes' theorem as a normalisation constant that is independent of the model parameters, so we can ignore it if we are interested only in the posterior probability distribution.

In traditional applications of Bayesian methods to problems in gravitational wave data analysis, we are trying to infer the parameters of the waveform based on the observed data from our detector. The uncertainty comes from the noise distribution in the detector, and the likelihood is based on the spectral density of instrumental noise. What we are interested in here is not the parameter estimation problem for a single source, but in inference about the underlying galaxy population based on the set of sources that we have detected. In that case, the uncertainty comes from the fact that a particular galaxy population does not predict uniquely the events we will see, but only the rate at which events occur. We will use Bayesian inference to quantify the statements we can make about the properties of galactic black holes, based on the number of EMRI sources of each type that LISA sees. The probability distribution for the set of events will be discussed in the next section.

\subsection{LISA event probability distribution}
\label{sec:probdist}
We now turn to the question of modelling the likelihood, $p(D|\vec{\lambda},H)$. The data, $D$, consists of the parameters of all of the events that LISA has observed. These parameters will not be measured exactly, but from the analysis of the LISA data stream we will obtain a posterior probability distribution for the parameters of each event, and for the number of events in the data set. We will discuss how these errors may be included in the analysis in the next section, but for the following discussion we shall assume that the parameters of each event are known exactly. For clarity in the following description, we also assume that the parameter space of possible events has been divided into a finite number of bins. Results can also be obtained straightforwardly in a continuum analysis by taking the limit as the bin sizes tend to zero. This limit will be discussed at the end of this section.

In a binned analysis, the data is the number of events in each of the bins. The number of events is determined by two things --- the number of events with particular parameters that occur in the Universe, and the sensitivity of LISA to events with those parameters. In a given galaxy, EMRIs with particular parameters will finish, in a plunge into the central black hole, at a certain rate. We can model the number of plunges that occur in a certain time (or the number of EMRIs that start in a certain time window) as a Poisson process with mean equal to that rate. It is the fact that any given model predicts only the rate of EMRIs that gives rise to uncertainty. Only inspirals that have started in the right range of times will be detected by LISA, so the probability distribution of observed events follows this Poisson distribution. The rate for systems with particular parameters, $\vec{\lambda}$, is the product of the number density of galaxies with those parameters, $N(\vec{\lambda})$, with the intrinsic rate at which EMRIs start in such galaxies, ${\cal R}(\vec{\lambda})$.

The sensitivity of LISA can be characterised by a completeness function, ${\cal C}(\vec{\lambda})$. A given event will only be detected if it has sufficient signal-to-noise ratio (SNR) to be identified in the LISA data. The completeness is the probability that an event with particular parameters is detected and will depend on the data analysis pipeline. It can either be computed in advance or by using the cleaned LISA data stream via numerical injection. One reasonable model for ${\cal C}(\vec{\lambda})$ is that all events with SNR $\rho >\rho_{\rm th}$ will be detected $100\%$ of the time, and those with $\rho <\rho_{\rm th}$ will be detected $0\%$ of the time. This is a good model if an SNR cut is imposed on events included in a follow-up analysis of the type described here. One of the parameters in $\vec{\lambda}$ is the time remaining until plunge, $t_{\rm pl}$, at the start of the LISA observation. For fixed values of the other parameters, an SNR cut defines an allowed range of values for $t_{\rm pl}$. If a given EMRI is too close to plunge at the time when LISA starts taking data, then insufficient SNR will be accumulated before the object plunges for a detection. Conversely, if the EMRI is too far from plunge, the gravitational radiation will be weak, and so even over the full lifetime of LISA the SNR accumulated will be insufficient for detection. This range of times defines  the {\it observable lifetime} of a source, $\tau(\vec{\lambda})$. Using the SNR cut model for completeness, we can eliminate $t_{\rm pl}$ as a parameter and replace ${\cal C}(\vec{\lambda})$ by $\tau(\vec{\lambda})$. The function $\tau(\vec{\lambda})$ can be computed in advance, and depends only on the properties of the detector. It was computed for circular, equatorial inspirals in~\cite{gairEMRIastro}, using an SNR threshold of $\rho_{\rm th}=30$. We will use this simplified model for the completeness in the present paper. With a more complicated completeness function, we can still eliminate the time-to-plunge parameter by defining an effective observable lifetime as an integral of the completeness function over $t_{\rm pl}$, $\tau = \int_0^\infty {\cal C}(\vec{\lambda}) {\rm d}t_p$

Typically, a given bin will contain events from several galaxies, but these galaxies will behave independently and the sum of two independent Poisson distributions is a Poisson distribution with mean equal to the sum of the means. The number of events in a given bin is therefore determined by a Poisson process with mean equal to the expected number of events in that bin under the particular model. We describe the observed data by a vector $\vec{n} = (n_1, n_2, ..., n_K)$ of the number of events $n_i$ in each bin $i$. The probability of the data under a model $H$ depending on parameters $\vec{\mu}$ is
\begin{equation}
p(\vec{n}|\vec{\mu},H) = \prod_{i=1}^K \frac{(r_i(\vec{\mu}))^{n_i} {\rm e}^{-r_i(\vec{\mu})}}{n_i!}
\label{binnedprob}
\end{equation}
where the model-dependent rate, $r_i$, in a given bin is equal to the integral of the rate over the bin, ${\cal B}_i$
\begin{equation}
r_i (\vec{\mu}) = \int_{{\cal B}_i} N(\vec{\lambda}|\vec{\mu}) \,{\cal R}(\vec{\lambda}|\vec{\mu}) \, \tau(\vec{\lambda}) {\rm d} \lambda
\end{equation}
As before, $N(\vec{\lambda}|\vec{\mu})$ denotes the number density of galaxies, such that $ \int_{{\cal B}_i} N(\vec{\lambda}|\vec{\mu}) {\rm d}\lambda$ is the number of galaxies that contribute to the events in bin $i$ when the model parameters are $\vec{\mu}$. In practice, we work with log-likelihoods, so that the total log-likelihood is equal to the sum of the log-likelihoods for each bin. More generally, $\tau(\vec{\lambda})$ can be replaced by ${\cal C}(\vec{\lambda})$ and $t_{\rm pl}$ included in $\vec{\lambda}$.

The completeness and observable lifetime depend only on the detector, but the intrinsic rates $N(\vec{\lambda}|\vec{\mu}) \, {\cal R}(\vec{\lambda}|{\mu})$ depend on the properties of the underlying black hole distribution through $\vec{\mu}$. It is this latter that we will try to measure using EMRI observations. The dependence of the EMRI rate in a given system, ${\cal R}(\vec{\lambda})$, on black hole properties is somewhat uncertain. Hopman~\cite{hopman09} found that ${\cal R}(\vec{\lambda})$ varied with the mass of the central black hole, $M$, as
\begin{equation}
{\cal R}=400 {\rm Gyr}^{-1} \, \left(\frac{M}{3\times 10^6 M_\odot}\right)^{-0.15} .
\label{hoprate}
\end{equation}
The derivation of this equation assumed (i) an isothermal (cuspy) distribution of stars around the black hole, and that (ii) the $M-\sigma$ relation holds at all galaxy and black hole masses. However, if low-mass black holes  ($10^4M_{\odot} < M < 10^5 M_{\odot}$) are hosted in dwarf galaxies, which are analogues of the Milky Way satellites \citep{svanwas2010}, one or both of these assumptions may break down. In particular, the central region of such galaxies is dominated by dark matter, and the density displays a shallow core. The $M-\sigma$ relation is unknown at low galaxy and black hole masses, which means that there is uncertainty in assumption (ii), but there is no robust alternative model that could be used insted. The validity of assumption (i), the density profile, can be assessed using the observational properties of galaxies. If the central density of stars, $n_h$, is constant then we can express the sphere of influence of the black hole as $r_h^3=3M/(4\pi\,m_*\,n_h)$, where $m_*$ is the typical stellar mass. The normalisation of the EMRI rate per galaxy (Equation 8 in \cite{hopman09}) scales then as $n_h\,M^{1/4}$ instead of $M^{-1/4}$.  Studies of the demography of black holes in low-mass galaxies will be beneficial to improve our understanding of the mass scaling of EMRI rates. Since we can hope that observational and theoretical work will improve our understanding of ${\cal R}(\vec{\lambda})$ before LISA flies, for the purpose of this work we will take ${\cal R}(\vec{\lambda})$ as known, and given by Eq.~(\ref{hoprate}). We will therefore present the results in terms of how well LISA can measure the mass function, $N(\vec{\lambda})$, which is rather poorly constrained for black holes in the LISA range. However, the uncertainties in the scaling of ${\cal R}(\vec{\lambda})$  must be borne in mind when interpreting the results.

\subsubsection*{Continuum limit}
If we take the limit where the size of all the bins tends to zero, then equation~(\ref{binnedprob}) becomes
\begin{equation}
p(\vec{\vec{\Lambda}} | \vec{\mu},H) = {\rm e}^{-N_{\mu}} \prod_{i=1}^{N_{\rm o}} r(\vec{\lambda_i}|\vec{\mu})
\label{contprob}
\end{equation}
where $\vec{\vec{\Lambda}} = \{\vec{\lambda_1}, \vec{\lambda_2},\cdots, \vec{\lambda_{N_{\rm o}}}\}$ is the vector of measured parameter values, $\vec{\lambda_i}$, for source $i$, $N_{\rm o}$ is the number of observed events and $N_{\mu}$ is the expected number of events under the model parameters $\vec{\mu}$. This continuum limit is of the form that we would expect --- we have a product over all of the {\it observed} events of the probability that the event would be seen under the model parameters $\vec{\mu}$, with a weighting factor of ${\rm e}^{-N_{\mu}}$ which penalises models that predict too few or too many events.

In the analysis for this paper we will use the binned event distribution, as it makes the generation of data and the inclusion of parameter errors more straightforward, and we will check that the size of bins used does not significantly affect the results. The final analysis of LISA data will most likely use the continuum limit, but the broad conclusions about the precision with which LISA will be able to measure the properties of the black hole distribution should not be sensitive to the use of binning.

\subsection{Allowing for parameter measurement errors}
\label{parerr}
In the preceding section we assumed that we took the parameters measured by LISA at face value when assigning sources to bins. However, noise in the detector has the effect that the measured parameters for a source, $\vec{\lambda_{\rm o}}$, will differ from the true parameters, $\vec{\lambda_{\rm t}}$. If the error distribution is narrow compared to the size of the bins then it should be safe to ignore parameter errors in the analysis, as the majority of events will be assigned to the correct bin. This is one advantage of using binning. We will look at this further when we discuss our results in Section~\ref{sec:res}. In the continuum limit, the approximation of ignoring errors will also be OK provided that the rate predicted by the model does not change significantly over the typical width of the measured parameter distribution.

If the error distribution is broad compared to the scale over which the predicted rate varies, then parameter errors must be folded into the analysis, as outlined below.

\subsubsection{Continuum analysis}
The treatment of errors in the continuum limit is relatively straightforward theoretically. Our data is the output of the detector, $\vec{s}$, which is a combination of $N_{\rm o}$ signals, $\vec{h}_i$, with parameters $\vec{\lambda_i}$ and instrumental noise, $\vec{n}$. The probability that we would observe this data given our model and model parameters $\vec{\mu}$ is the integral over all possible parameters values, $\{\vec{\lambda}_i\}$, for the sources, of the product of the probability that the model Universe would yield these events (equation~(\ref{contprob})) with the probability that the noise in the detector would equal $\vec{s}-\sum_i \vec{h}_i$, i.e.,
\begin{equation}
p(D|\vec{\mu},H) = {\rm e}^{-N_{\mu}} \int\int\cdots\int \left[p\left(n = \vec{s}-\sum_i \vec{h}_i (\vec{\lambda}_i)\right) \prod_{i=1}^{N_{\rm o}} r(\vec{\lambda_i}|\vec{\mu}) \right] {\rm d}^k \vec{\lambda}_1  {\rm d}^k \vec{\lambda}_2 \cdots  {\rm d}^k \vec{\lambda}_{N_{\rm o}}
\label{ContErrs}
\end{equation}
where we use $k$ to denote the dimensionality of the parameter space for a signal. The first term inside the square bracket is the usual posterior probability distribution for the source parameters as computed from the LISA data. If the sources are well separated in parameter space, this is the product of the posterior probabilities for each individual source, but in the case of confused sources this is a multi-source posterior. We note that the above equation takes the form of an integral over the multi-source posterior probability distribution, which is exactly the type of problem for which Monte Carlo integration was designed. If we assume that the posterior pdf has been computed using Markov Chain Monte Carlo (MCMC) techniques, using appropriately uninformative priors, then the above integral can be evaluated directly as a sum over the MCMC samples of the rate for the sample parameters (Eq.~(\ref{contprob})). This follow-up, including errors, can therefore be done relatively cheaply once the initial source identification and characterisation has been completed.
To be consistent, the posteriors on the parameters of each individual source should be obtained by integrating over the parameters, $\vec{\mu}$, of the model for the Universe. These posteriors will therefore have to be recomputed in post-processing and will depend on both the particular population model used and on the other sources present in the data set. However, this calculation can also be done cheaply without repeating the MCMC sampling and, in practice, if the rate function varies slowly over the size of a typical error box for an EMRI event, then the posteriors should not change significantly. The rate function $r(\vec{\lambda_i}|\vec{\mu})$ plays the role of a prior on the waveform parameters $\vec{\lambda_i}$. In sampling literature the parameters that determine this prior are usually referred to as hyperparameters and any prior on $\vec{\mu}$ as a hyperprior.

In the above we have assumed that the number of sources is fixed and known. In practice, the survey will not be complete and we will have some distribution over the number of events that we have detected. The expression is similar in that case, but the integral will also extend over the number of observed events, $N_{\rm o}$.

While this work was being completed, we became aware of work on a related problem being carried out concurrently~\cite{mandelMS} in the context of LIGO. The focus in~\cite{mandelMS} was on the evaluation of the integral~(\ref{ContErrs}) including error distributions on the source parameters. 
Here we concentrate on what we can learn about a particular model for the underlying population of black holes harbouring EMRIs. While it will be important to include parameter measurement uncertainties in the actual analysis of LISA data, the details of these errors should not affect our general conclusions about what LISA will be able to achieve. For this reason we ignore errors in our analysis, but show that the inclusion of errors in the generation of the source population does not significantly alter the conclusions. We leave a fuller analysis to a future paper.

\subsubsection{Binned analysis}
In a binned analysis, parameter errors will cause sources to appear in the wrong bin. Errors can be folded into the analysis in two ways, depending on whether the data is binned according to the intrinsic source parameters, or according to the observed parameters.

\paragraph{Binning according to intrinsic parameter values}
A LISA observation of a given event will give a maximum a-posteriori set of parameters, and a distribution of the possible parameter values. Each event can then be assigned to one of a number of bins, with certain probabilities. If this is done for each event in turn, we obtain a set of possible intrinsic event distributions with relative probabilities. Each distribution can be analysed and a sum of the posteriors can be computed, weighted according to these probabilities. This approach has the advantage that it makes use of the actual measured errors for each source, and therefore gives proper relative weighting to the events. The disadvantage is that the number of possible intrinsic event distributions grows rapidly with the number of observed sources, $N_{\rm o}$. If each source can spread into $\sim N_{\rm b}$ bins, the number of possible distributions that we would have to consider would grow like $N_{\rm b}^{N_{\rm o}}$.

\paragraph{Binning according to observed parameter values}
Binning according to observed parameter values produces a single data set to analyse. The probability distribution of observed events can be computed using the expected error distribution for a source with given parameters $\vec{\lambda}$. We show in appendix~\ref{app:errdist} that when an underlying Poisson process with mean $\beta$ spreads into multiple bins, indexed by $i$, with known probabilities, $p_i$, then the observed distribution of events in each of the bins is drawn from an independent Poisson process with mean $p_i \beta$. This result is straightforward to derive and probably well known, but we include the result in the appendix for completeness. This property means that we can fold parameter errors into the analysis just by modifying the rate in each bin. In~\cite{plowmanSMBH}, the LISA ``error kernel'' for supermassive black hole mergers was defined in essentially the same way. The disadvantage of this approach is that it assigns average errors to the events, and does not make use of the measured error distributions. For this reason, we advocate using the continuum analysis when accounting for parameter measurement errors, as the posterior pdf required to evaluate the integral~(\ref{ContErrs}) will have already been computed. The continuum analysis also makes full use of all the information available in the LISA observation. In the present work, we employ the binning analysis because posterior pdfs are not readily available.

\subsection{Black hole mass function}
We require a model for the mass function of black holes, $N(\vec{\lambda})$, in the mass range $10^4 M_{\odot} < M < 10^7 M_{\odot}$, where LISA is sensitive to EMRI events. In this mass range the total mass function of black holes, including both active and quiescent systems, is poorly constrained. The mass function of quiescent  black holes is estimated in the literature by coupling the empirical correlations found between black hole mass and host properties (bulge mass, luminosity and velocity dispersion, see \citep[and references therein]{marconi2004,gultekin2009}) with the distribution of galaxies as a function of these properties:
\begin{equation}
\frac{{\rm d}n}{{\rm d}M}=\frac{{\rm d}n}{{\rm d}X}\frac{{\rm d}X}{{\rm d}M},
\label{eq:MFBH}
\end{equation}
where $X$ is either the velocity dispersion, $\sigma$, the bulge mass, $M_{\rm bulge}$, or the bulge luminosity $L_{\rm bulge}$. 

However, most of the quiescent black holes masses that led to the determination of these correlations between $M$ and $X$ are above $10^7 M_{\odot}$.  Additionally, even assuming that the correlation between black hole mass and bulge luminosity holds at all masses, the luminosity function of galaxies is also limited at faint fluxes, adding to the uncertainty \citep[see the detailed discussion in][]{GreeneHo}.

For instance, \cite{shethSDSS} estimate the velocity dispersion function for early-type galaxies in the Sloan Digital Sky Survey, in a flux limited sample. In order to obtain the complete distribution, including both early and late type galaxies, it is necessary to take several steps (convert the luminosity function of galaxies which are not early types into a distribution of circular velocities, and then convert circular velocity to velocity dispersion using empirical correlations), as measurements of the velocity dispersion are available only for galaxies which are early-types.

Among the three properties mentioned above ($\sigma$, $M_{\rm bulge}$, $L_{\rm bulge}$) the best constrained galaxy distribution comes from the luminosity function (the number density of galaxies per unit $L_{\rm bulge}$). Extrapolating both the $M-L_{\rm bulge}$ correlation, and the luminosity function of galaxies, \cite{GreeneHo} determine a black hole mass function that appears to be flat at $M<3\times 10^6 M_{\odot}$. This is likely an upper limit to the mass function at low black hole masses.  \cite{GreeneHo}  compare the mass function of quiescent black holes thus determined to the mass function of active black holes (clearly, a lower limit to the total black hole mass function). They find that the mass function of active black holes turns over at $M<3\times 10^6 M_{\odot}$, leading to a decreasing number density of small black holes. EMRIs could therefore be critical to constrain the total mass function of black holes (active and inactive) at low masses. The mass function at  $M<10^5 M_{\odot}$ is also a critical test for models of black hole formation \citep{volonteri2008,svanwas2010}. 

\begin{figure}[ht]
\includegraphics[width=0.5\textwidth,keepaspectratio=true]{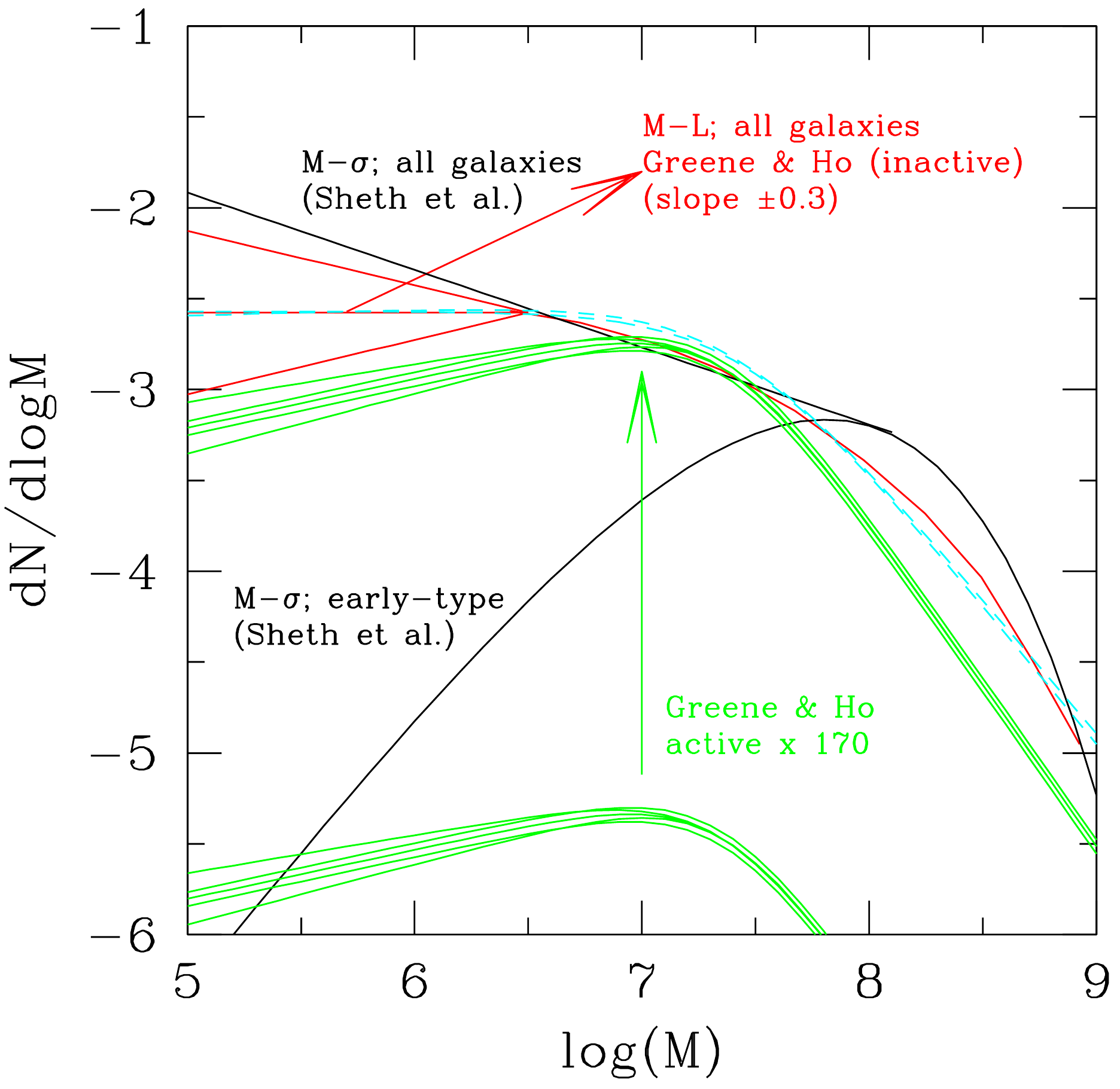}
\caption{Mass function of black holes taken from Greene and Ho~\cite{GreeneHo} and Sheth et al.~\cite{shethSDSS}, divided into type as labelled. The dashed cyan curves are fits to the Greene and Ho data, of the form given by the ansatz in Eq.~(\ref{GHfit}). We show the two fits described in the text --- one with all four parameters allowed to vary freely 
and one in which the function was forced to be flat for low masses by setting $\alpha=0$. The two fits lie almost on top of each other, but the curve that is slightly higher for $M=10^7M_{\odot}$ is the fit with $\alpha \neq 0$.
 }
\label{fig:GH}
\end{figure}

In this paper we adopt the mass function of `inactive' black holes given in Greene and Ho~\cite[long-dashed line in their Figure 11a]{GreeneHo}. It is relatively flat for low black hole masses, but turns over at $M \sim 10^7M_{\odot}$ and then decreases for larger black hole masses. Such a functional form is well fit by a simple ansatz of the form
\begin{equation}
\frac{{\rm d}n}{{\rm d}\log M} = \frac{A (M/M_*)^{\alpha}}{1 + B(M/M_*)^\beta} .
\label{GHfit}
\end{equation}
We find that with $M_*=3\times10^6M_{\odot}$, the best fit to the Greene and Ho data has parameters
\begin{equation}
A = 0.00284 \mbox{ Mpc}^{-3},\qquad \alpha = 0.0311, \qquad B = 0.041, \qquad \beta = 1.5105 .
\end{equation}
If we force the slope to be flat at low masses by setting $\alpha = 0$, we instead find the parameters
\begin{equation}
A = 0.00269 \mbox{ Mpc}^{-3},\qquad \alpha = 0, \qquad B = 0.031, \qquad \beta = 1.5262 .
\end{equation}
In Figure~\ref{fig:GH}, we show the black-hole mass function from Greene and Ho for inactive and active galaxies along with the two fits given above. We also show the black hole mass function obtained from the Sloan Digital Sky Survey for early-type galaxies, and the corresponding mass function inferred for all galaxies via the procedure described above~\cite{shethSDSS}. These curves are included to illustrate the degree of uncertainty in the literature about the mass-function in the LISA range. It is clear that a relatively precise measurement of the slope of the mass function would significantly enhance our understanding of the low-end of the mass function.

If LISA could probe all black hole masses, we could hope to determine all four parameters in a fit of the form~(\ref{GHfit}) from LISA observations. However, LISA is only sensitive to gravitational waves from black holes in the range $10^4 M_{\odot} < M < 10^7 M_{\odot}$. It is clear from Figure~\ref{fig:GH} that in this range the mass function is relatively flat. For that reason we choose to adopt a simpler prescription for the mass function in the range of interest for LISA, namely a simple power law ${{\rm d}n}/{{\rm d}\log M} \propto M^\alpha$.

\subsection{Markov Chain Monte Carlo}
To recover the posterior probability distribution using Eq.~(\ref{bayes}) we make use of Markov Chain Monte Carlo techniques (MCMC), which provide an efficient way to explore posterior probability distributions. For the redshift-independent model we consider below, it is relatively straightforward to evaluate the posterior directly without using Monte Carlo integration. However, for higher dimensional problems the gain from MCMC methods becomes increasingly significant. We used MCMC methods to obtain all of the results reported in this paper to demonstrate the general framework.

The procedure for MCMC is to construct a sequence of points in the parameter space $\{\vec{x}_i\}$. The initial point, $\vec{x}_0$, is chosen at random from the prior. At a subsequent iteration, $i$, a new point $\vec{y}$ is drawn from a proposal distribution $q(\vec{y} | \vec{x}_i)$ and we evaluate the Metropolis-Hastings ratio
\begin{equation}
R = \frac{\pi(\vec{y}) p(D|\vec{y},H) q(\vec{x}_i | \vec{y})}{\pi(\vec{x}_i) p(D|\vec{x}_i,H) q(\vec{y} | \vec{\lambda}_i)}
\end{equation}
A random sample, $u$, is then drawn from a uniform distribution $u\in U[0,1]$ and if $u < R$, the move is accepted and we set $\vec{x}_{i+1}=\vec{y}$, otherwise the move is rejected and we set $\vec{x}_{i+1}=\vec{x}_{i}$. If $R > 1$, the move is always accepted, but if $R<1$ there is still a chance that the chain will move to the new point. If the points in the sequence are generated via this algorithm, then the distribution of points visited by the chain follows the posterior posterior probability distribution of the parameters.

\section{Results}
\label{sec:res}
In order to simplify the calculation for this first analysis, we make various assumptions about the parameter space of EMRIs. We assume that all EMRIs are circular, equatorial inspirals into black holes with a fixed spin and that prograde or retrograde inspirals are equally likely. We also ignore sky position and orientation dependence of the LISA response, and use position and orientation averaged SNRs. These assumptions allow us to use the observable lifetime functions tabulated in~\cite{gairEMRIastro} and mean that the only parameters that we are interested in for each event are the mass of the black hole, $M$, and the redshift of the source, $z$. While this considerably simplifies the current analysis, the framework outlined in the previous section does not rely on these assumptions and so the present analysis could be easily extended to include more parameters as required. In~\cite{gairEMRIastro} results were presented for three different values of the fixed spin of the central black holes, $a=0,0.5,0.9$, and for two different sets of assumptions about the performance of the LISA detector. The ``optimistic'' assumptions were that the LISA mission lasted for 5 years and had full functionality for the whole time. With full functionality, the LISA response is equivalent to two independent right-angle interferometers, but if there is a failure on the satellite LISA can still operate with the equivalent response of a single right-angle interferometer. The ``pessimistic'' assumptions were that LISA was operating in this failure mode for the whole mission, and the mission only lasted for 2 years. We will take $a=0$ and the optimistic LISA initially, but will describe how the results change when we take $a=0.9$ or the pessimistic LISA in Section~\ref{varyLISA}.

As described in the previous section, each LISA measurement will have an error in the mass and redshift determination. These can be folded into the analysis, but for this first computation we ignore the errors and assume that the parameters are measured perfectly. The typical error we expect in a measurement of the {\emph redshifted} mass, $M(1+z)$, is $O(10^{-4})$, which is almost certainly ignorable. However, the redshift must be inferred from the distance, in which we expect an error of $O(10^{-2})$, and this propagates into a comparable error in $M$. Further error arises from uncertainties in the cosmology used to map between the measured luminosity distance and the redshift. Over a large population of events, we expect these errors to average out, and they should also be unimportant if the scale over which the event rate varies is larger than the scale of the typical errors. These arguments justify the omission of errors in our first analysis, but we will verify in Section~\ref{errres} below that our results do not change significantly when errors are included in the generation of the observed event distribution.

\subsection{Redshift-independent mass function}
\begin{figure}[ht]
\begin{tabular}{cc}
\includegraphics[width=0.35\textwidth,keepaspectratio=true]{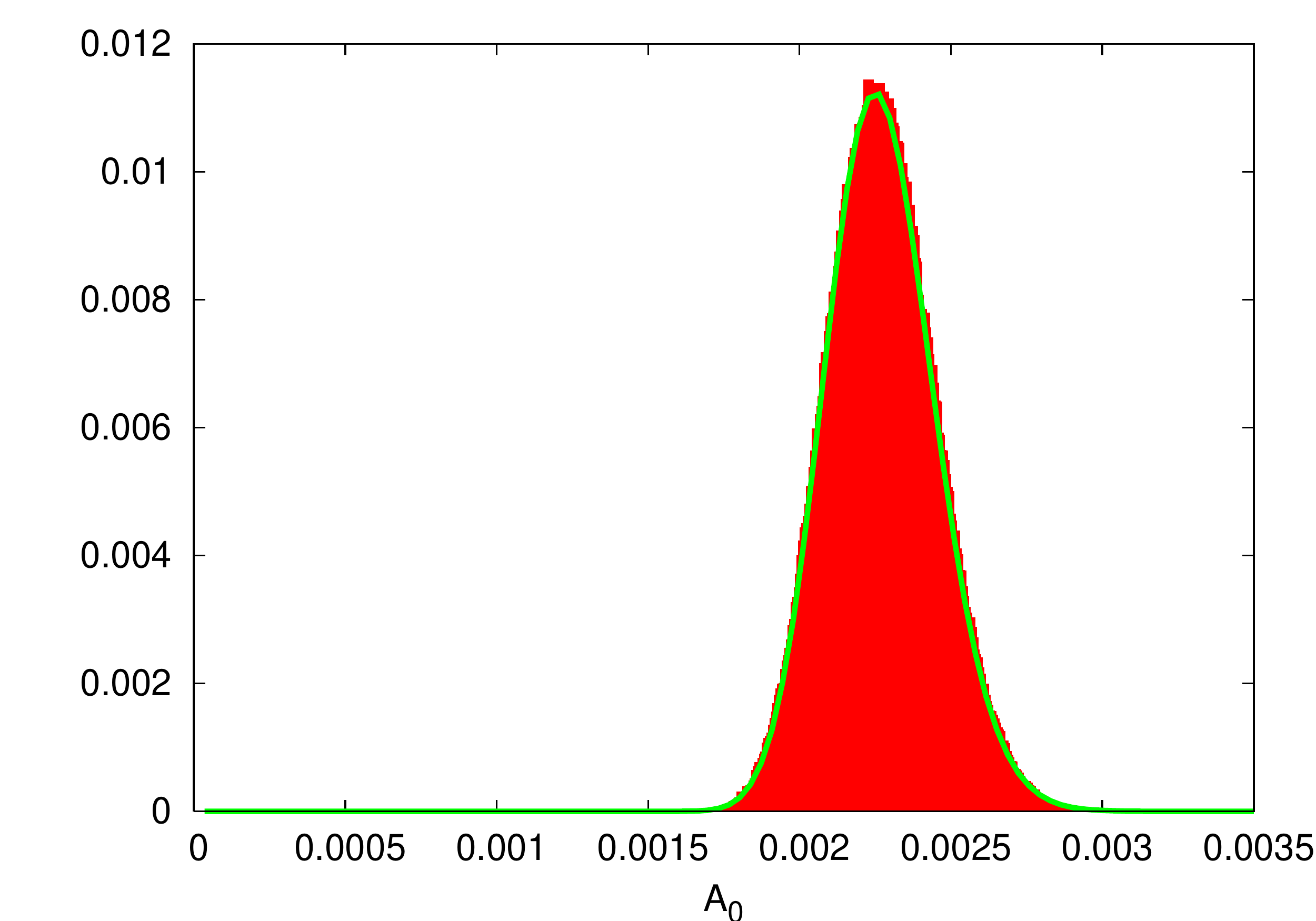}&
\includegraphics[width=0.35\textwidth,keepaspectratio=true]{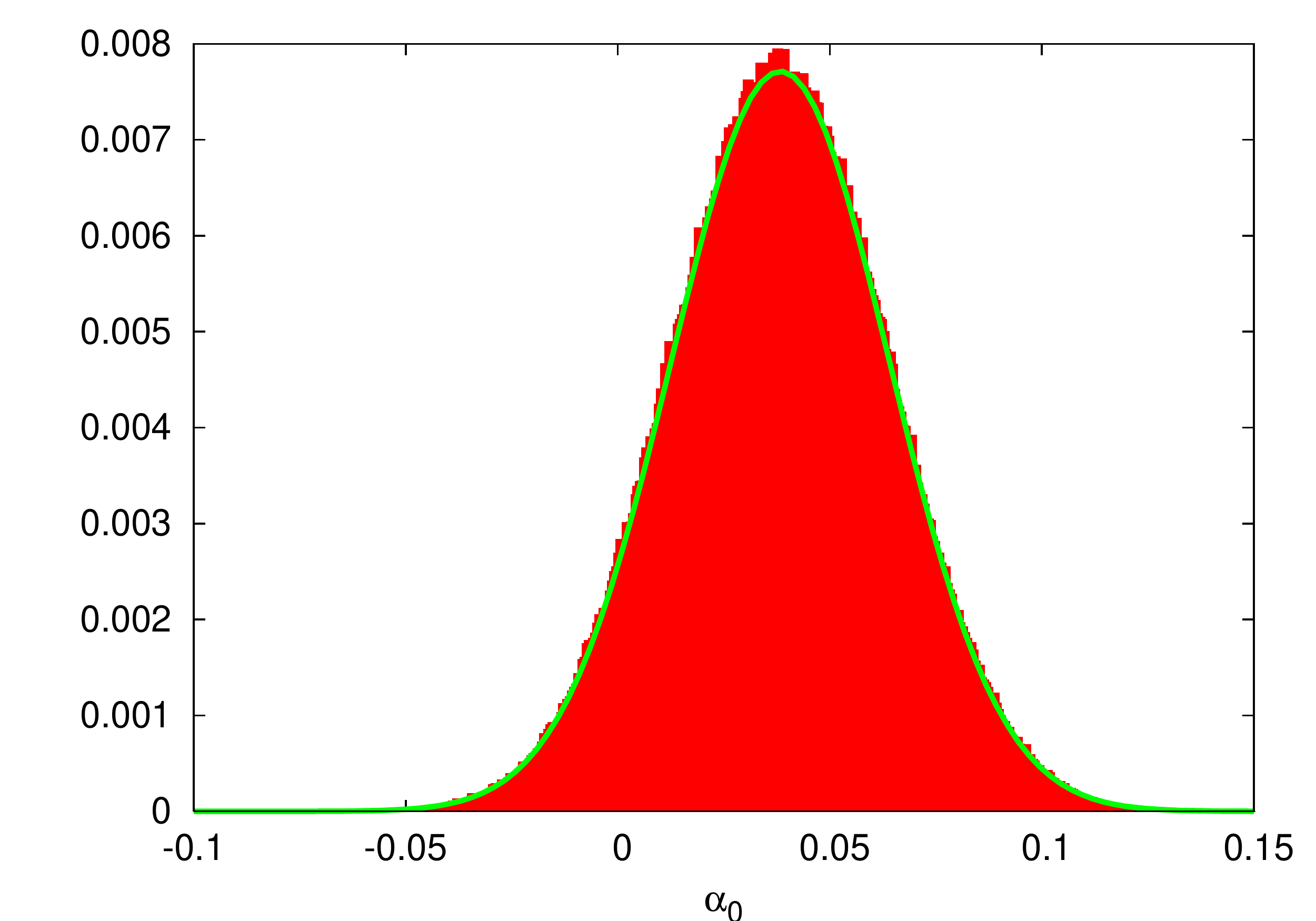}
\end{tabular}
\caption{Distributions of the mass function parameters $A_0$ (left) and $\alpha_0$ (right) that could be inferred from the set of LISA events. The solid green lines are the best-fit Gaussians to $\ln A_0$ and $\alpha_0$, obtained via fitting a quadratic to the logarithm of the distribution.}
\label{nonevolv}
\end{figure}

\begin{figure}[ht]
\includegraphics[width=0.5\textwidth,keepaspectratio=true]{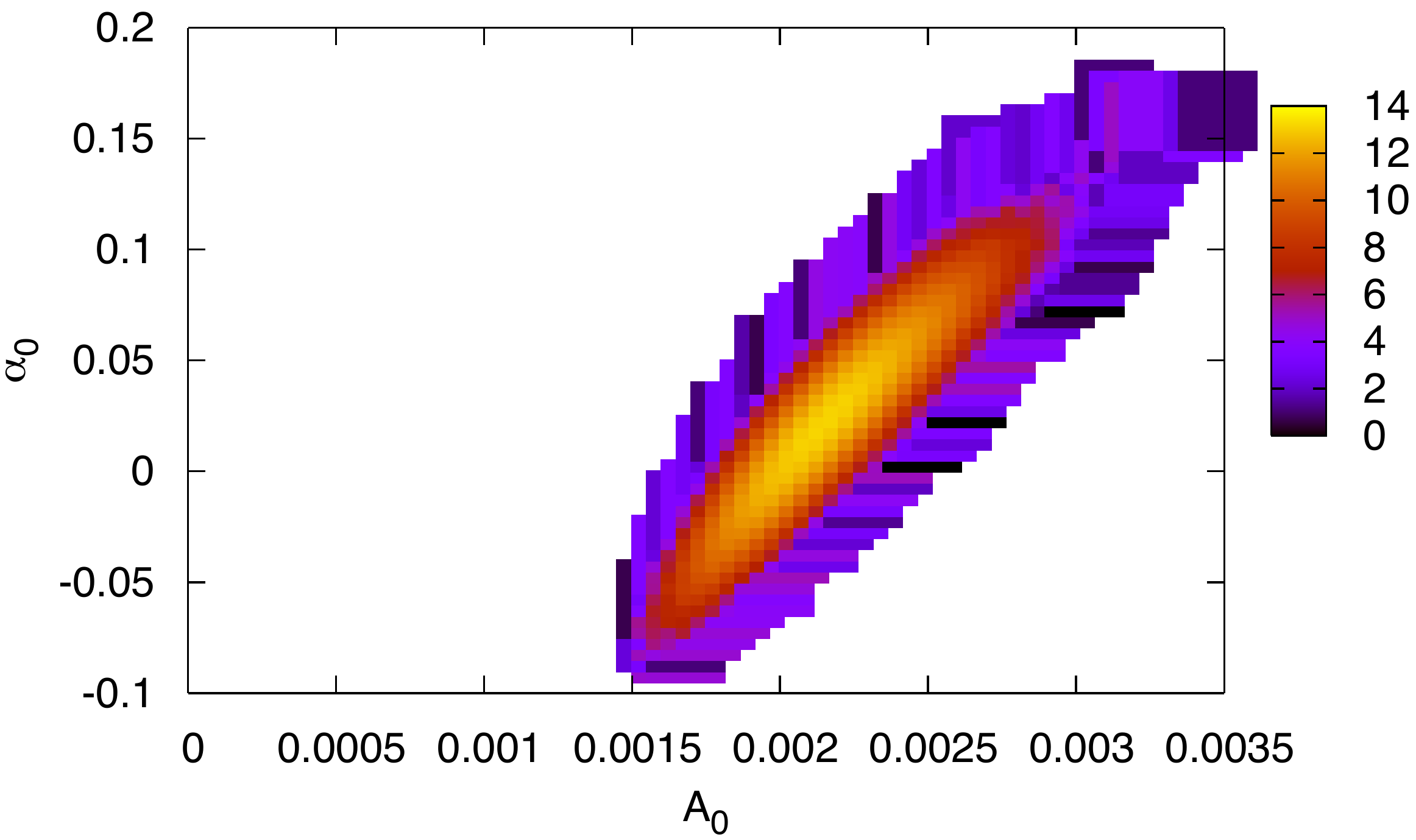}
\caption{Correlation between the values of $A_0$ and $\alpha_0$ inferred from the set of LISA events.}
\label{nonevolvCorr}
\end{figure}

As a test case, we took the mass function of black holes in the LISA mass range, $10^4 M_{\odot} < M < 10^7 M_{\odot}$ to be
\begin{equation}
\frac{{\rm d}n}{{\rm d} \log M} = A_0 \left(\frac{M}{M_*}\right) ^{\alpha_0} .
\end{equation}
with $M_* = 3\times10^6M_{\odot}$. 
We generated data sets from the probability distribution described in Section~\ref{sec:probdist}, using the values $A_0=0.002$Mpc$^{-3}$ and $\alpha_0=0$. These data sets were then explored using MCMC techniques to recover posterior probability distributions for $A_0$ and $\alpha_0$. For a two dimensional problem of this nature, MCMC techniques are not necessary for posterior recovery, particularly when the likelihood takes such a simple analytic form. However, numerical integration is still required to marginalise the posteriors and MCMC techniques provide a quick and efficient method for doing that. Using MCMC methods in this simple problem also allowed an easier extension to more complicated models, such as the redshift-dependent mass function that we will describe later. The recovered 1D posterior distributions for a typical case are shown in Figure~\ref{nonevolv}, and the corresponding 2D posterior (correlation plot) is shown in Figure~\ref{nonevolvCorr}. These posteriors are typical of what we would obtain from analysis of the LISA data. Both $A_0$ and $\alpha_0$ will be measured relatively precisely, and we see that the true parameters, $A_0 = 0.002$Mpc$^{-3}$, $\alpha_0=0$, are consistent with the mean and width of these posteriors. Figure~\ref{nonevolvCorr} indicates that $A_0$ and $\alpha_0$ are correlated. This correlation corresponds to keeping the number of black holes with $M\sim5\times10^5M_{\odot}$--$10^6M_{\odot}$ approximately constant. LISA is most sensitive to events of this type, and any black hole distribution that predicts approximately the correct number of such events will be reasonably consistent with the data.

\begin{figure}[ht]
\begin{tabular}{ccc}
\includegraphics[width=0.33\textwidth,keepaspectratio=true]{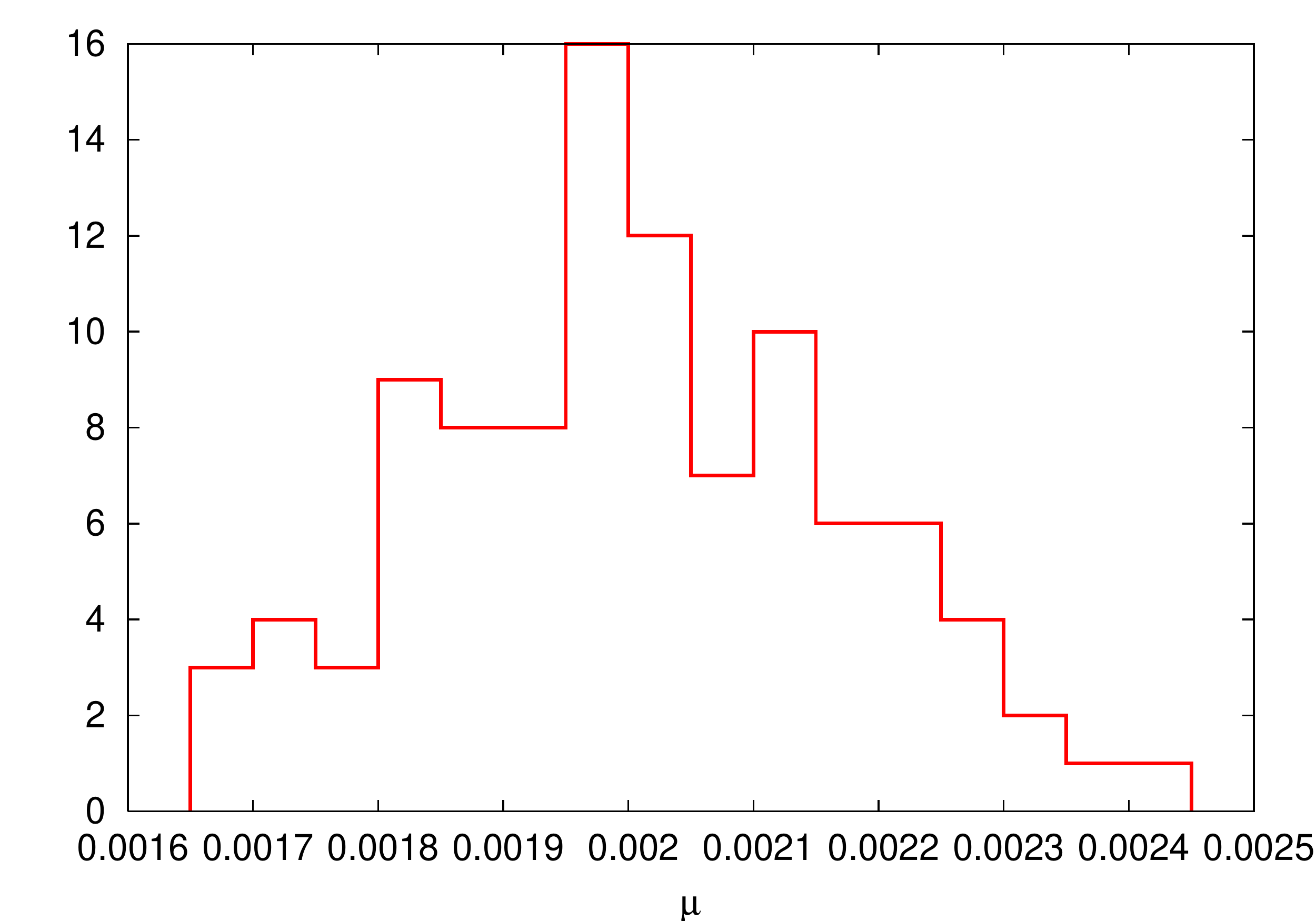}&
\includegraphics[width=0.33\textwidth,keepaspectratio=true]{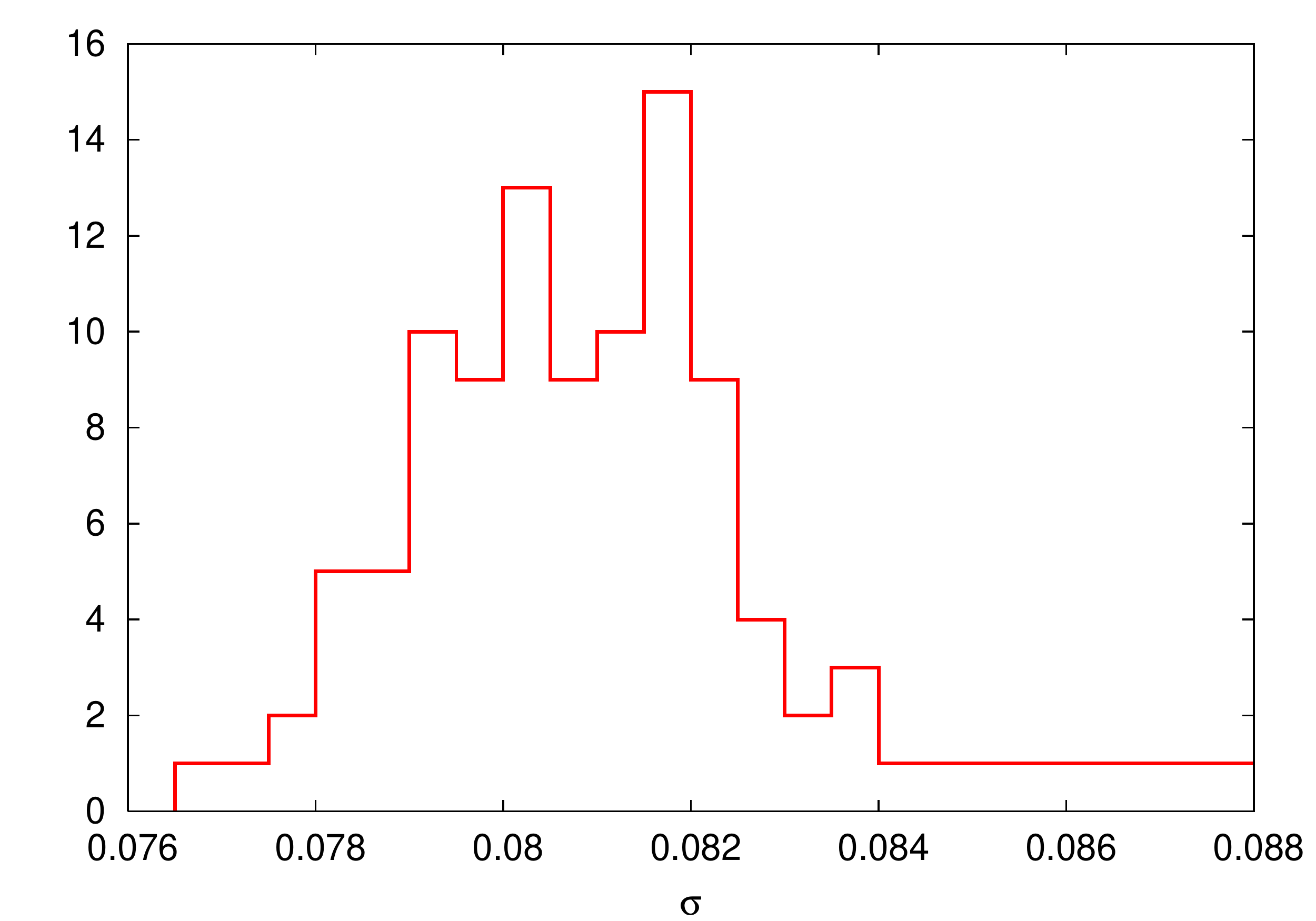}&
\includegraphics[width=0.33\textwidth,keepaspectratio=true]{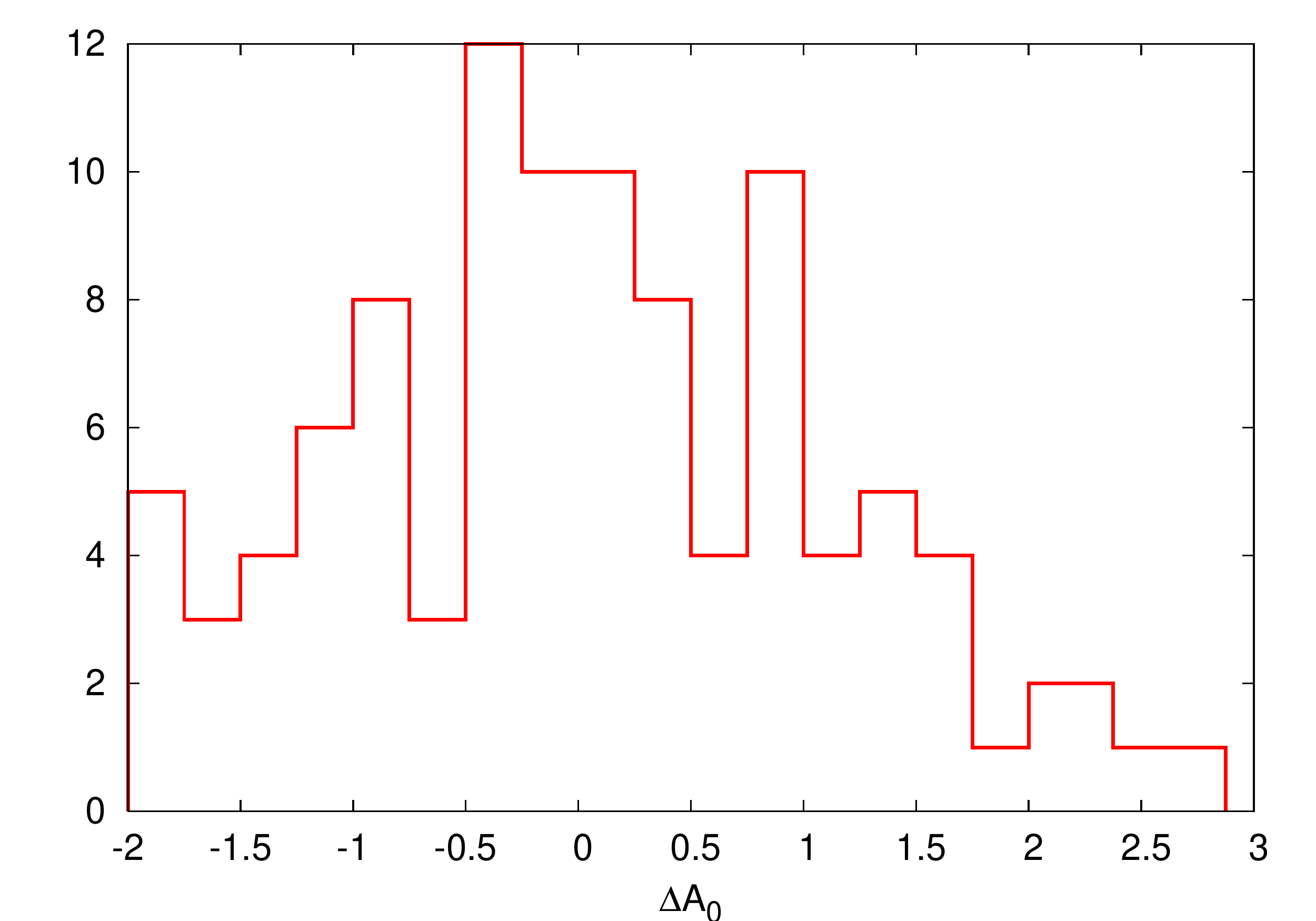}\\
\includegraphics[width=0.33\textwidth,keepaspectratio=true]{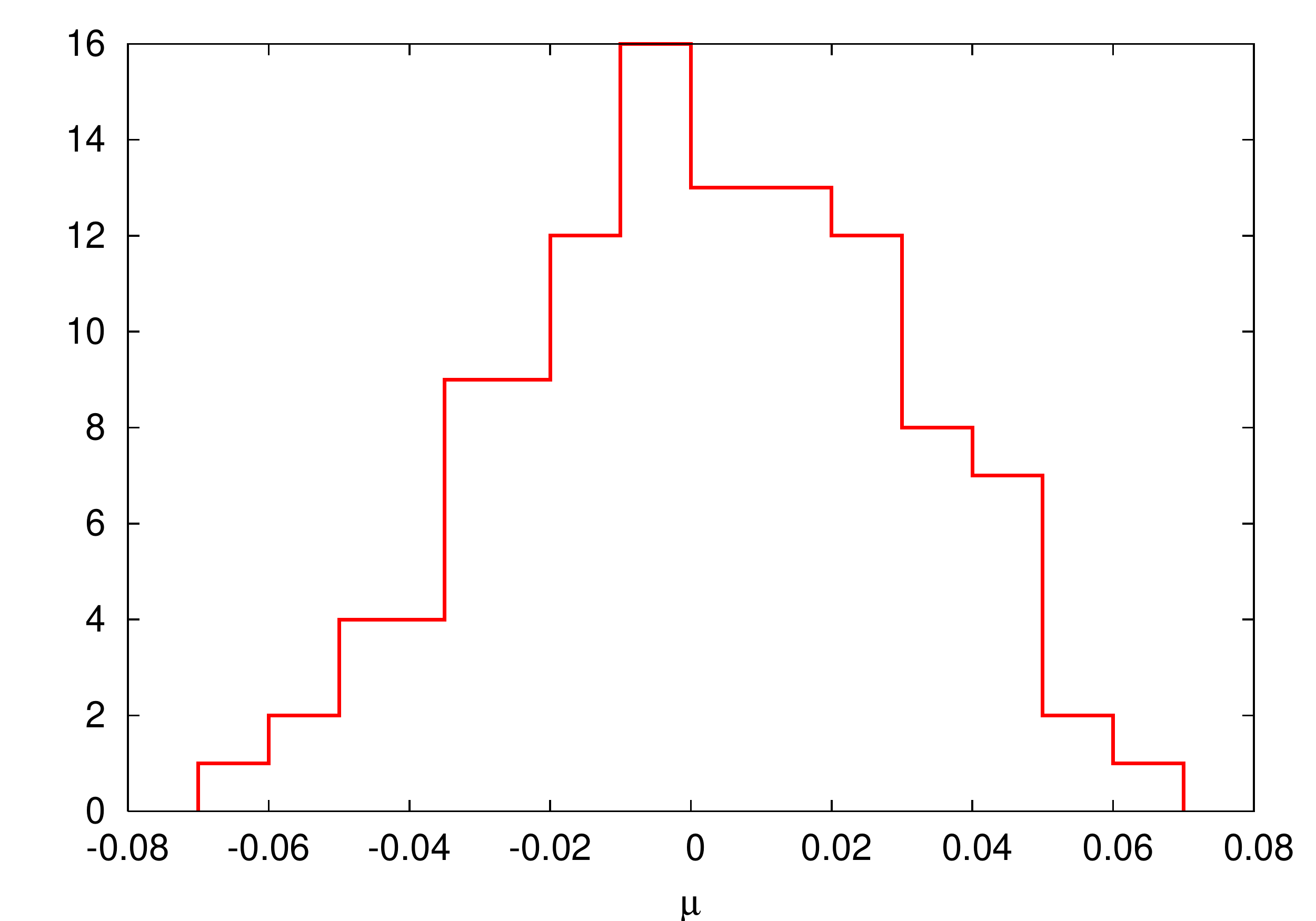}&
\includegraphics[width=0.33\textwidth,keepaspectratio=true]{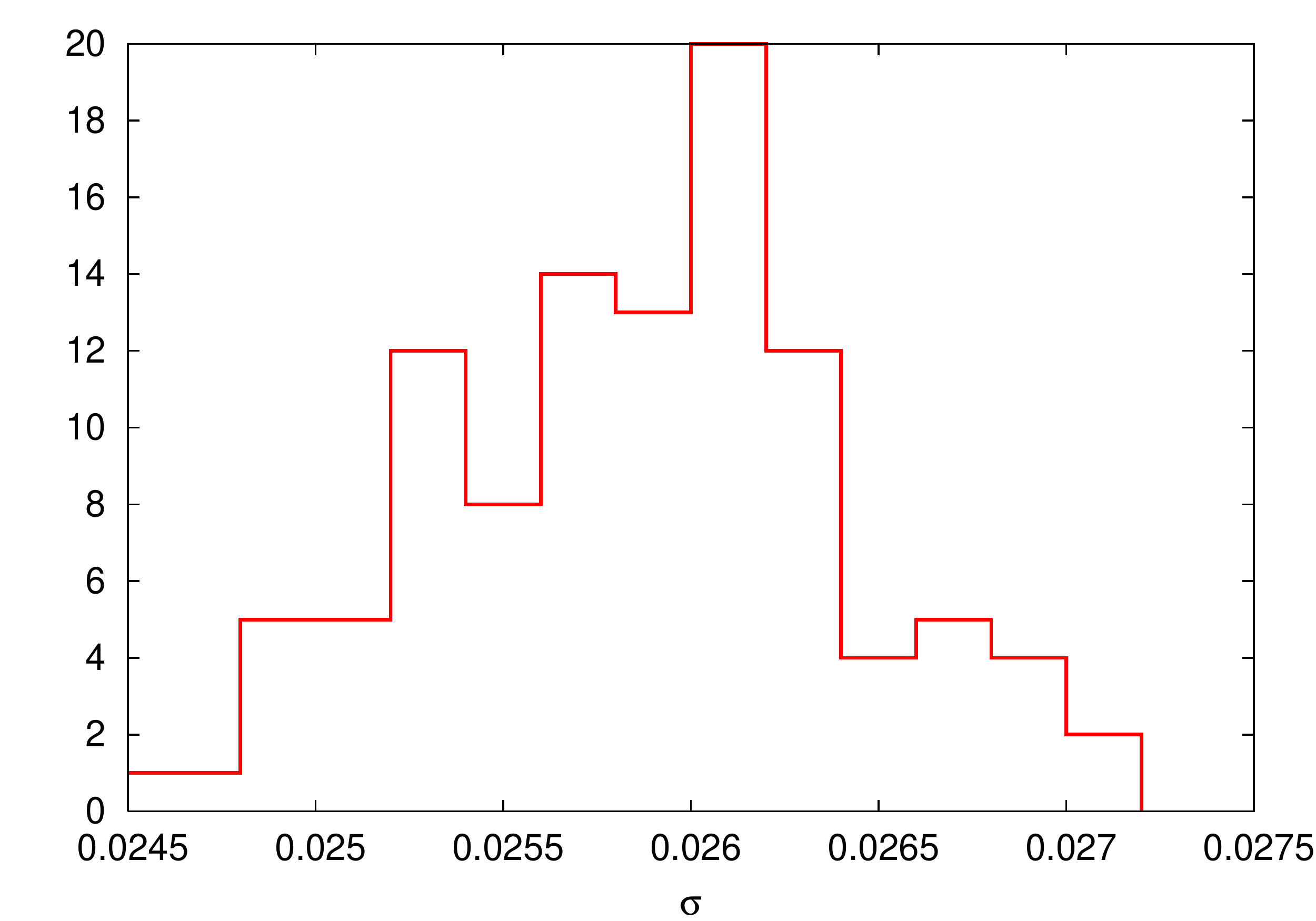}&
\includegraphics[width=0.33\textwidth,keepaspectratio=true]{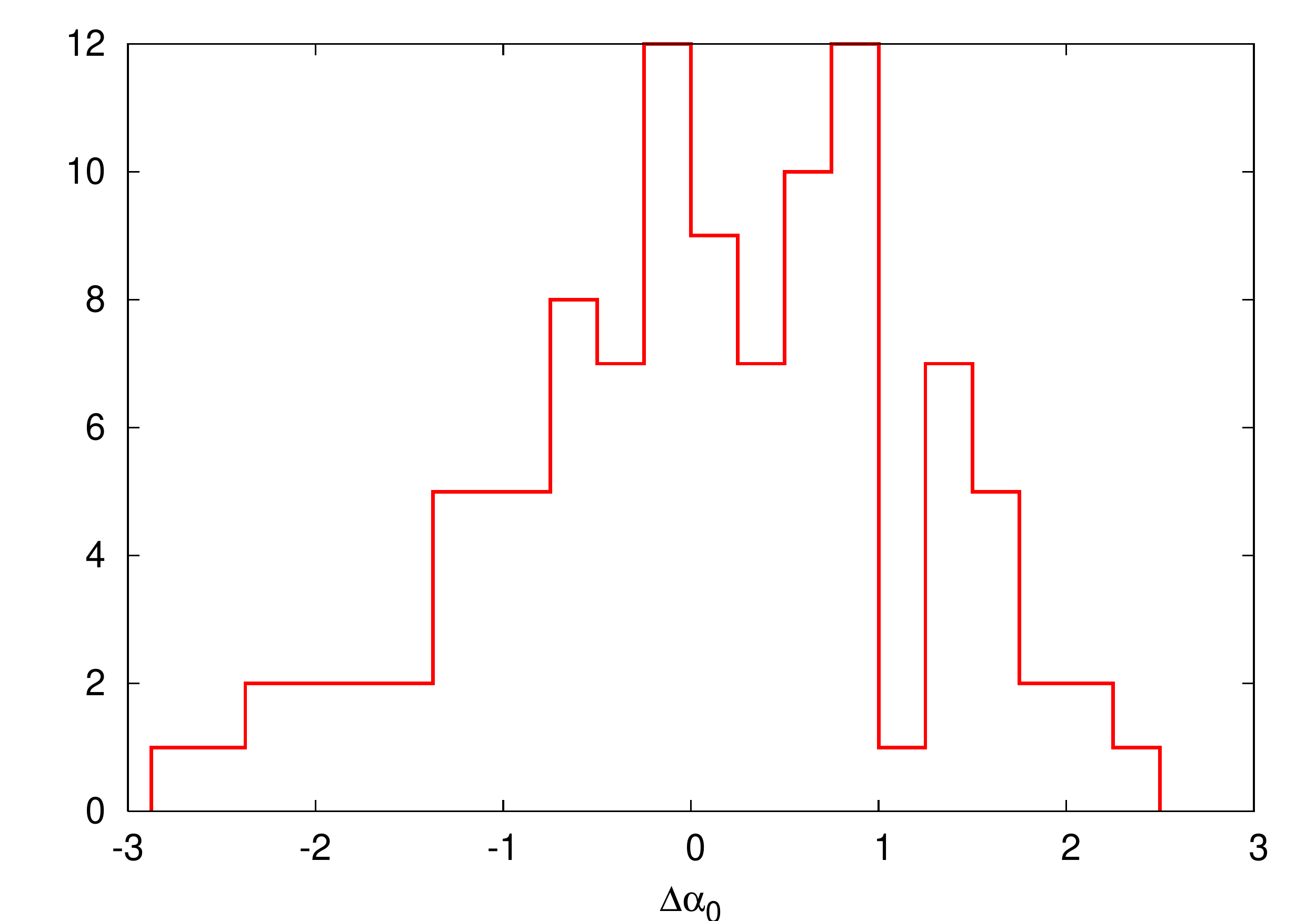}
\end{tabular}
\caption{Distribution of the mean (left) and standard deviation (middle) of Gaussian fits to the posterior for $\ln A_0$ (upper row) and $\alpha_0$ (lower row) over $100$ realisations of the set of observed EMRI events. In the right hand figures in each row we show the distribution of the ``error'' in the mean of the distributions, i.e., $\Delta A_0 = (\mu-A_0)/\sigma$ or  $\Delta \alpha_0 = (\mu-\alpha_0)/\sigma$.}
\label{MCresAalp}
\end{figure}

We find that the posterior distributions for both $\ln A_0$ and $\alpha_0$ are well fit by Gaussians, which can be found by using a linear least squares fit of a quadratic to the logarithm of the distribution. The best-fit Gaussians are also shown in~Figure~\ref{nonevolv}. These Gaussian fits provide a useful way to characterize different realisations of the set of EMRI events, drawn from the same underlying galaxy black hole distribution. For a given realisation we can compute the posterior, obtain a Gaussian fit of the form $B\exp\{-(x-\mu)^2/2\sigma^2\}$ and record the mean, $\mu$, and standard deviation, $\sigma$ of the Gaussian. We can also compute the ``error'' in the mean, namely $(\mu - X)/\sigma$, where $X$ is the value of the parameter ($A$ or $\alpha$) used to generate the underlying distribution. We would expect the true value to lie within one or two standard deviations of the mean, and so this last quantity should be small. In 
Figure~\ref{MCresAalp} we show the distribution of these three quantities for both $A_0$ and $\alpha_0$ over $100$ realisations of the measured EMRI distribution.

The distribution of the Gaussian means resembles the posterior of the parameter in each case, and the distribution of the errors looks approximately Normal. This is what we would expect, since we are using the same model to construct the data set and to analyse the data. The more interesting quantity is the standard deviation, since this is a measure of the accuracy to which we can measure the parameters of the underlying distribution. We see that the distribution of standard deviations is fairly narrow for both $\ln A_0$ and $\alpha_0$, and we deduce that, for this choice of $A_0$ and $\alpha_0$, we would be able to measure $\ln A_0$ to a precision of $0.08$, and $\alpha_0$ to a precision of $0.025$. In subsequent sections when we consider the effects of measurement errors, bin size and different choices for $A_0$ and $\alpha_0$, we will use $\sigma$ to characterise the measurement precision.

\subsubsection{Inclusion of parameter errors}
\label{errres}
As mentioned earlier, the mass and redshift of each source that LISA detects will be subject to measurement errors. In Section~\ref{parerr}, we described how these errors could be included in the analysis in a consistent manner, given properly sampled posterior probability distributions (pdfs) for each of the detected sources. In the current analysis, we wish to avoid the overhead of recovering such pdfs and have been taking the parameter values of each source at face value. In order to verify that our results are insensitive to these errors, we also carried out an analysis in which an error was added to the parameters ($M$ and $z$) of each source after they were generated, but before they were assigned to a particular bin. We then carried out the analysis of the data set including errors in the same way as before.

\begin{figure}[ht]
\begin{tabular}{cc}
\includegraphics[width=0.33\textwidth,keepaspectratio=true]{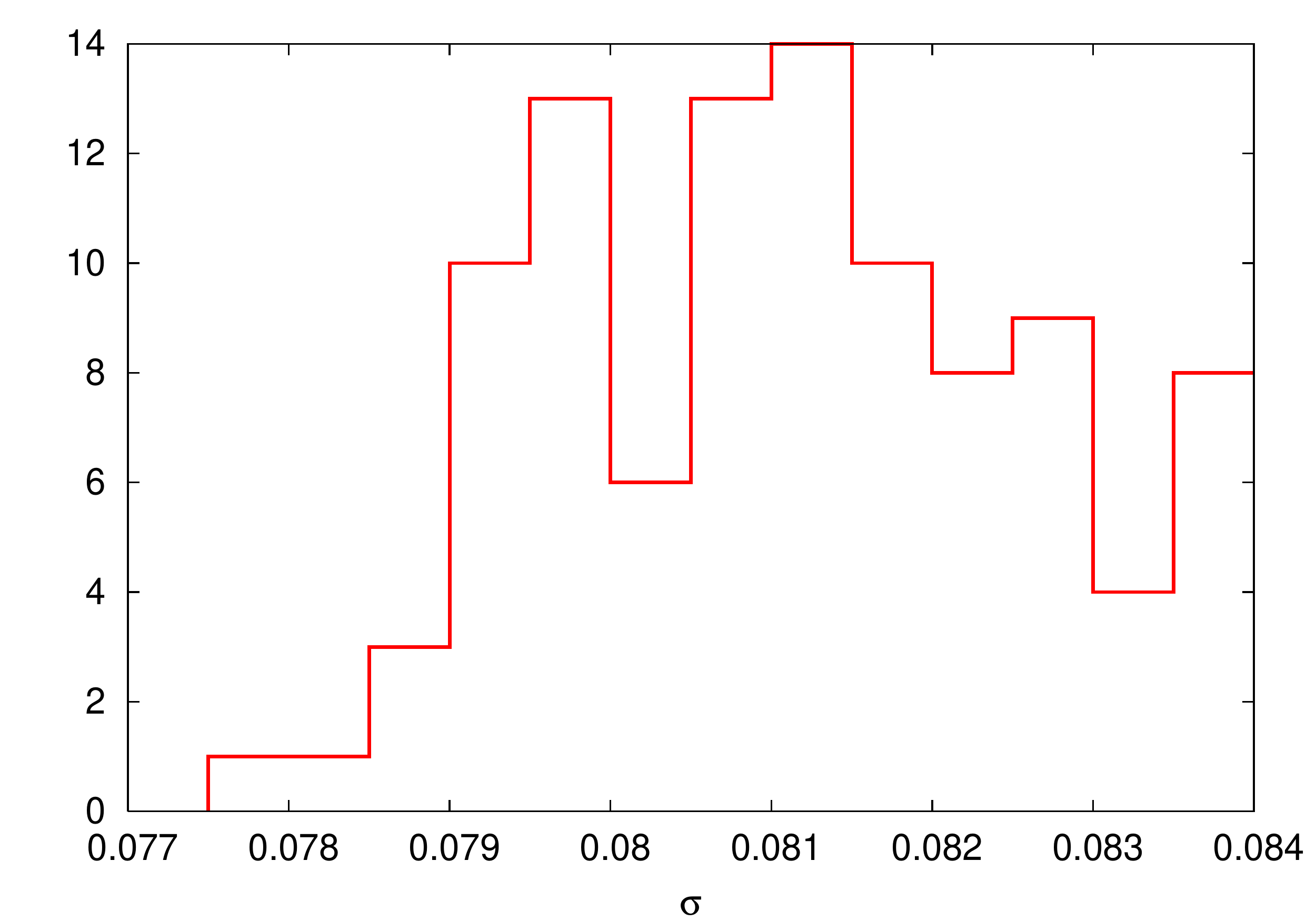}&
\includegraphics[width=0.33\textwidth,keepaspectratio=true]{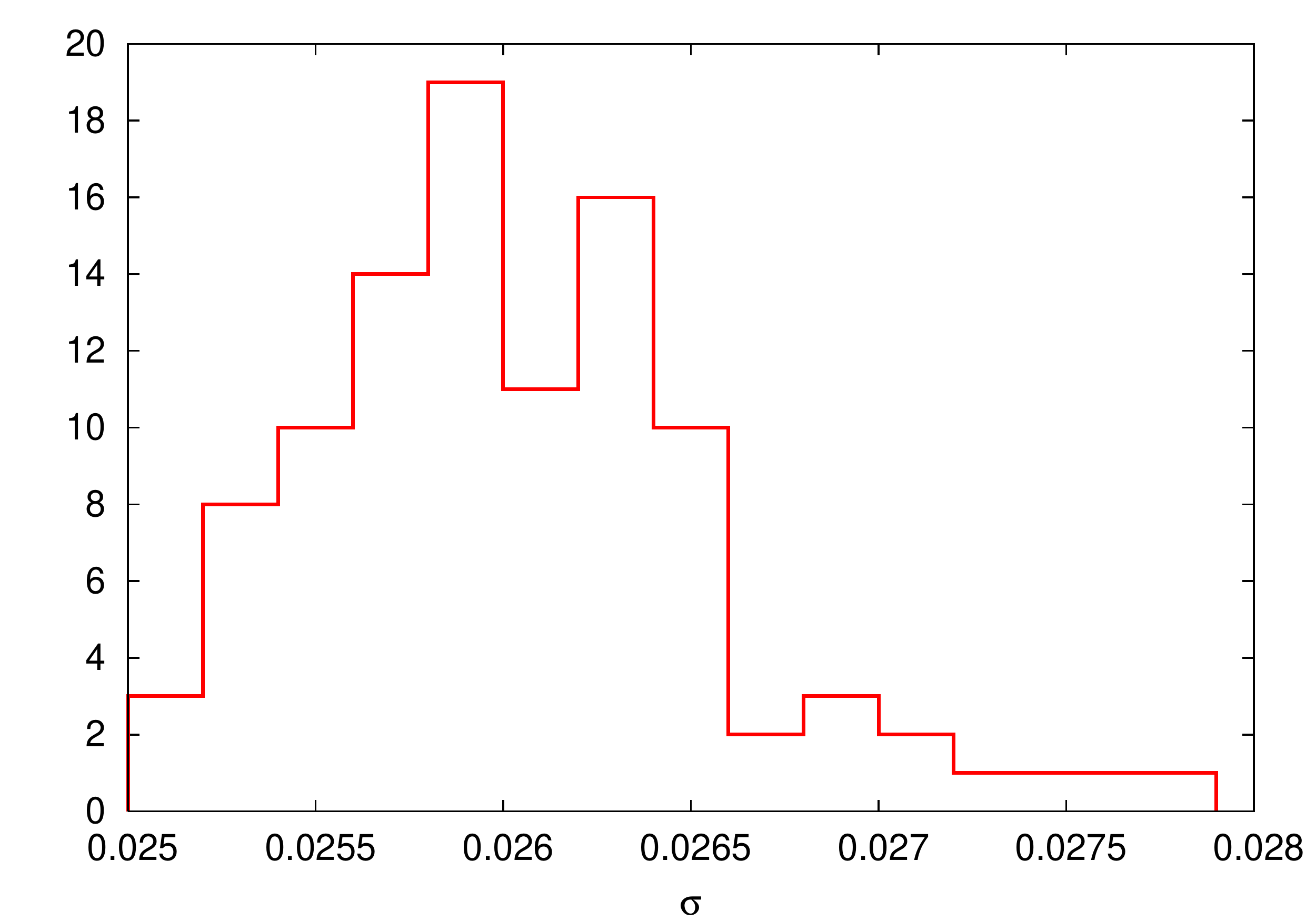}
\end{tabular}
\caption{Distribution of the standard deviation of a Gaussian fit to the posterior for $\ln A_0$ (left) and $\alpha_0$ (right) over $100$ realisations of the set of observed EMRI events, but now including measurement errors in the construction of the observed event set.}
\label{MCresWithErrAalp}
\end{figure}

Following~\cite{MacLeod08}, we took the error in the redshift to be $\Delta(\ln z) = 0.05z$ for a source at redshift $z$. The error in the redshifted total mass of the source, $M_z$, will be small~\cite{AK} in all cases, but we need the intrinsic mass of the source which is $M_z/(1+z)$. The error in redshift will therefore propagate into the mass also, and so we included a mass error of the same magnitude,  $\Delta(\ln M) = 0.05z$. The errors for each source were drawn from Gaussians with these widths. Over 100 realisations of the LISA event distribution, we computed Gaussian fits to the posteriors. We found that the distribution of the standard deviation, mean and ``error'' of these fits were indistinguishable from the results presented in Figure~\ref{MCresAalp} when measurement errors were ignored. In Figure~\ref{MCresWithErrAalp} we show the distribution of the standard deviations found in this way. These results indicate that it is reasonable to ignore parameter errors when analysing the results in this way and so in all subsequent parts of this paper we will ignore parameter errors when generating the observed EMRI distribution.

\subsubsection{Dependence on bin size}
\begin{figure}[ht]
\begin{tabular}{cc}
\includegraphics[width=0.33\textwidth,keepaspectratio=true]{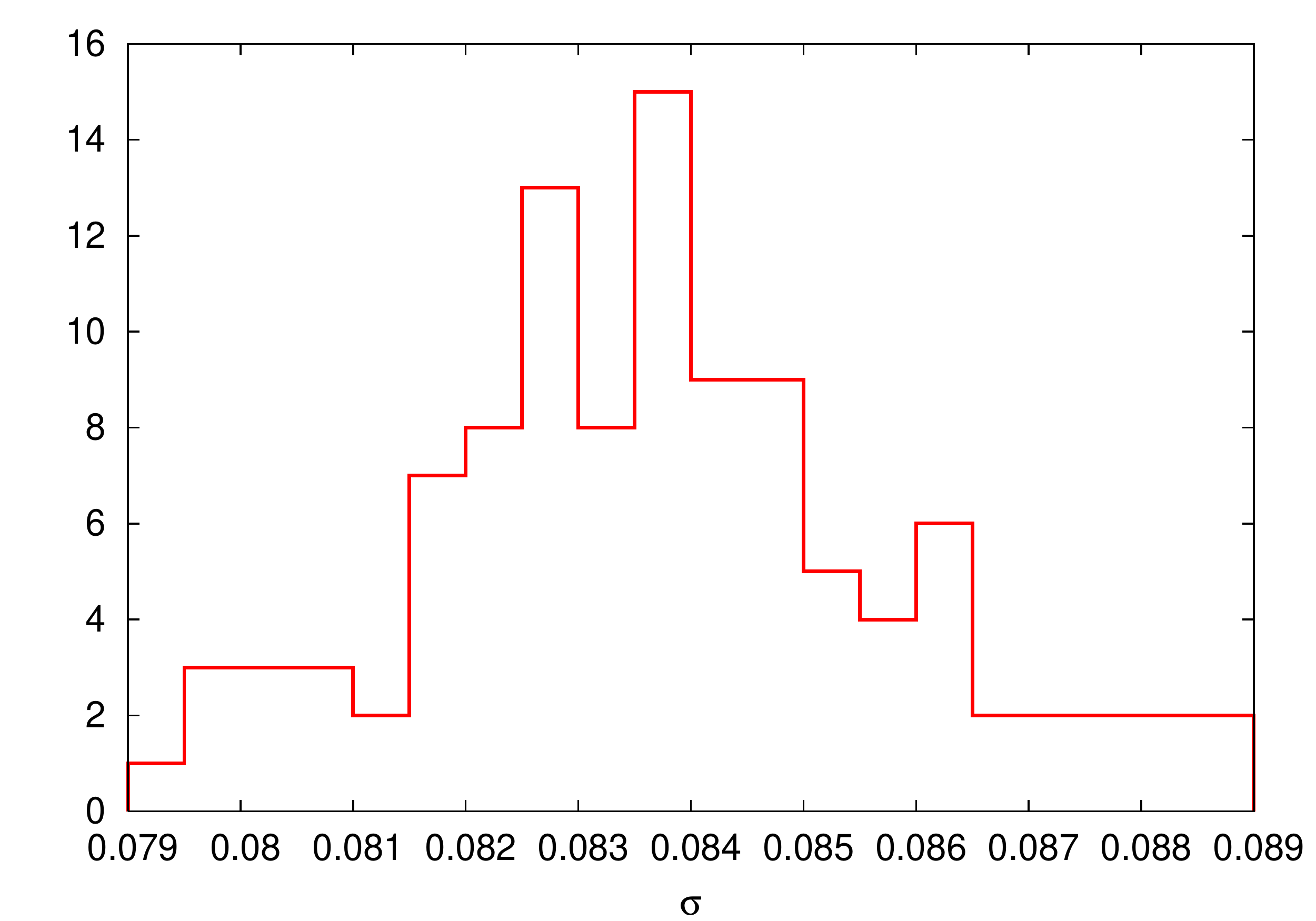}&
\includegraphics[width=0.33\textwidth,keepaspectratio=true]{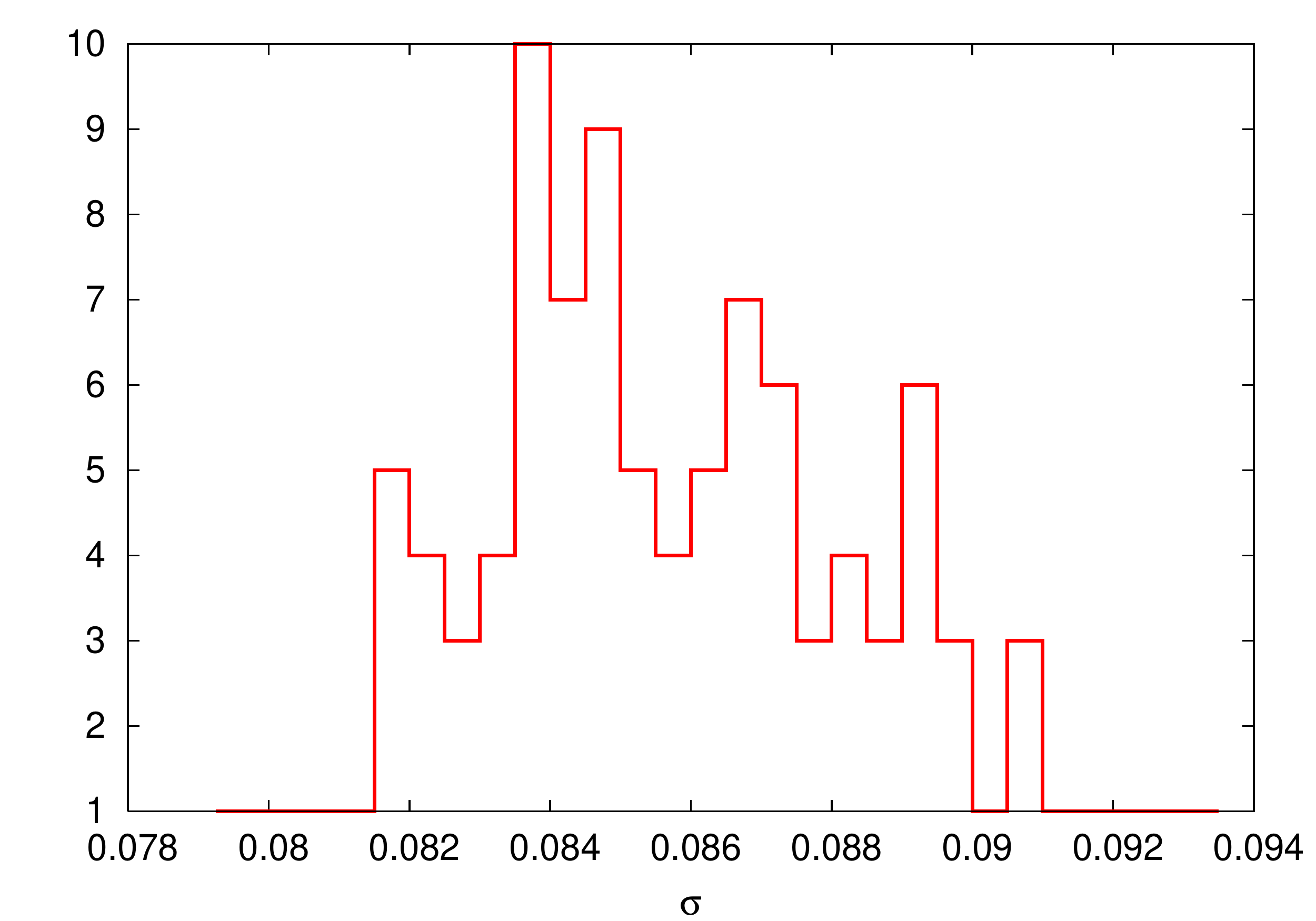}\\
\includegraphics[width=0.33\textwidth,keepaspectratio=true]{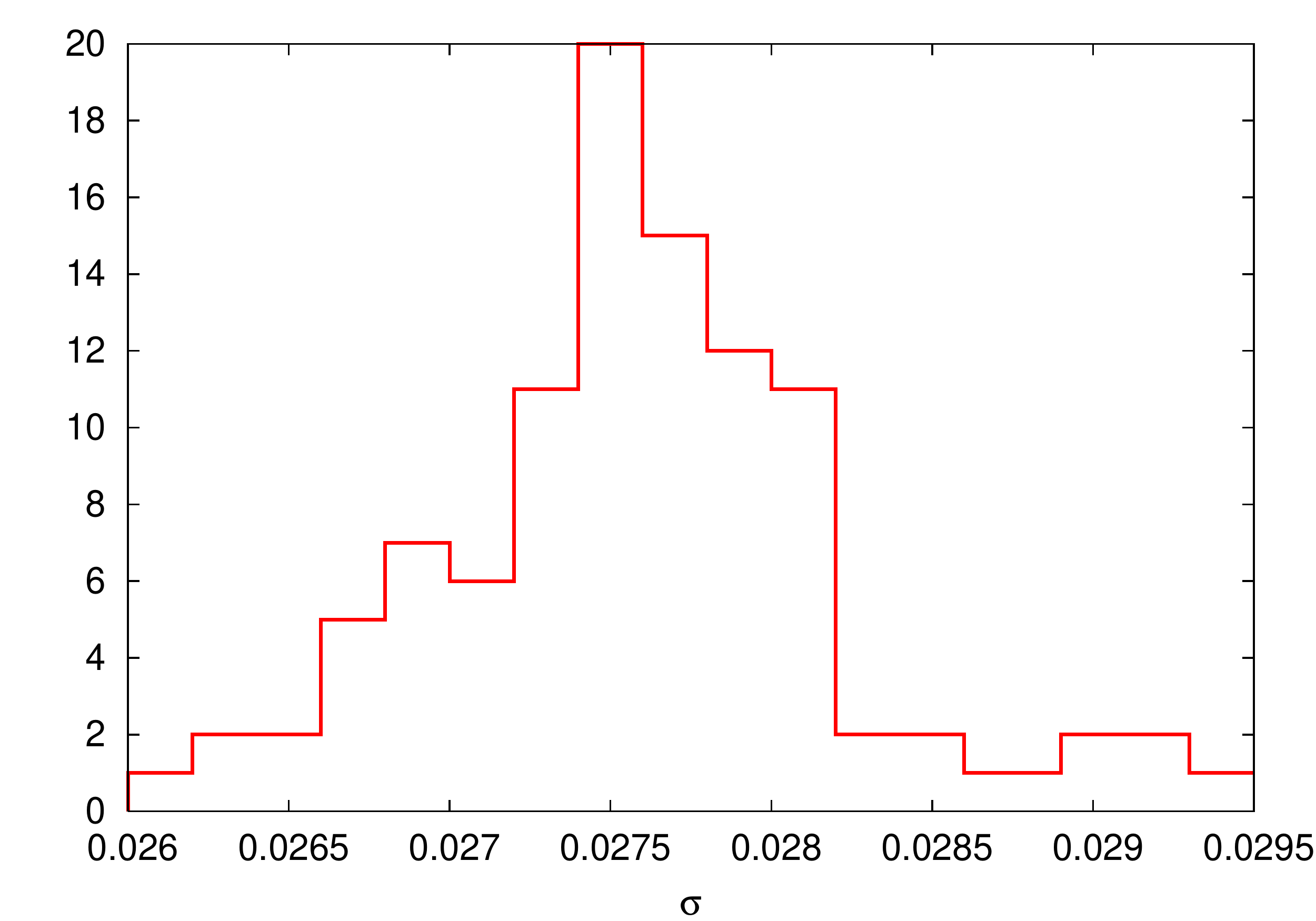}&
\includegraphics[width=0.33\textwidth,keepaspectratio=true]{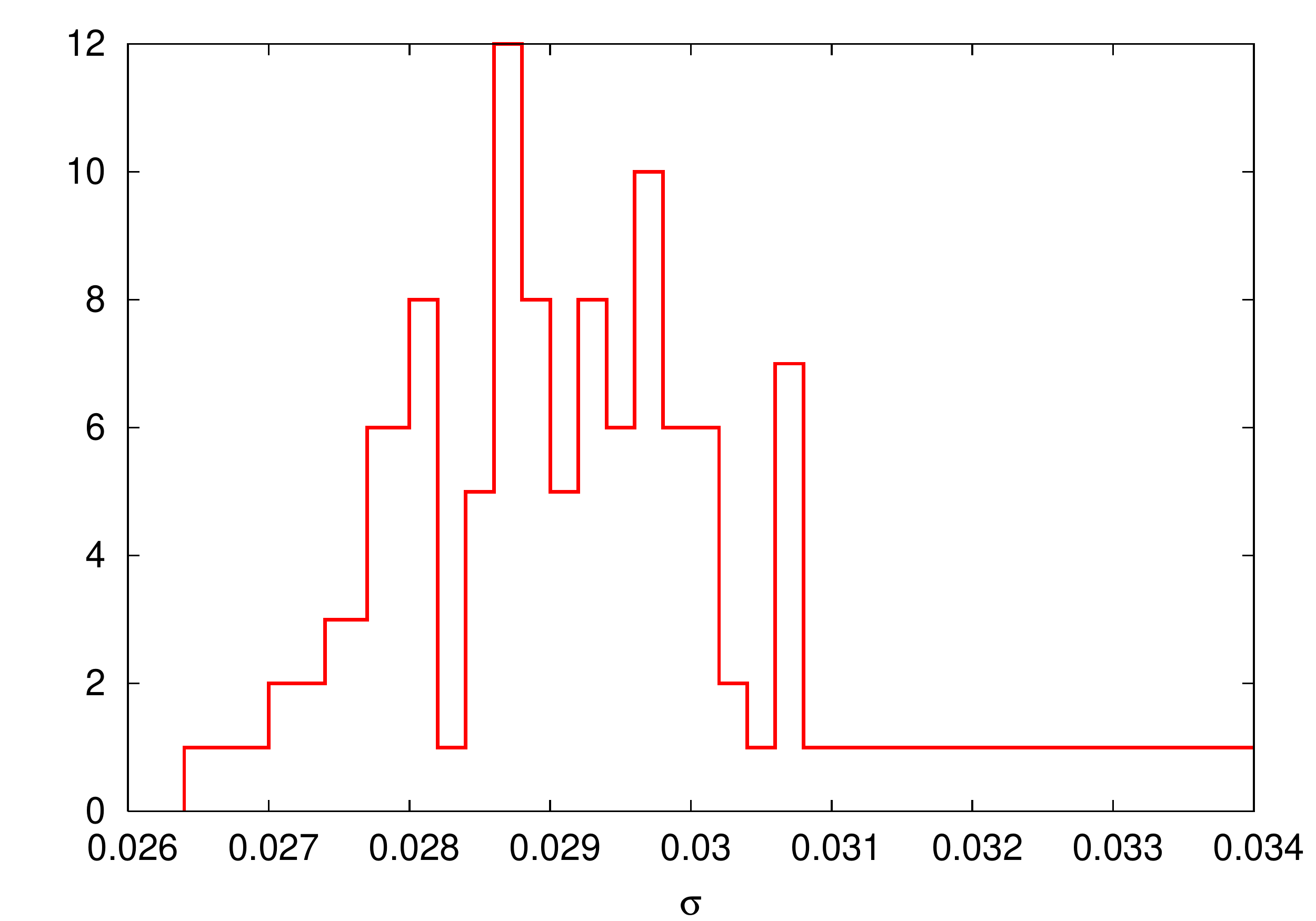}
\end{tabular}
\caption{Distribution of standard deviations of best-fit Gaussians to the recovered posterior over $100$ realisations of the set of observed EMRI events for $\ln A_0$ (top plots) and $\alpha_0$ (bottom plots). These results were obtained using bin sizes one half (left panels) or one quarter (right panels) of the bin size used in the original analysis.}
\label{MCresBinWidth}
\end{figure}

In the preceding section we took bins of particular size in mass and redshift space, but to verify the consistency of the results we explored how things changed as we varied the size of the bins used. We repeated the analysis using bins of one half and one quarter of the width used in the original analysis. In Figure~\ref{MCresBinWidth} we show the distribution of standard deviations of the best-fit Gaussians when using these smaller bin sizes. These do not show particularly significant differences from the original analysis, but the typical widths of the posteriors are slightly larger when using smaller bin sizes. This arises because the number of events in a typical bin is smaller when the bin size is reduced, so if the bins are too small, the generation of the LISA event distribution that we search tends to be noisier. If the bins are too large then we start to lose the ability to discriminate models. Through experimentation we found bin sizes of $\delta z \approx 0.1$ and $\delta \ln M \sim 0.2$ worked well and we used these bin sizes to obtain the results presented elsewhere in this paper.

\subsubsection{Dependence on parameters of observed distribution}
\begin{figure}[ht]
\begin{tabular}{cc}
\includegraphics[width=0.5\textwidth,keepaspectratio=true]{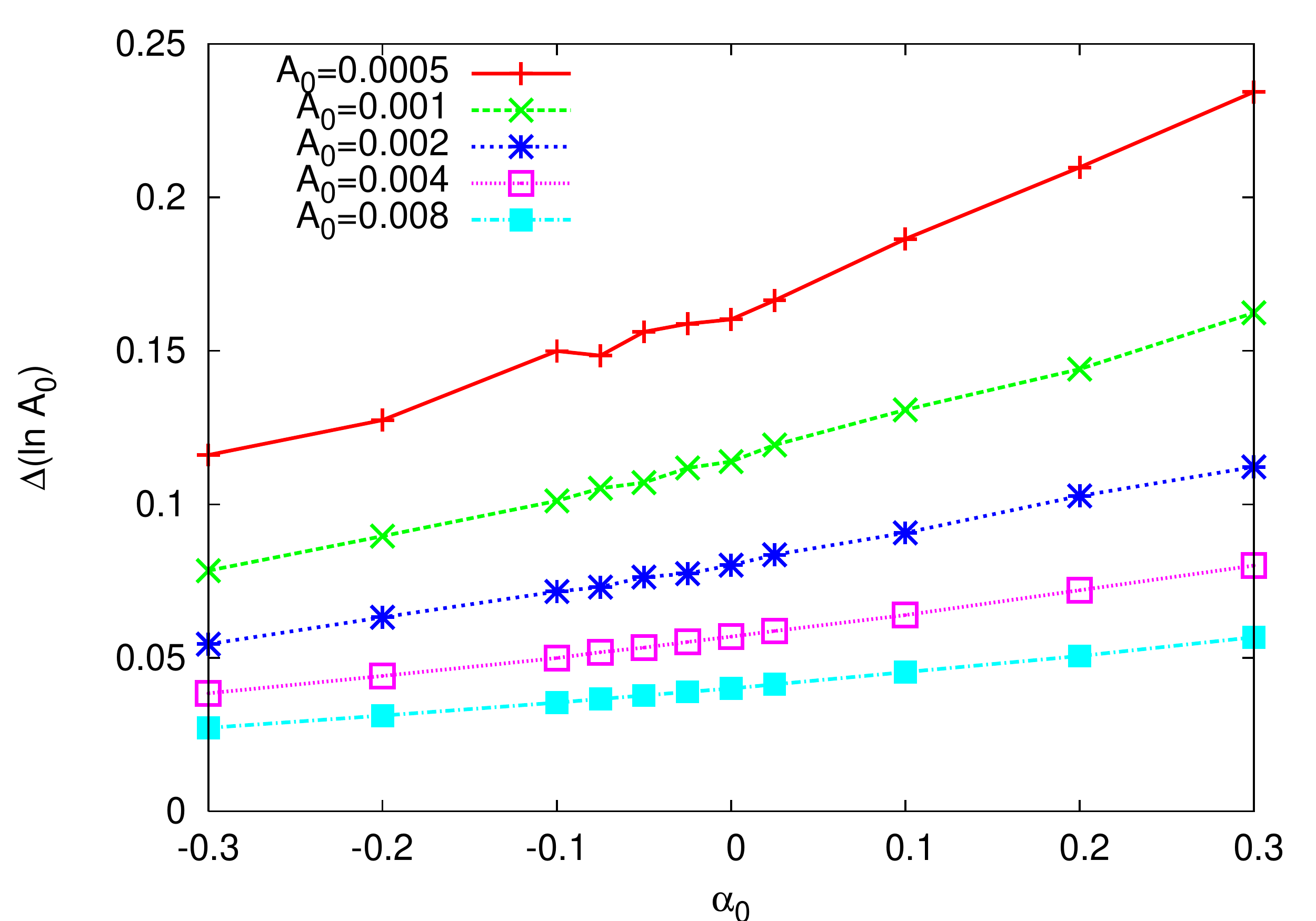}&
\includegraphics[width=0.5\textwidth,keepaspectratio=true]{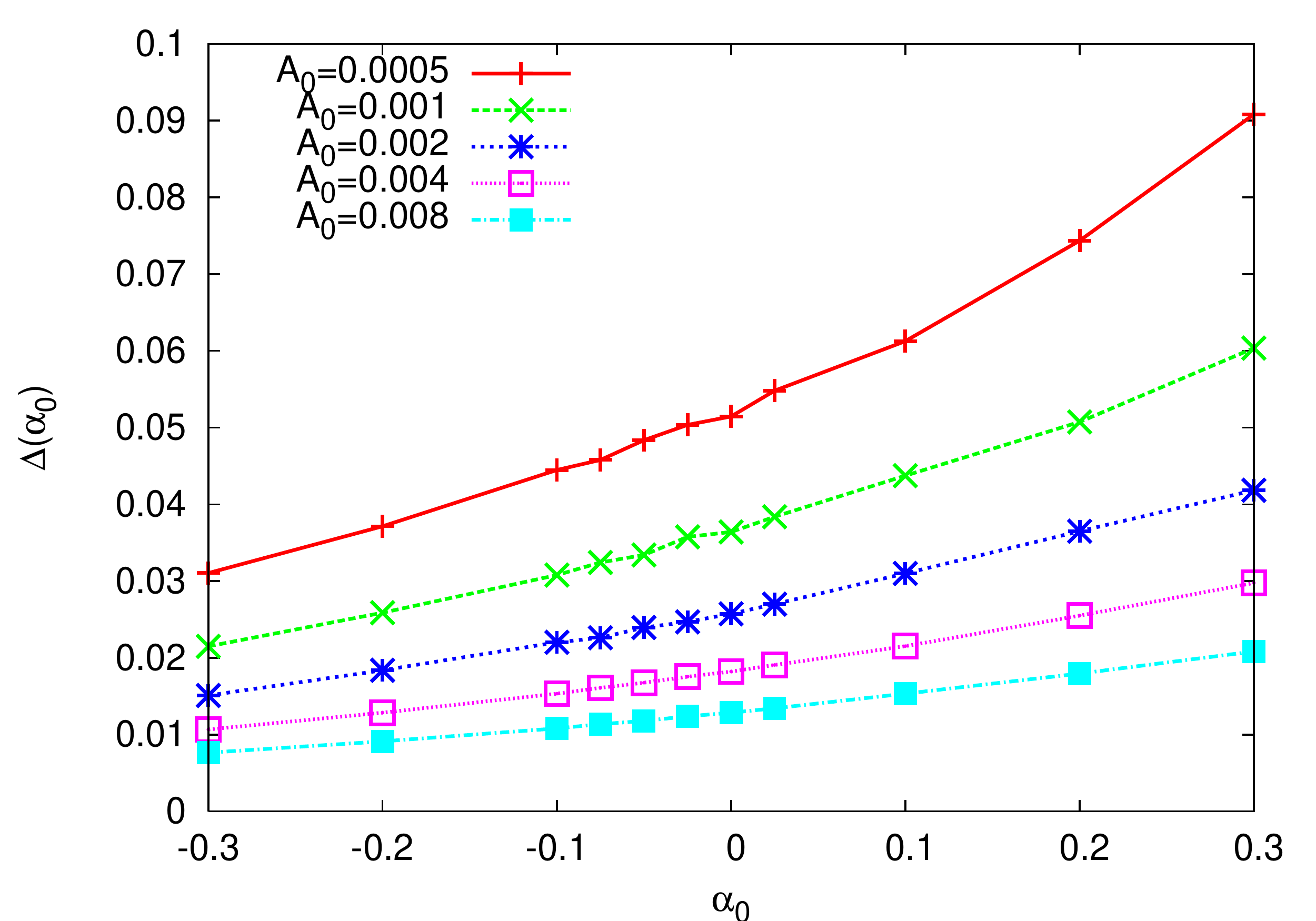}
\end{tabular}
\caption{Typical parameter estimation error as function of the parameters of the underlying distribution. We show the error in $\ln A_0$ (left panel) and in $\alpha_0$ (right panel) as a function of the value of $\alpha_0$, for various choices of $A_0$, as labelled in the key.}
\label{resVarypar}
\end{figure}

\begin{figure}[ht]
\begin{tabular}{cc}
\includegraphics[width=0.5\textwidth,keepaspectratio=true]{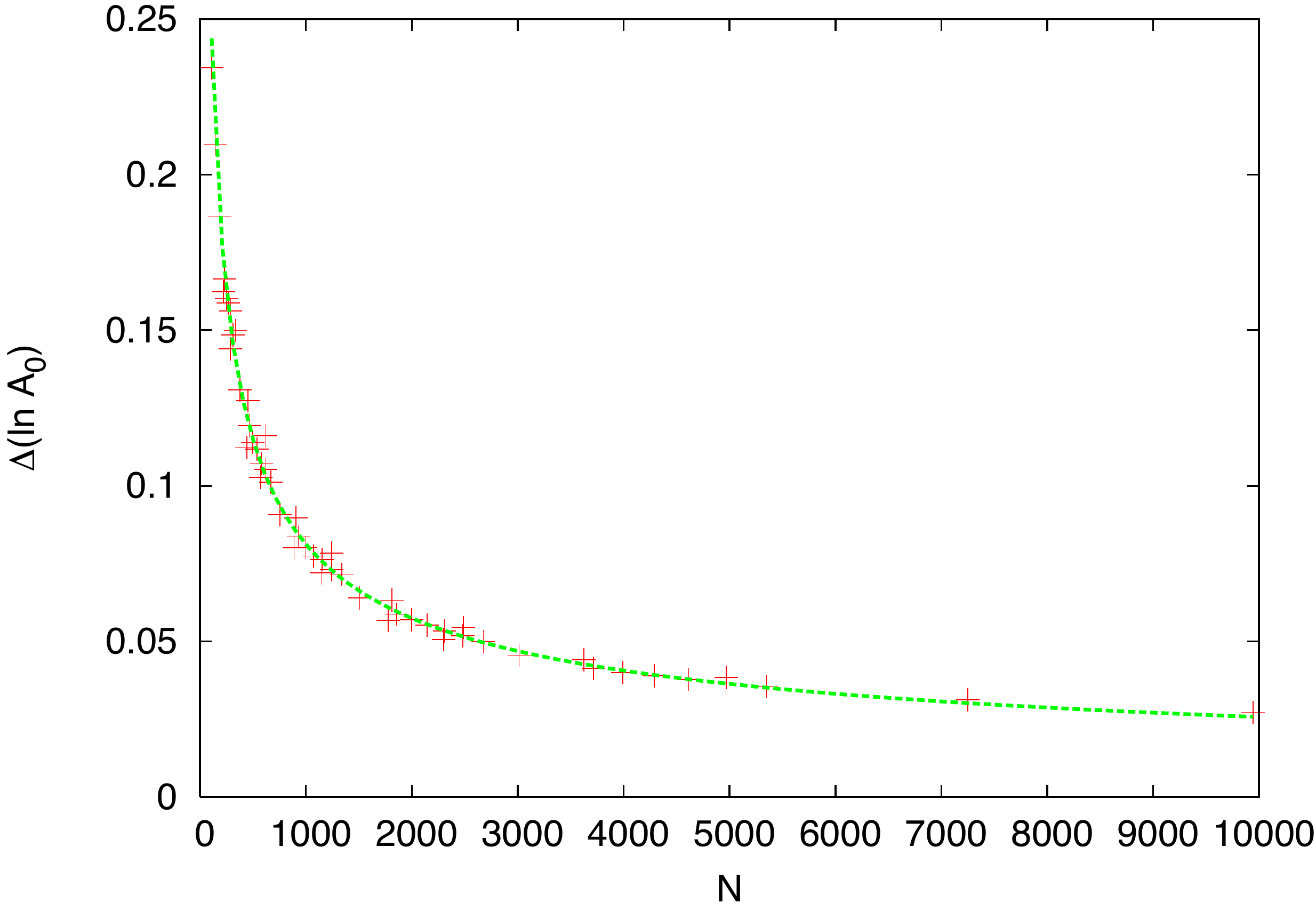}&
\includegraphics[width=0.5\textwidth,keepaspectratio=true]{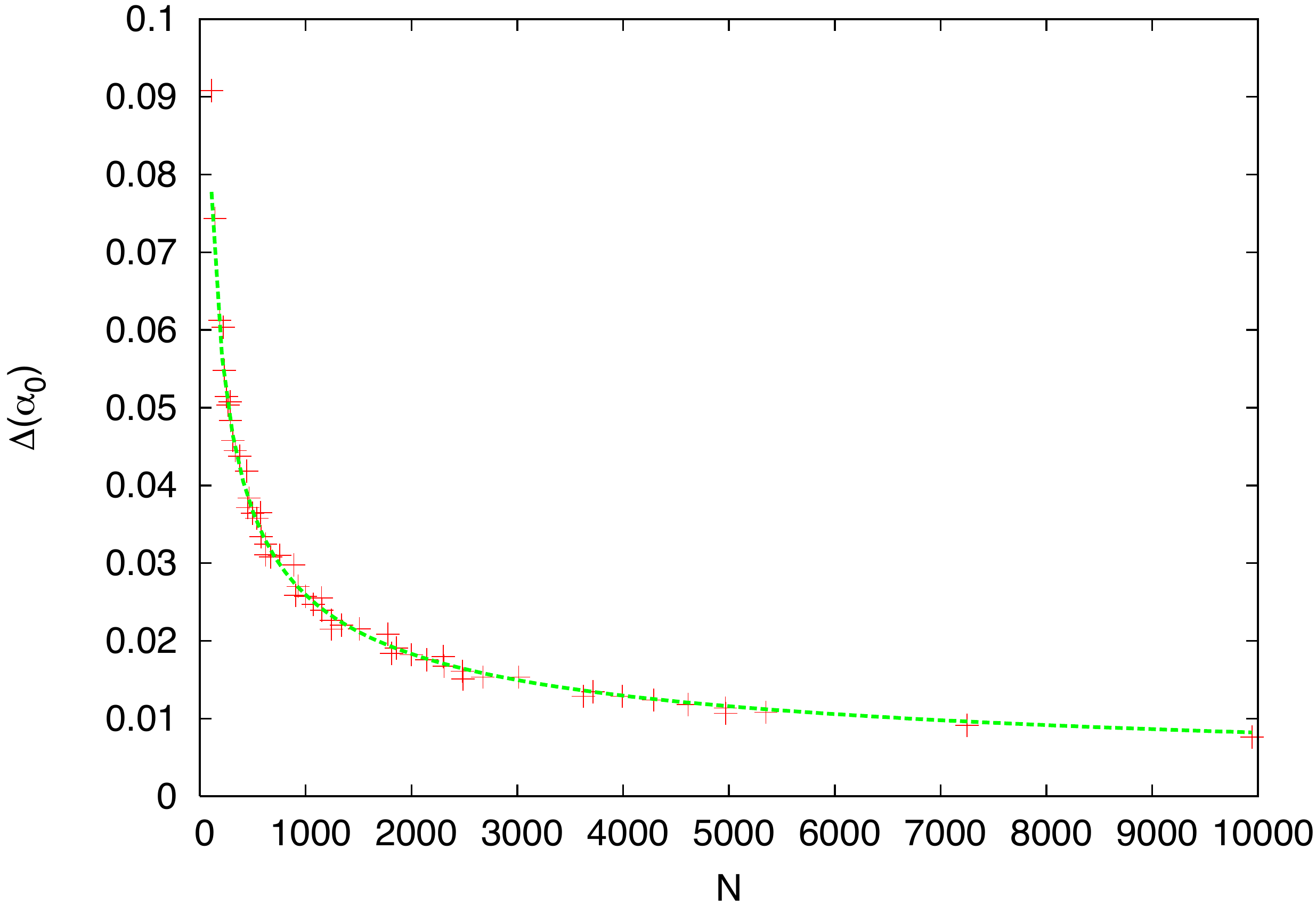}
\end{tabular}
\caption{The same results that are shown in Figure~\ref{resVarypar}, but now plotted as a function of the total number of events that would be observed for the particular model considered. The left panel shows the error in $\ln A_0$, and the right panel shows the error in $\alpha_0$. We also show best-fit curves of the form $kN_{\rm obs}^{-1/2}$.}
\label{ErrorvN}
\end{figure}

To explore the dependence of the error in the recovered parameters on the parameters of the underlying distribution from which the observed events are drawn, we repeated the computation for a range of choices of $A_0$ and $\alpha_0$. In Figure~\ref{resVarypar} we show how the typical parameter errors vary with these choices. The parameter errors were estimated by averaging the standard deviations of the best-fit Gaussians over 10 realisations of the observed EMRI events. If the relative number of different types of events are equal, then the total number of events observed, $N_{\rm obs}$, plays the role of the signal to noise ratio squared, and we expect the error in the recovered parameters to scale with one over the signal-to-noise ratio. Therefore, we would expect the accuracy with which we can measure the parameters to scale roughly as $N_{\rm obs}^{-1/2}$. The parameter $A_0$ is an overall normalisation and so we would expect the change in the number of events to completely explain the dependence of the error estimate on this parameter. The slope, $\alpha_0$, also affects the relative numbers of events of different types and therefore there could be other effects in that case. We can factor out the dependence on $N_{\rm obs}$ by plotting the errors as a function of the number of events. In Figure~\ref{ErrorvN} we replot all of the data shown in Figure~\ref{resVarypar}, but now with $N_{\rm obs}$ on the horizontal axis. We also show best-fit lines of the form $k N_{\rm obs}^{-1/2}$. The value of $N_{\rm obs}$ for each case can be obtained from fits given in~\cite{gairEMRIastro}.

It is clear from Figure~\ref{ErrorvN} that the change in $N_{\rm obs}$ between different models explains virtually all of the change in the precision with which we can determine the model parameters. The best-fit curves give us the nice rule of thumb that $\Delta(\ln A_0) \approx 0.08 \sqrt{1000/N_{\rm obs}}$ and $\Delta (\alpha_0) \approx 0.025 \sqrt{1000/N_{\rm obs}}$ when we make optimistic assumptions about LISA and assume that all the central black holes have spin $a=0$. In the next section we will see what happens when we relax the latter two assumptions.

\subsubsection{Dependence on LISA performance and black hole spin}
\label{varyLISA}
\begin{figure}[ht]
\begin{tabular}{cc}
\includegraphics[width=0.5\textwidth,keepaspectratio=true]{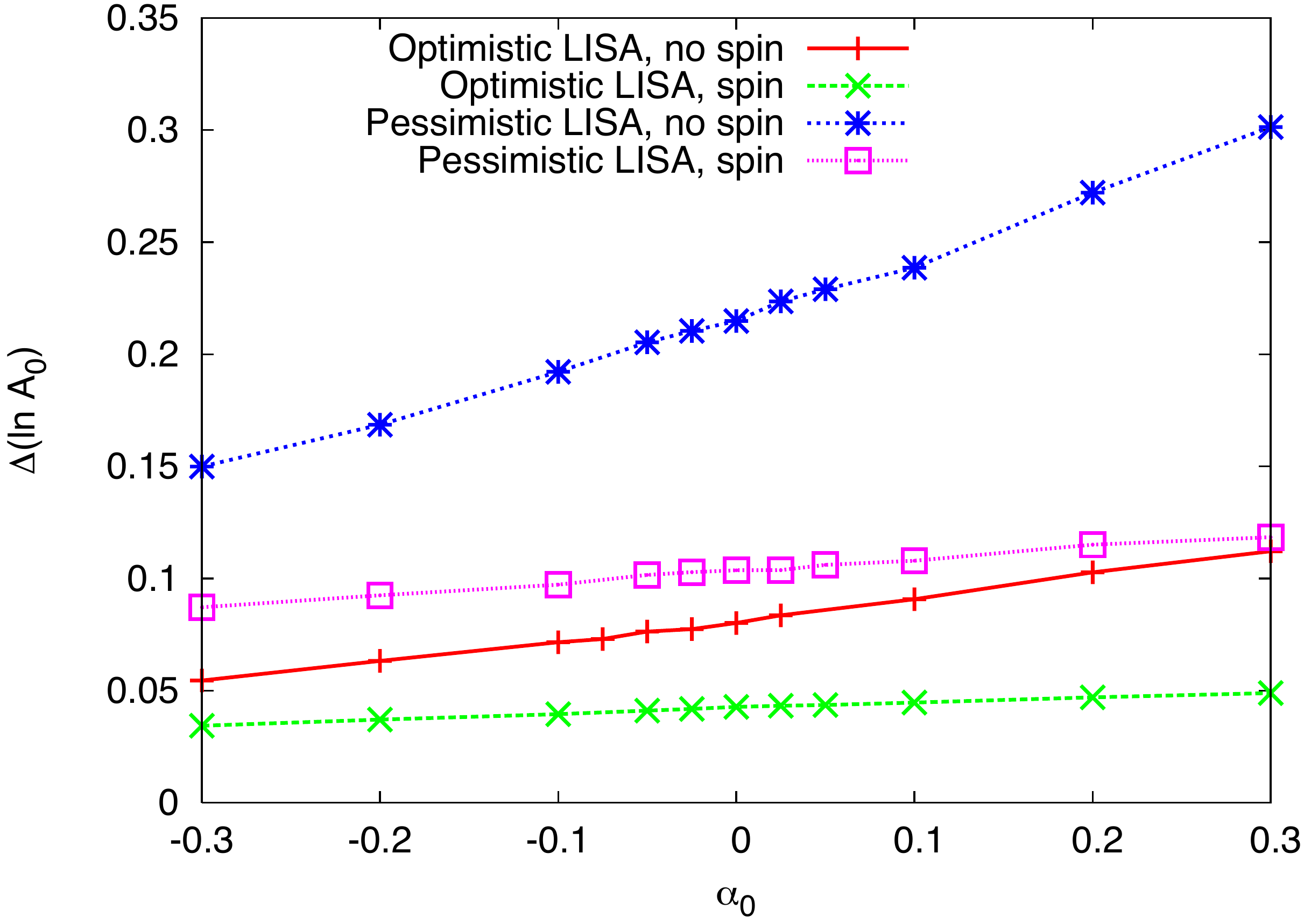}&
\includegraphics[width=0.5\textwidth,keepaspectratio=true]{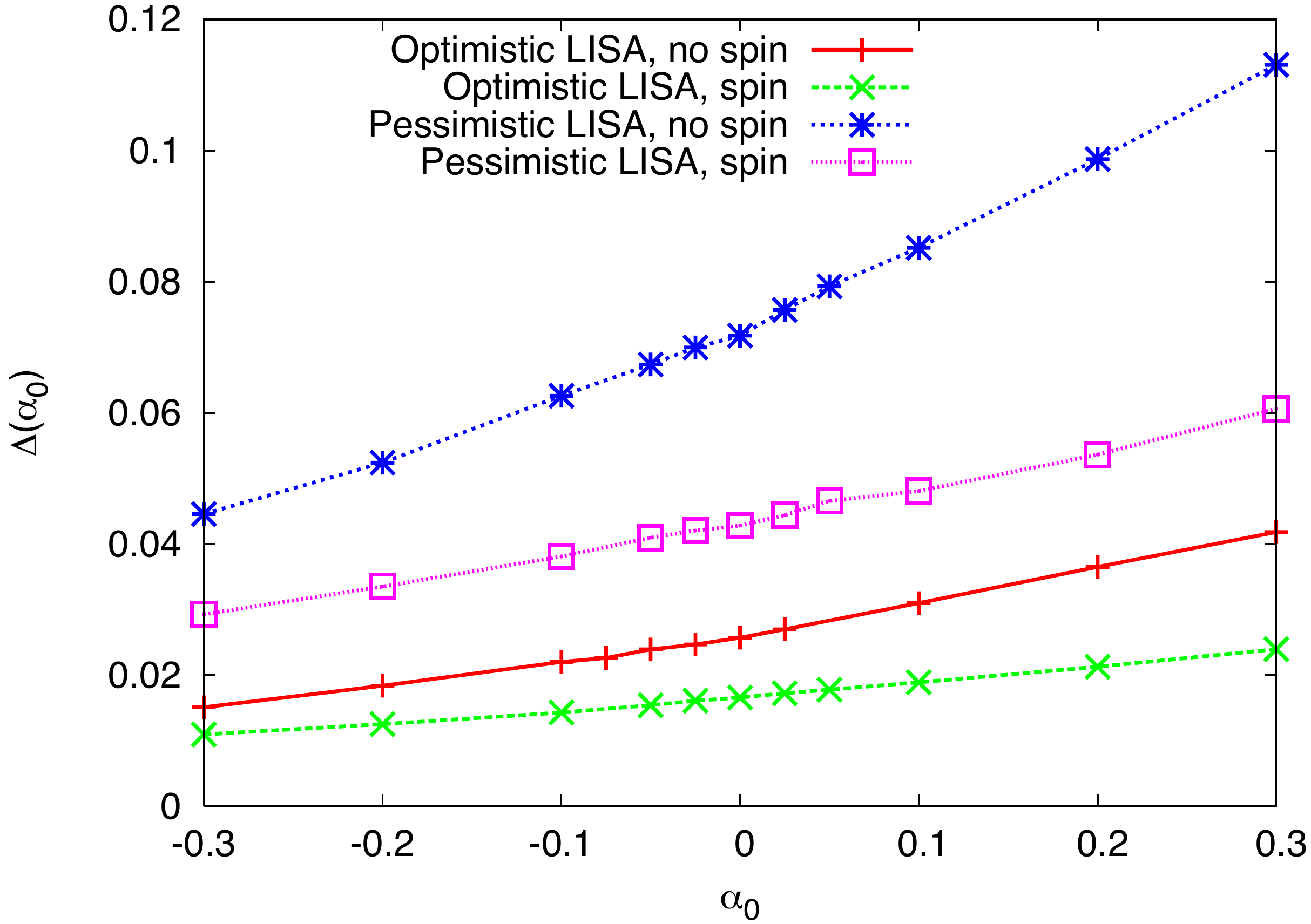}
\end{tabular}
\caption{As Figure~\ref{resVarypar}, but now showing results for various assumptions about LISA and about the spin of the central black holes of the EMRI sources. In all cases we have taken $A=0.002$Mpc$^{-3}$ and different curves correspond to different assumptions, as labelled in the key. Cases labelled ``spin'' assumed all black holes have spin $a=0.9$.}
\label{resAllLISA}
\end{figure}

All of the preceding results have used optimistic assumptions about the LISA detector and assumed that all the black holes into which EMRIs are falling have spin $a=0$. To relax these assumptions we need only to change the observable lifetime function, since we are not including spin as a parameter in the model. In~\cite{gairEMRIastro}, observable lifetime functions were also given for a pessimistic LISA configuration, and for a case in which all black holes had spin $a=0.9$. In Figure~\ref{resAllLISA} we show how the precision with which we can measure the parameters of the underlying population change when we use these different assumptions. We plot the results for optimistic LISA assumptions and spin $a=0$ for comparison.

Once again, we find that much of the dependence on $A_0$ and $\alpha_0$ can be explained by the change in the number of observed events that the model predicts. We can find best-fits to the dependence of the errors on $N_{\rm obs}$ for each of the different sets of assumptions. As for the $a=0$, optimistic LISA case, this dependence is well fit by $\Delta(\ln A_0) = k_{A_0} \sqrt{1000/N_{\rm obs}}$ and  $\Delta(\alpha_0) = k_{\alpha_0} \sqrt{1000/N_{\rm obs}}$. The parameters $k_{A_0}$ and $k_{\alpha_0}$ for all cases are given in Table~\ref{fitNoz}.
It is clear that the difference between optimistic LISA and pessimistic LISA comes primarily from the change in the number of events we are likely to observe, but if black holes tend to have significant spins we will be able to determine the model parameters more precisely, even with the same number of observed events. This arises because LISA can see spinning black holes over a wider range of redshifts and masses.

\begin{table}
\begin{tabular}{|c|cc|cc|}
\hline
&\multicolumn{2}{c|}{$a=0$}&\multicolumn{2}{c|}{$a=0.9$}\\
&Optimistic LISA&Pessimistic LISA&Optimistic LISA&Pessimistic LISA\\\hline
$k_{A_0}$&0.08&0.084&0.054&0.053 \\
$k_{\alpha_0}$&0.025&0.028&0.021&0.022\\\hline
\end{tabular}
\caption{Parameters describing how the uncertainty in the measured parameters varies with the number of observed events. As described in the text, the dependence is well fit by $\Delta(X)=k_X \sqrt{1000/N_{\rm obs}}$. This table contains the $k_X$ parameters for the two different sets of assumptions about LISA and the two choices of black hole spin, $a$.}
\label{fitNoz}
\end{table}

\subsection{Redshift dependent mass function}
\begin{figure}[ht]
\begin{tabular}{cc}
\includegraphics[width=0.35\textwidth,keepaspectratio=true]{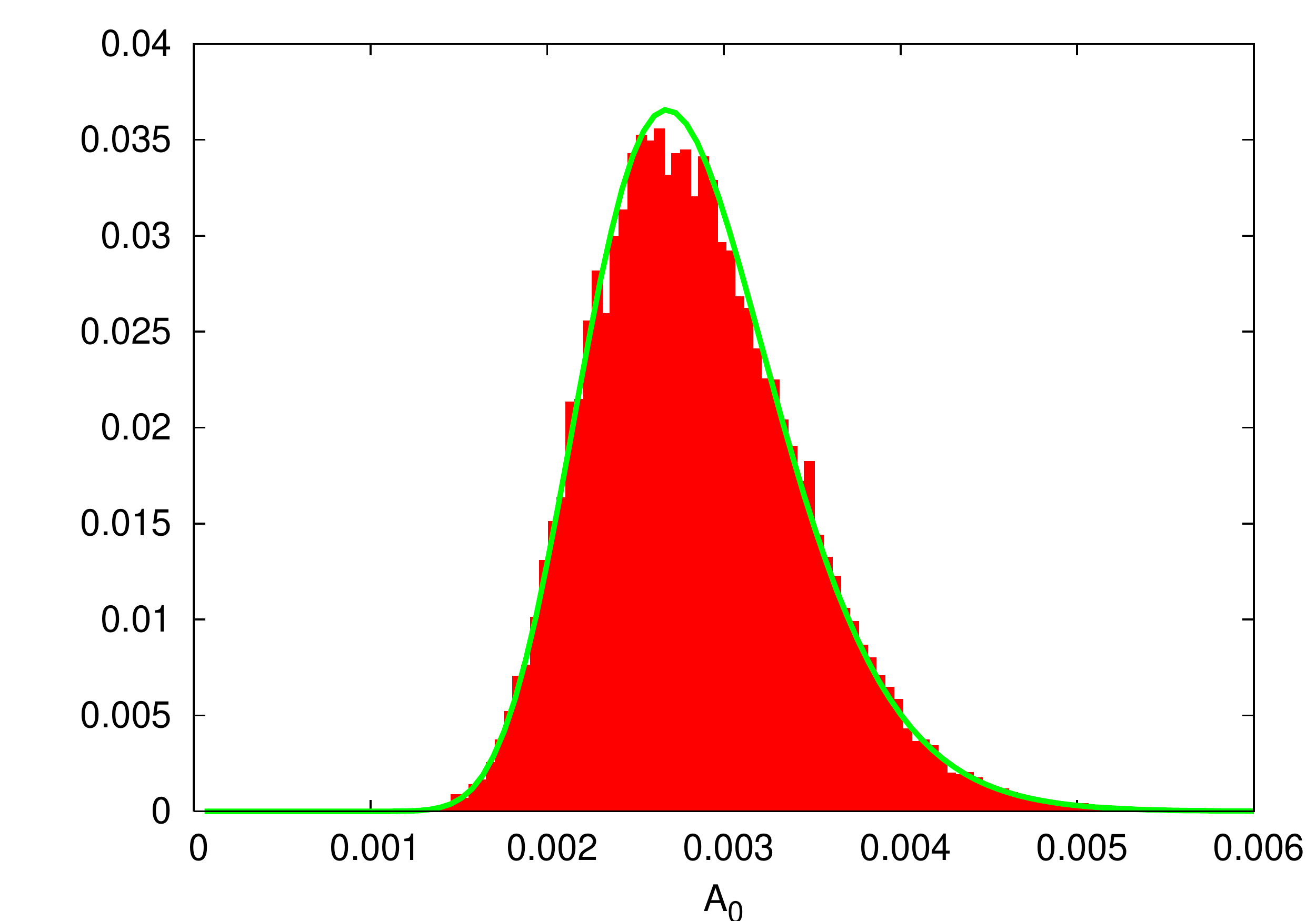}&
\includegraphics[width=0.35\textwidth,keepaspectratio=true]{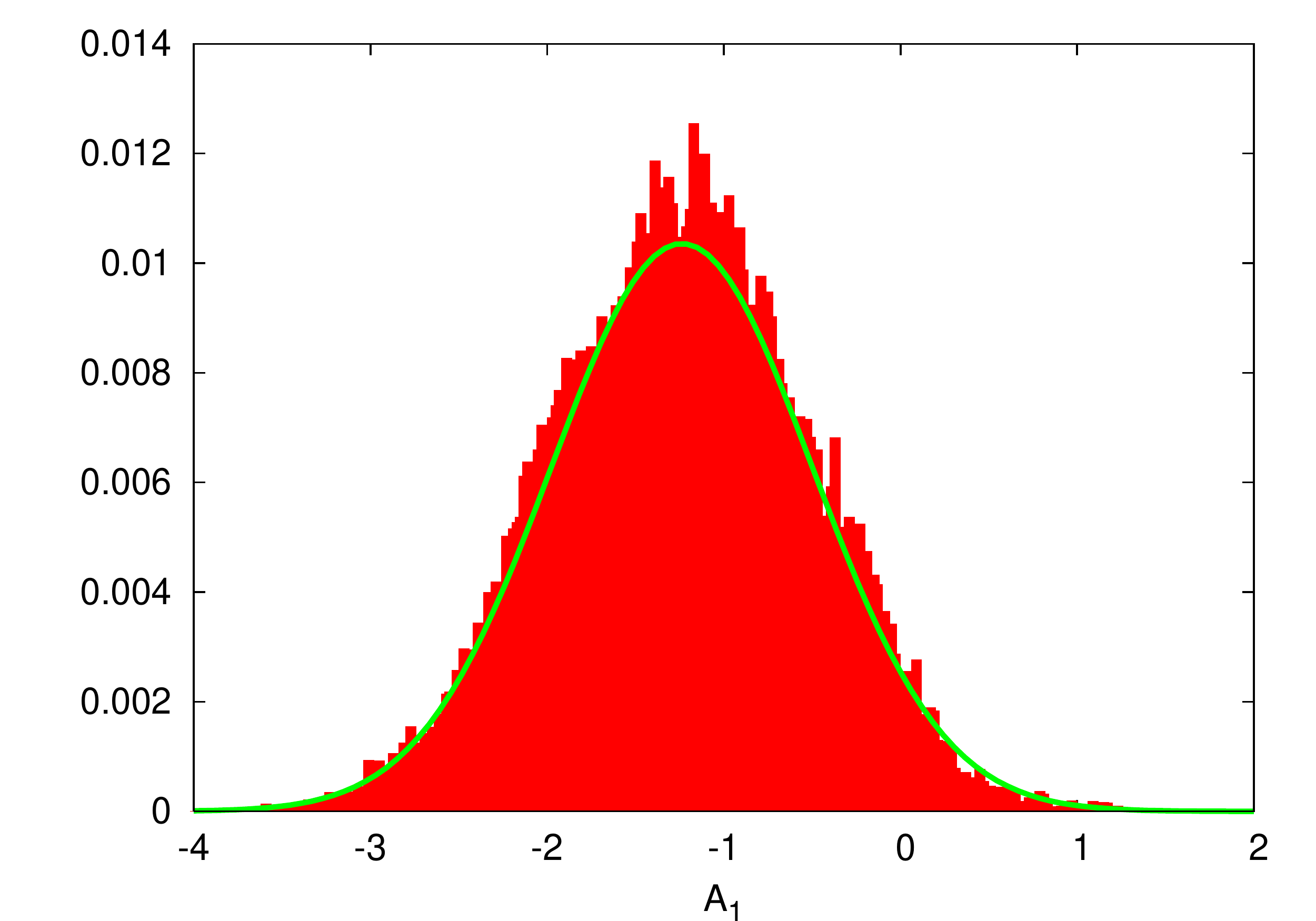}\\
\includegraphics[width=0.35\textwidth,keepaspectratio=true]{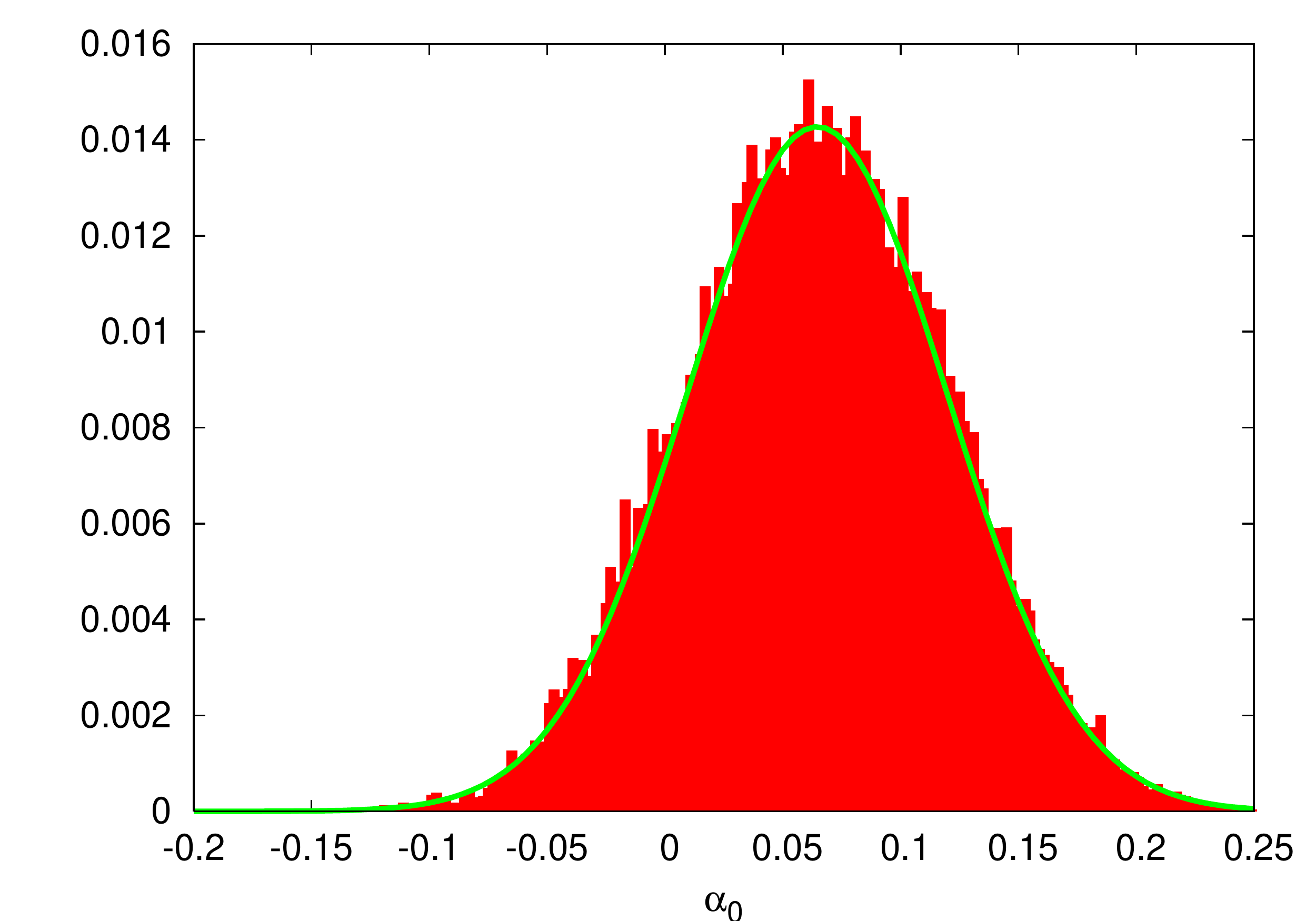}&
\includegraphics[width=0.35\textwidth,keepaspectratio=true]{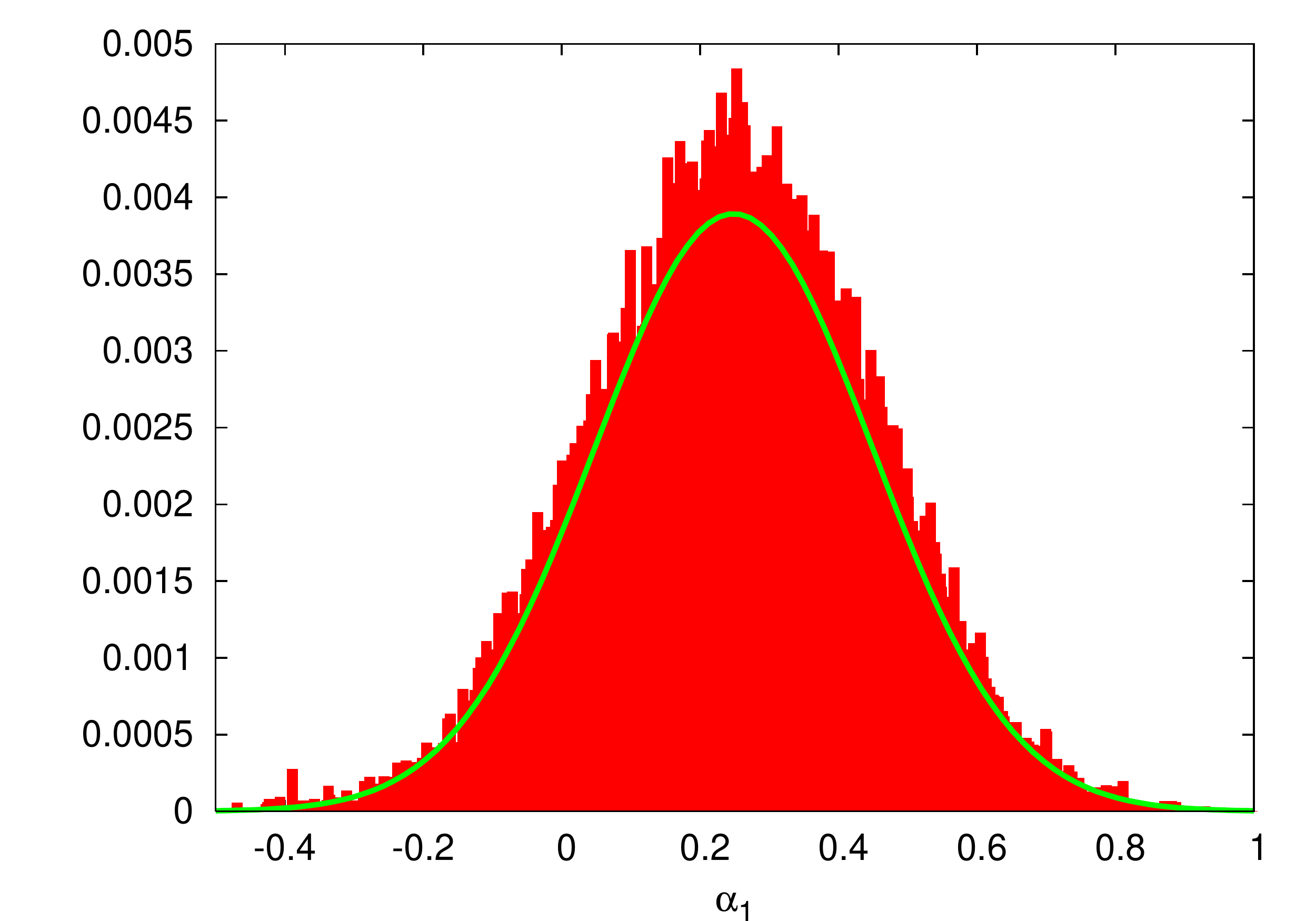}
\end{tabular}
\caption{As Figure~\ref{nonevolv}, but now for the redshift dependent mass function described in Eq.~(\ref{massfncev}). The four panels show, from left to right, the posterior distributions for $A_0$, $A_1$, $\alpha_0$ and $\alpha_1$.}
\label{evolv}
\end{figure}

\begin{figure}[ht]
\begin{tabular}{ccc}
\includegraphics[width=0.33\textwidth,keepaspectratio=true]{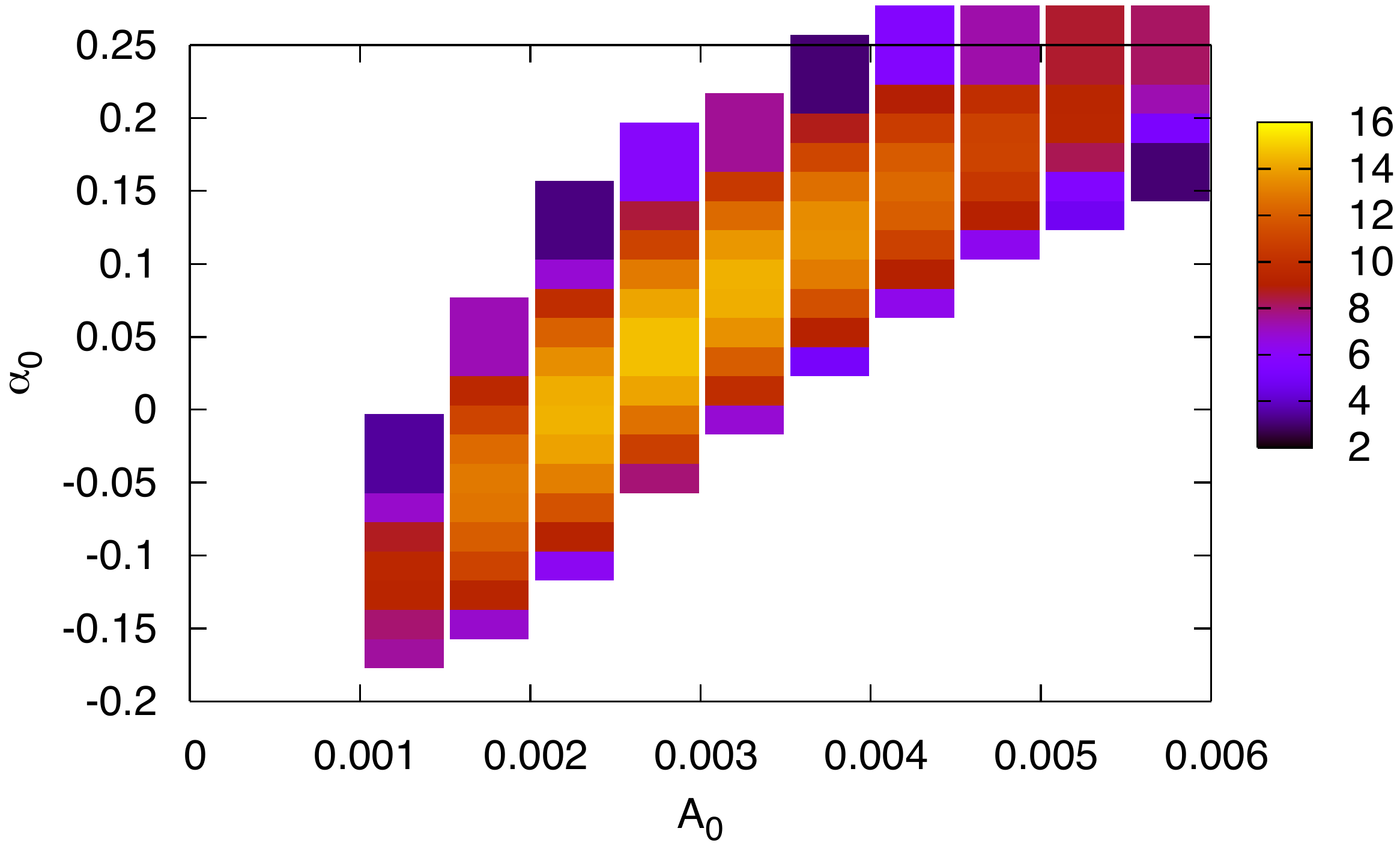}&
\includegraphics[width=0.33\textwidth,keepaspectratio=true]{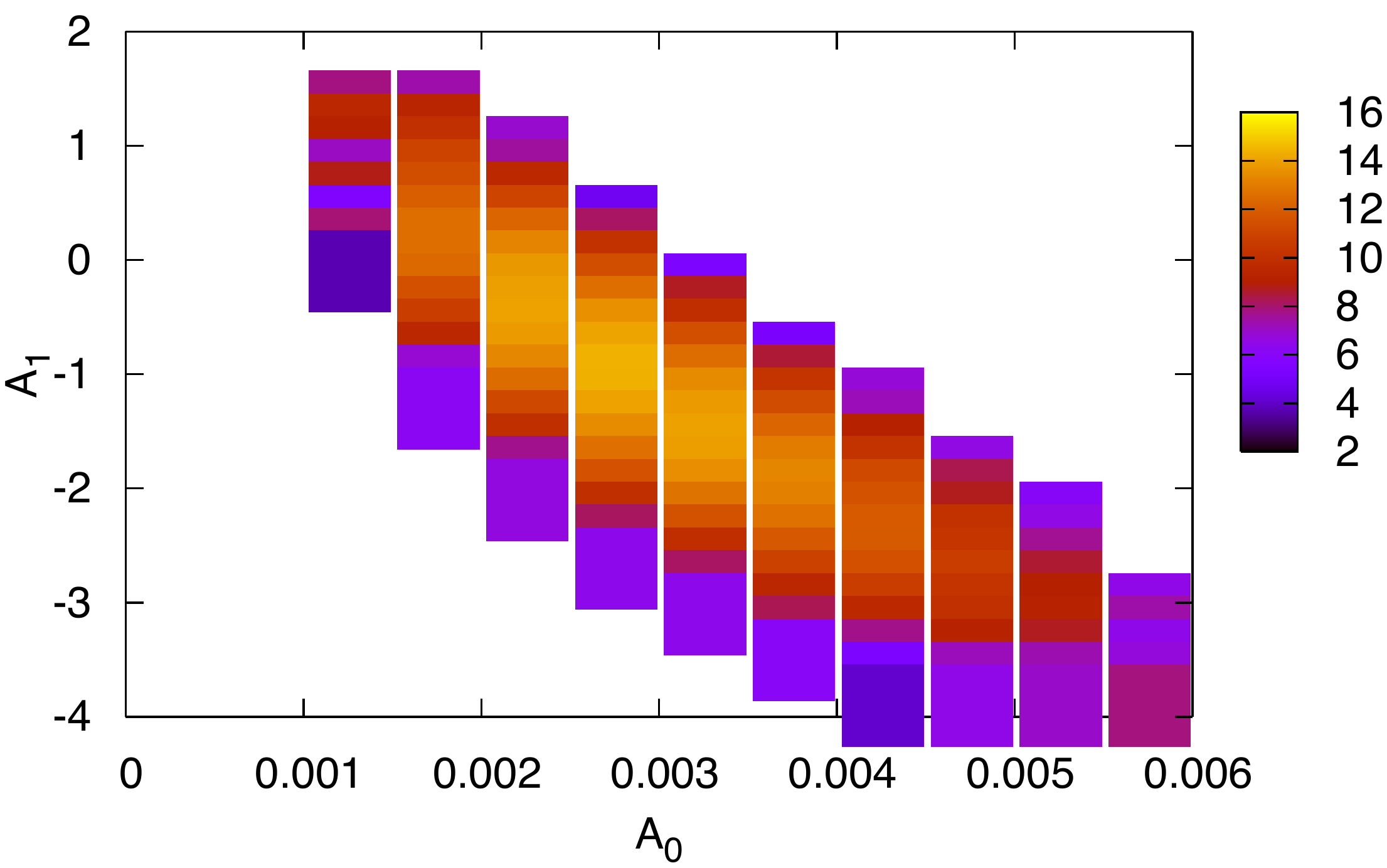}&
\includegraphics[width=0.33\textwidth,keepaspectratio=true]{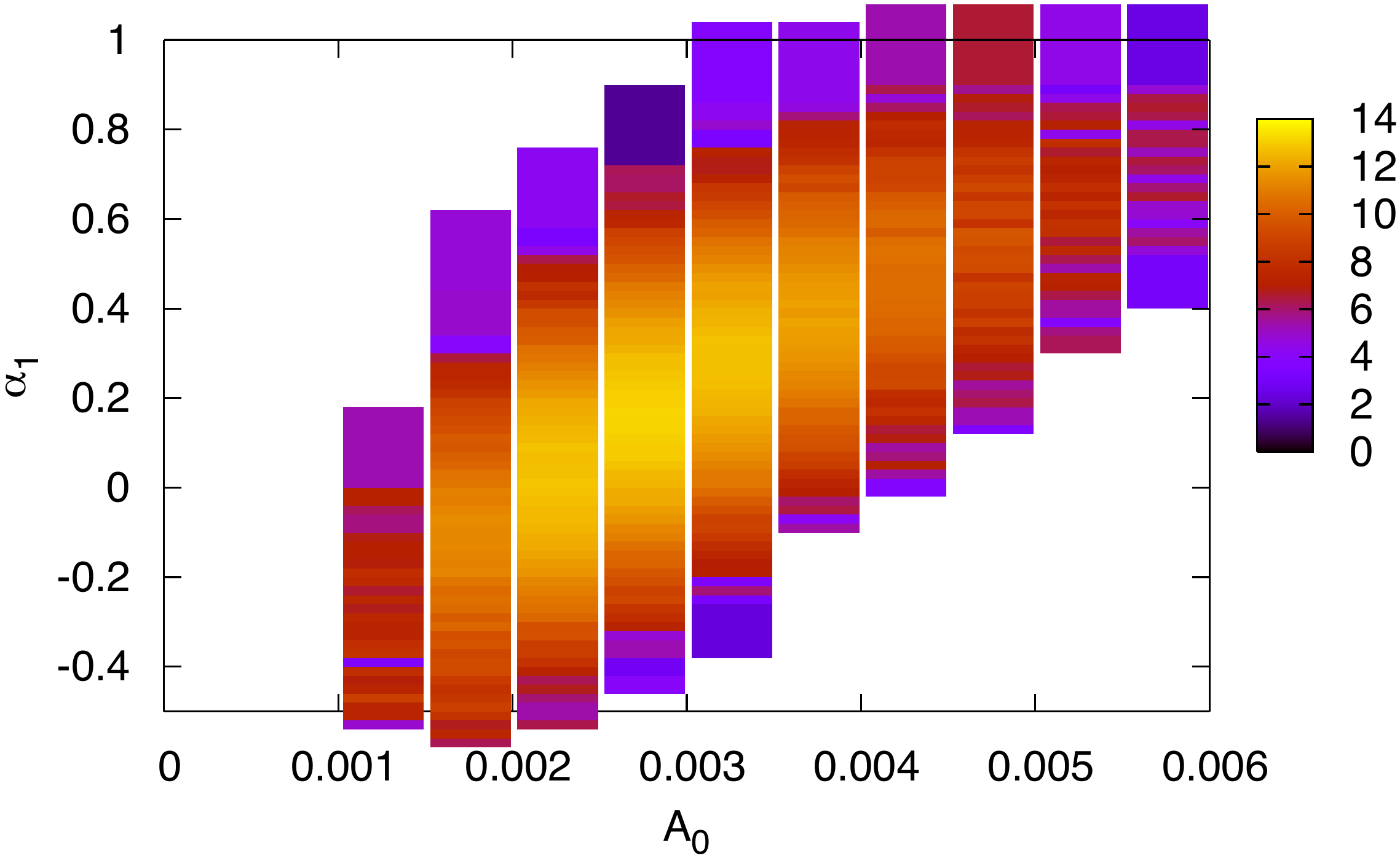}\\
\includegraphics[width=0.33\textwidth,keepaspectratio=true]{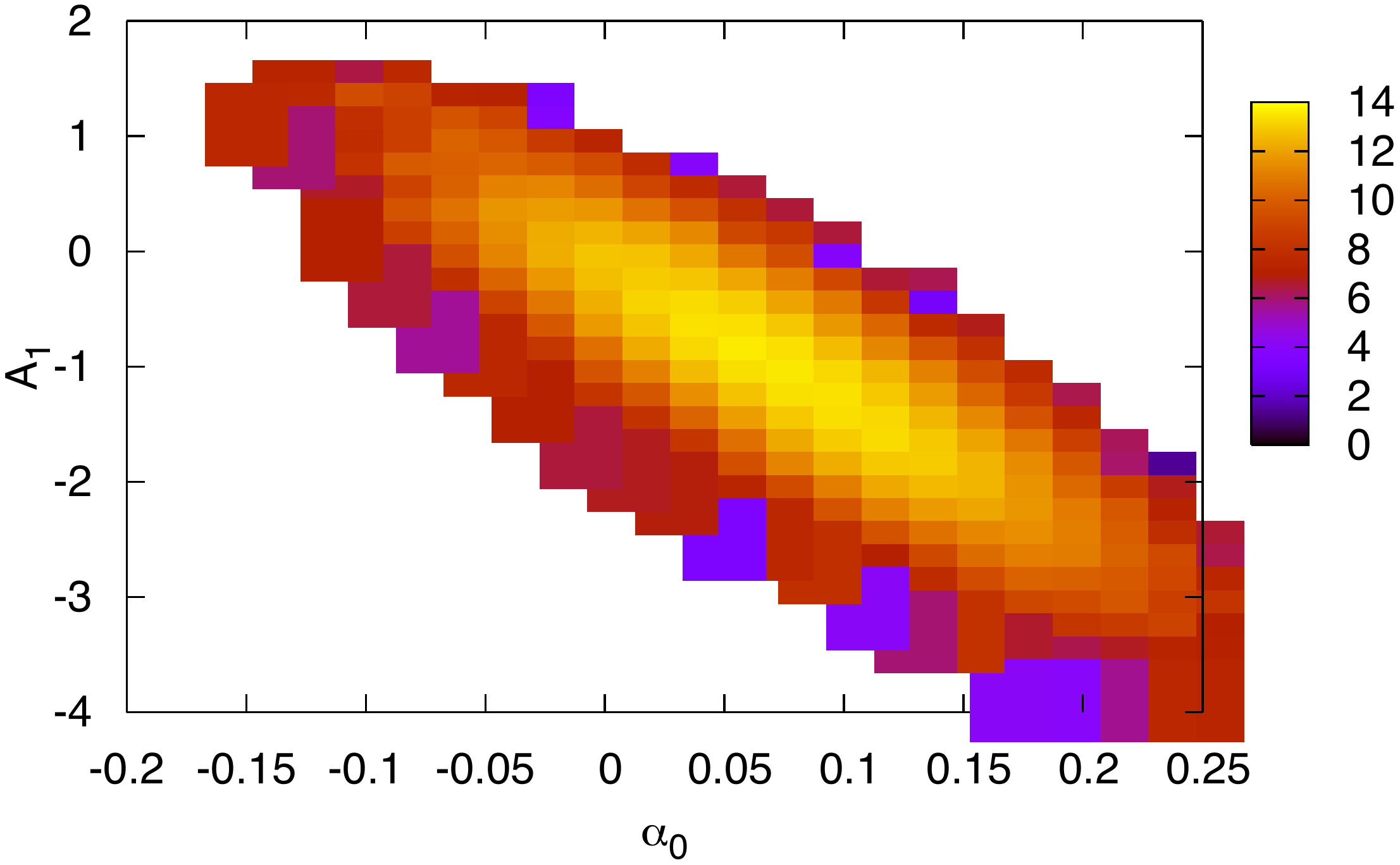}&
\includegraphics[width=0.33\textwidth,keepaspectratio=true]{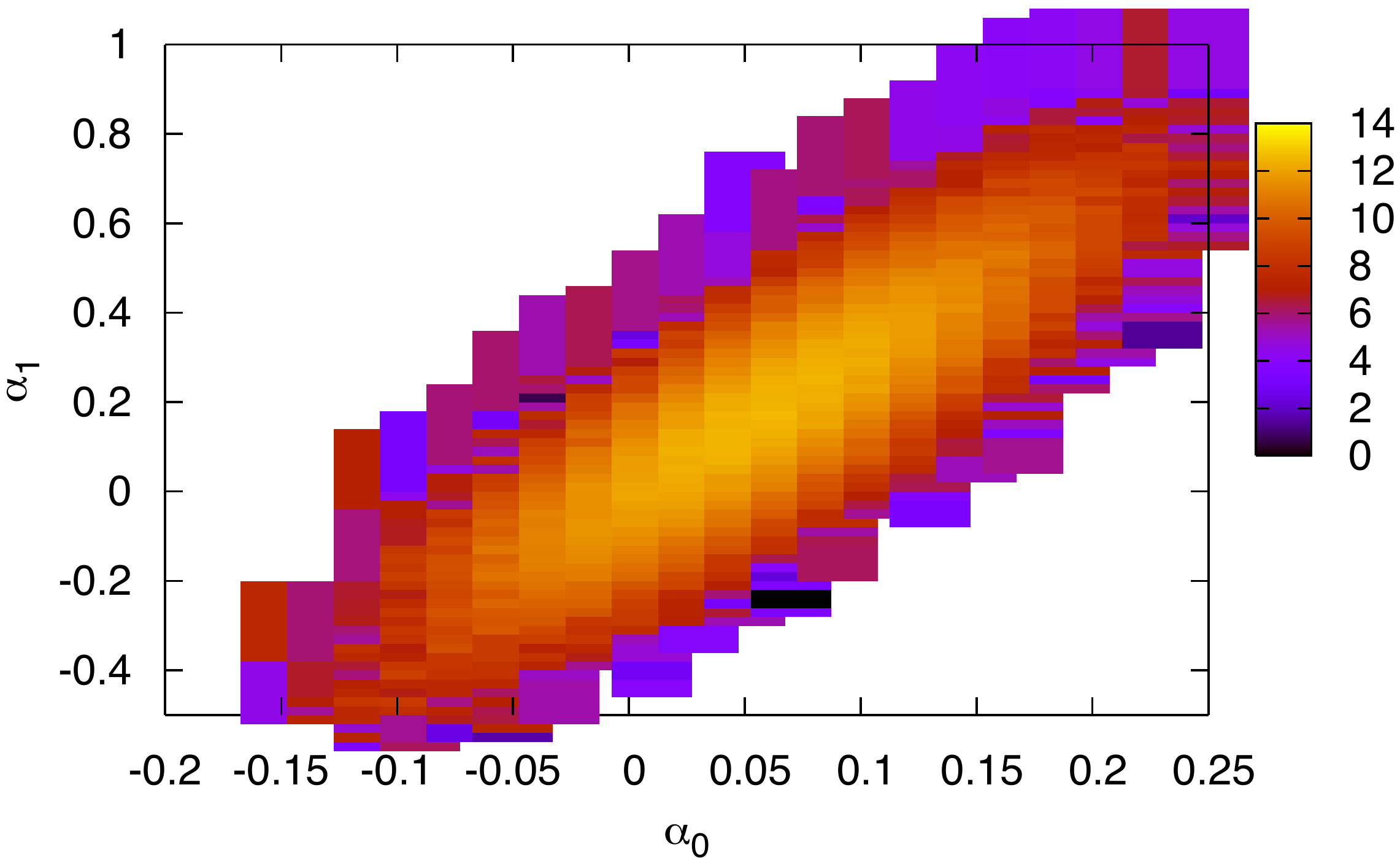}&
\includegraphics[width=0.33\textwidth,keepaspectratio=true]{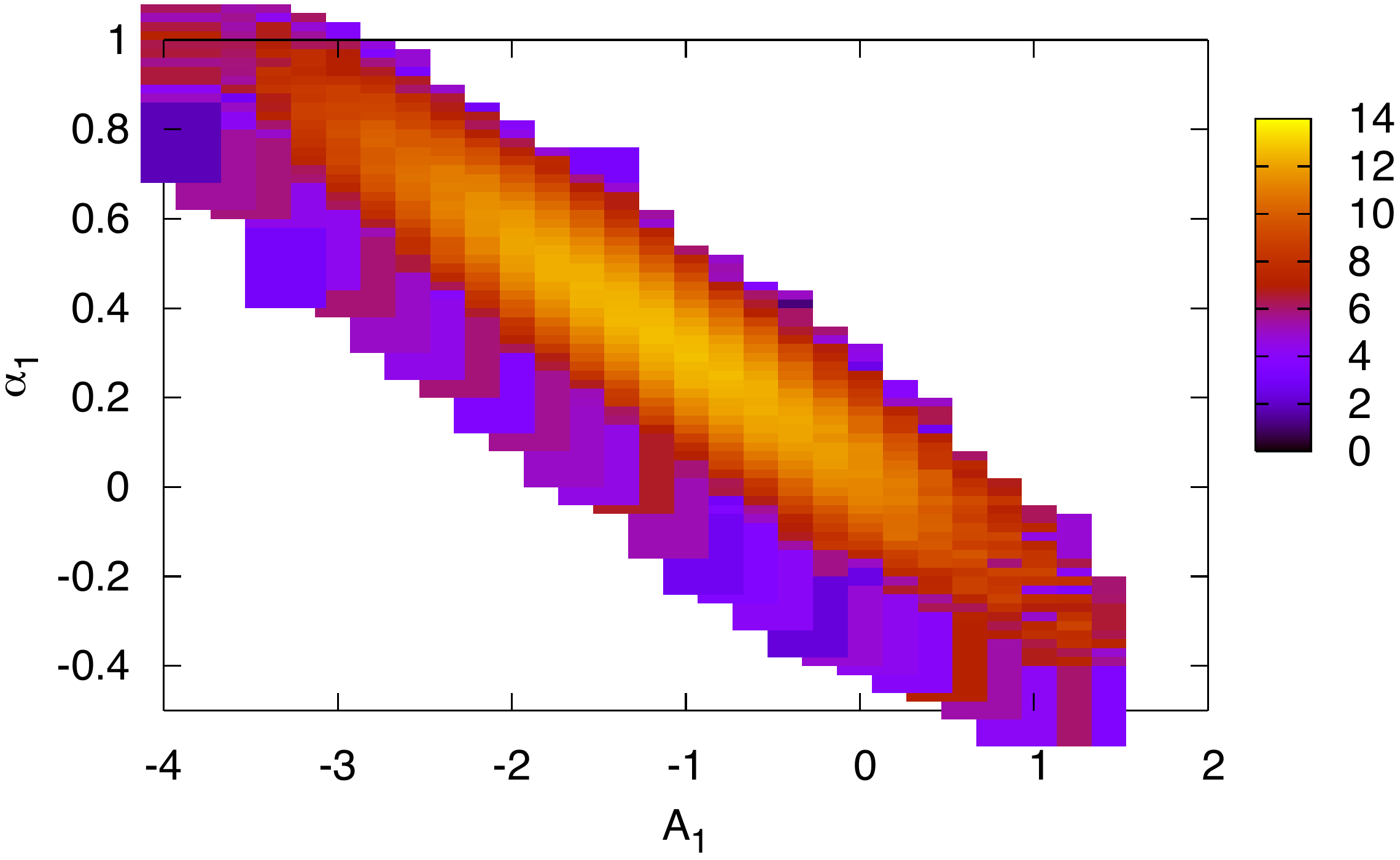}
\end{tabular}
\caption{As Figure~\ref{nonevolvCorr}, but now for the redshift dependent mass function described in Eq.~(\ref{massfncev}). The four panels show the correlations for $A_0/A_1$ (top left), $A_0/\alpha_0$ (top middle), $A_0/\alpha_1$ (top right), $A_1/\alpha_0$ (bottom left), $A_1/\alpha_1$ (bottom middle) and $\alpha_0/\alpha_1$ (bottom right).}
\label{evolvCorr}
\end{figure}

As an extension of the model, we modified the simple ansatz described above to allow for simple evolution of the mass function with redshift
\begin{equation}
\frac{{\rm d}n}{{\rm d} \log M} = A_0 (1+z)^{A_1} \left(\frac{M}{M_*}\right) ^{\alpha_0 - \alpha_1 z} .
\label{massfncev}
\end{equation}
with $M_*=3\times10^6M_{\odot}$ again. In this case, for our reference data set we took $A_0$ and $\alpha_0$ to have the same values as the non-evolving case, and additionally chose $A_1=\alpha_1=0$. In other words, we took the real Universe to have no evolution in mass-function, but allowed the model to include mass function evolution. The distributions in $A_1$ and $\alpha_1$ then tell us how much evolution there would have to be for LISA EMRI observations to detect it. The recovered distributions for these four parameters are shown for a typical example in Figure~\ref{evolv} and the corresponding 2D posteriors are shown in Figure~\ref{evolvCorr}. We can fit Gaussians to the distributions of $\ln A_0$, $A_1$, $\alpha_0$ and $\alpha_1$ in this case as well. The best fit Gaussians are also shown in Figure~\ref{evolv}. It is clear that the Gaussian approximation again provides a good estimate of the mean and width of the distribution and hence the precision to which we can measure the various model parameters. As in the redshift-independent case we find that the true parameters are consistent with the recovered posteriors in all cases, and there are strong correlations between the model parameters. The interpretation of these correlations is the same as before, i.e., that they correspond to keeping the number of events of the type to which LISA is most sensitive approximately constant.

\begin{figure}[ht]
\begin{tabular}{cc}
\includegraphics[width=0.33\textwidth,keepaspectratio=true]{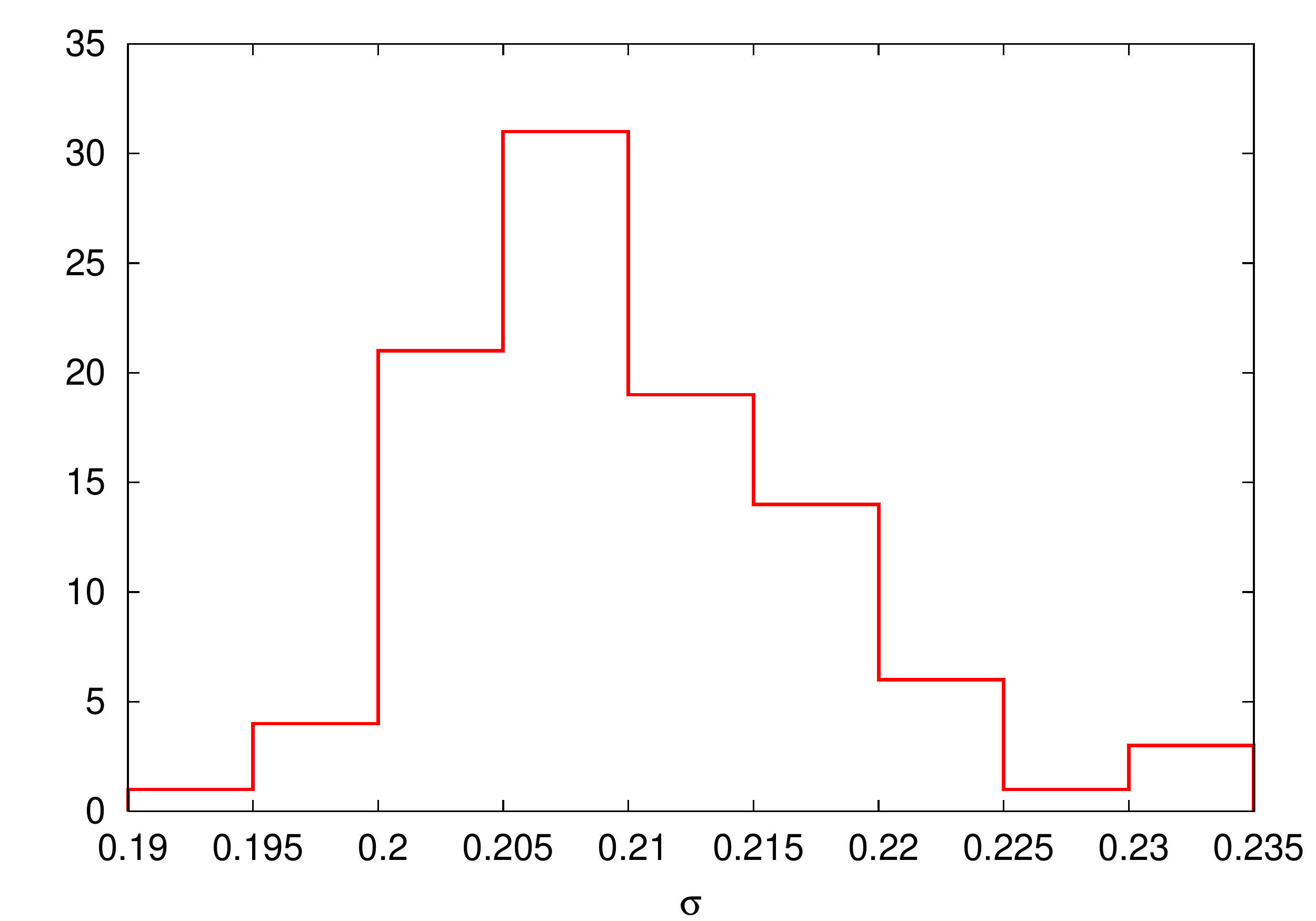}&
\includegraphics[width=0.33\textwidth,keepaspectratio=true]{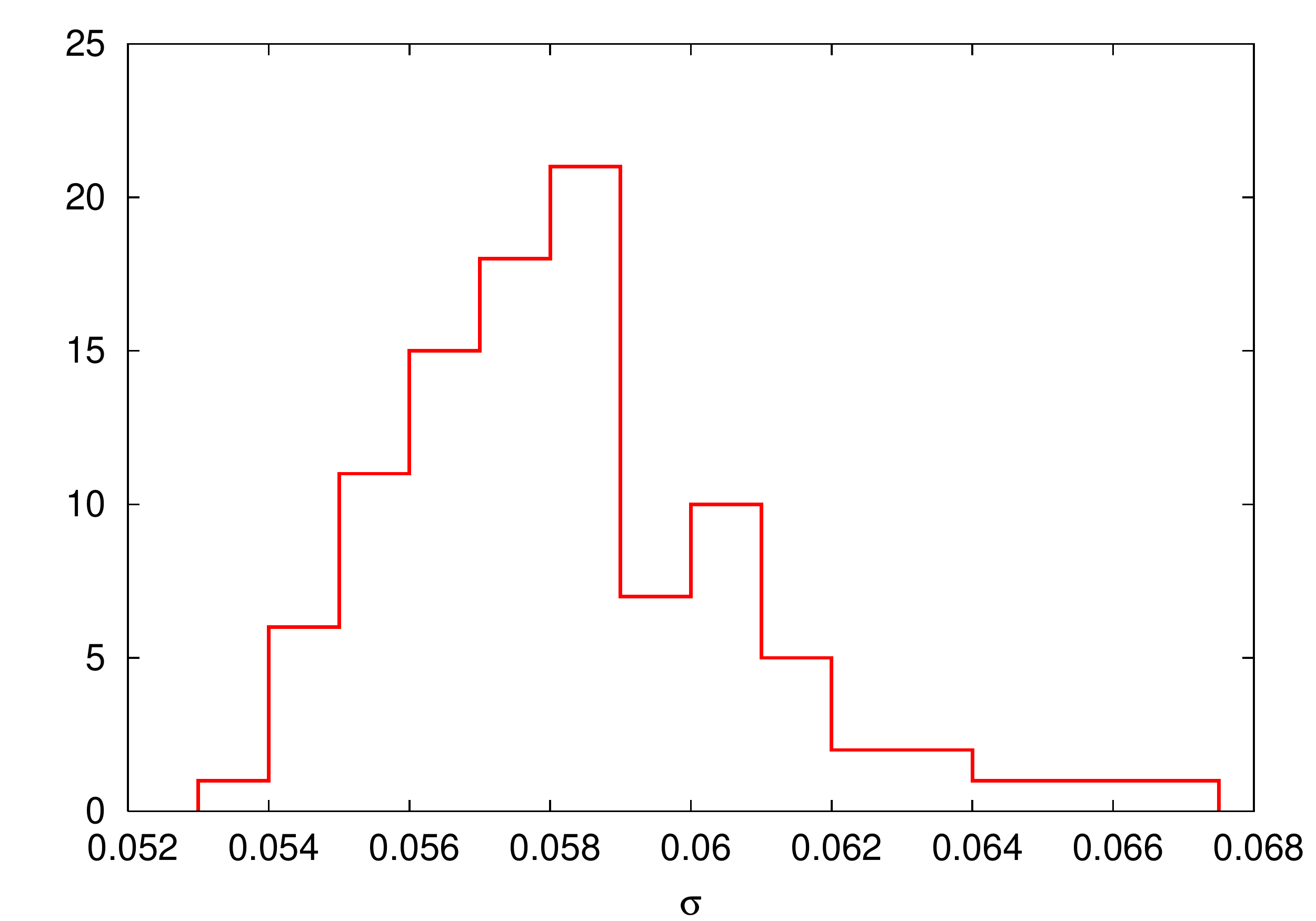}\\
\includegraphics[width=0.33\textwidth,keepaspectratio=true]{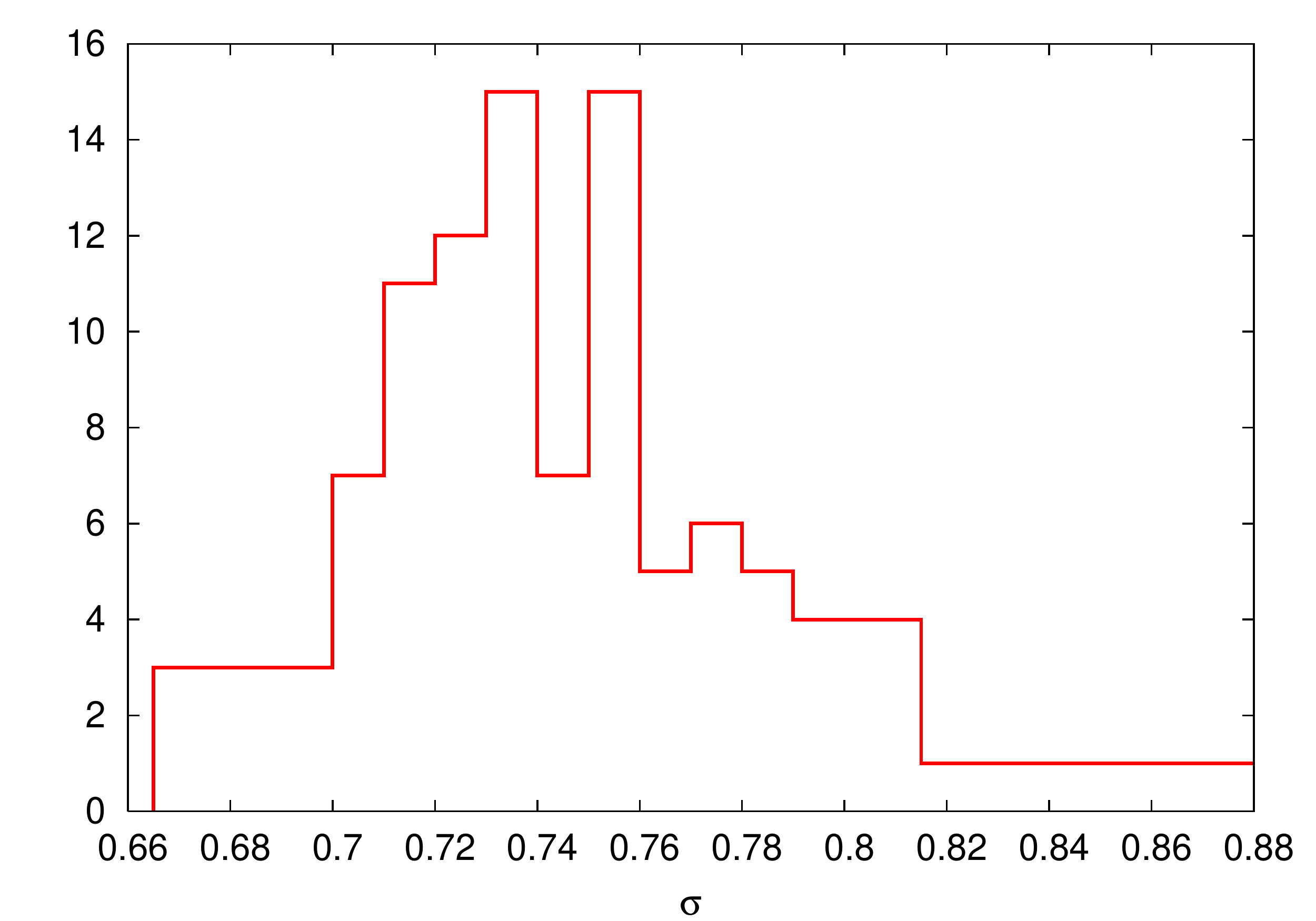}&
\includegraphics[width=0.33\textwidth,keepaspectratio=true]{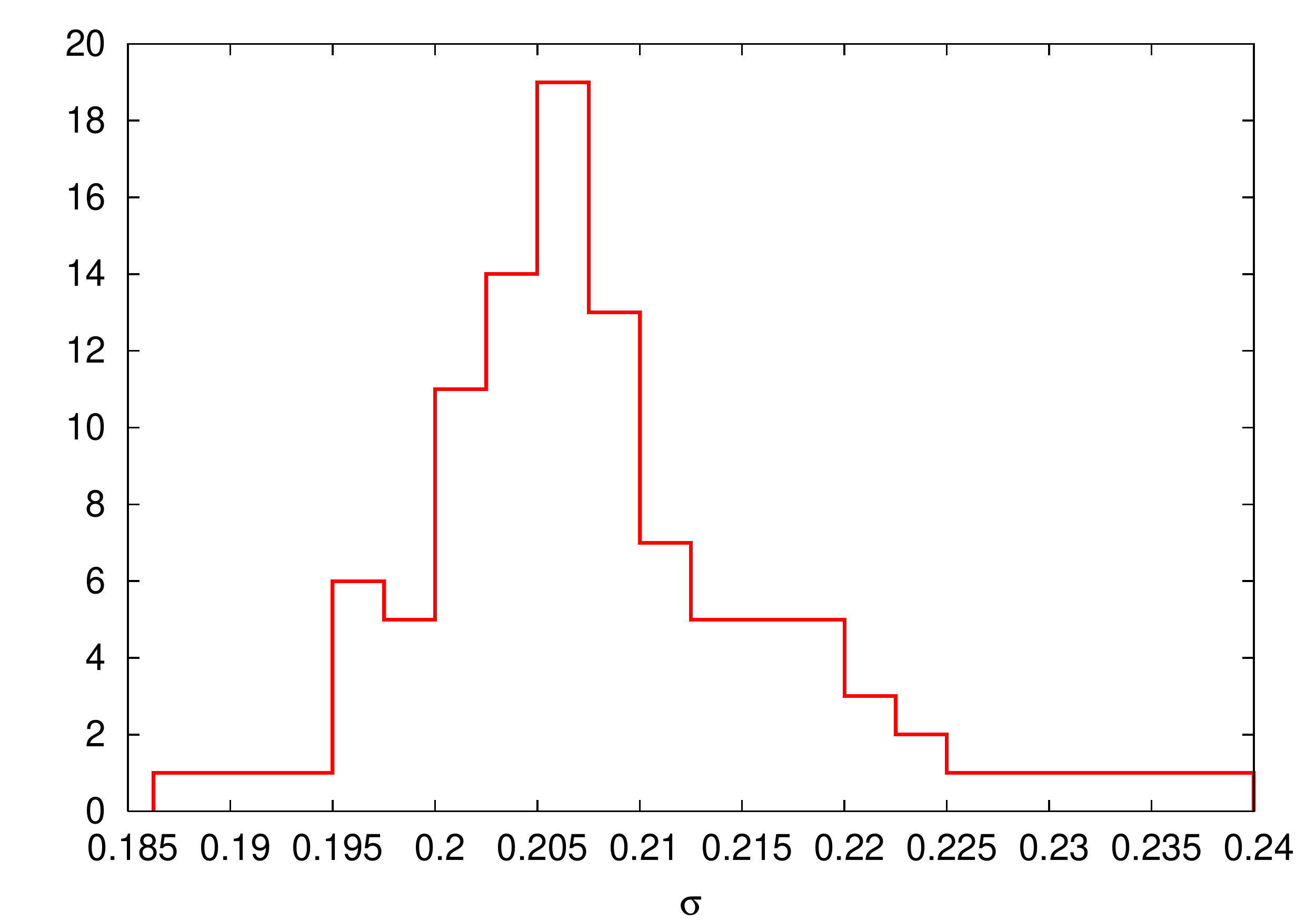}
\end{tabular}
\caption{Distribution of standard deviations of the best-fit Gaussians to the posteriors, computed over $100$ realisations of the set of observed EMRI events. The plots are for $\ln A_0$ (top left), $\alpha_0$ (top right), $A_1$ (bottom left) and $\alpha_1$ (bottom right).}
\label{4parMCres}
\end{figure}

In Figure~\ref{4parMCres} we show the distribution of the Gaussian widths over $100$ realisations of the set of observed EMRI events. We do not show the Gaussian means or the ``errors'' as the former were consistent with the recovered posteriors and the latter were consistent with a Normal distribution, as we would expect and as we saw in the redshift independent case. Unsurprisingly, we find that the addition of extra parameters in the model decreases the precision to which each of the model parameters can be measured, with typical errors in $\ln A_0$ of $\sim0.2$ and in $\alpha_0$ of $\sim 0.055$. The errors in the evolution parameters are typically $\sim 0.75$ for $A_1$ and $\sim0.2$ for $\alpha_1$. These latter errors are quite large, which suggests that we would not be able to say with any confidence that there was an evolution in the mass function. This conclusion will depend on the actual amount of evolution, but these results suggest that  the true $|A_1|$ would have to be at least $0.75$ or the true $|\alpha_1|$ would have to be at least $0.2$ in order for a redshift evolution to be apparent from the data.

\subsubsection{Dependence on parameters of observed distribution}
\begin{figure}[ht]
\begin{tabular}{cc}
\includegraphics[width=0.5\textwidth,keepaspectratio=true]{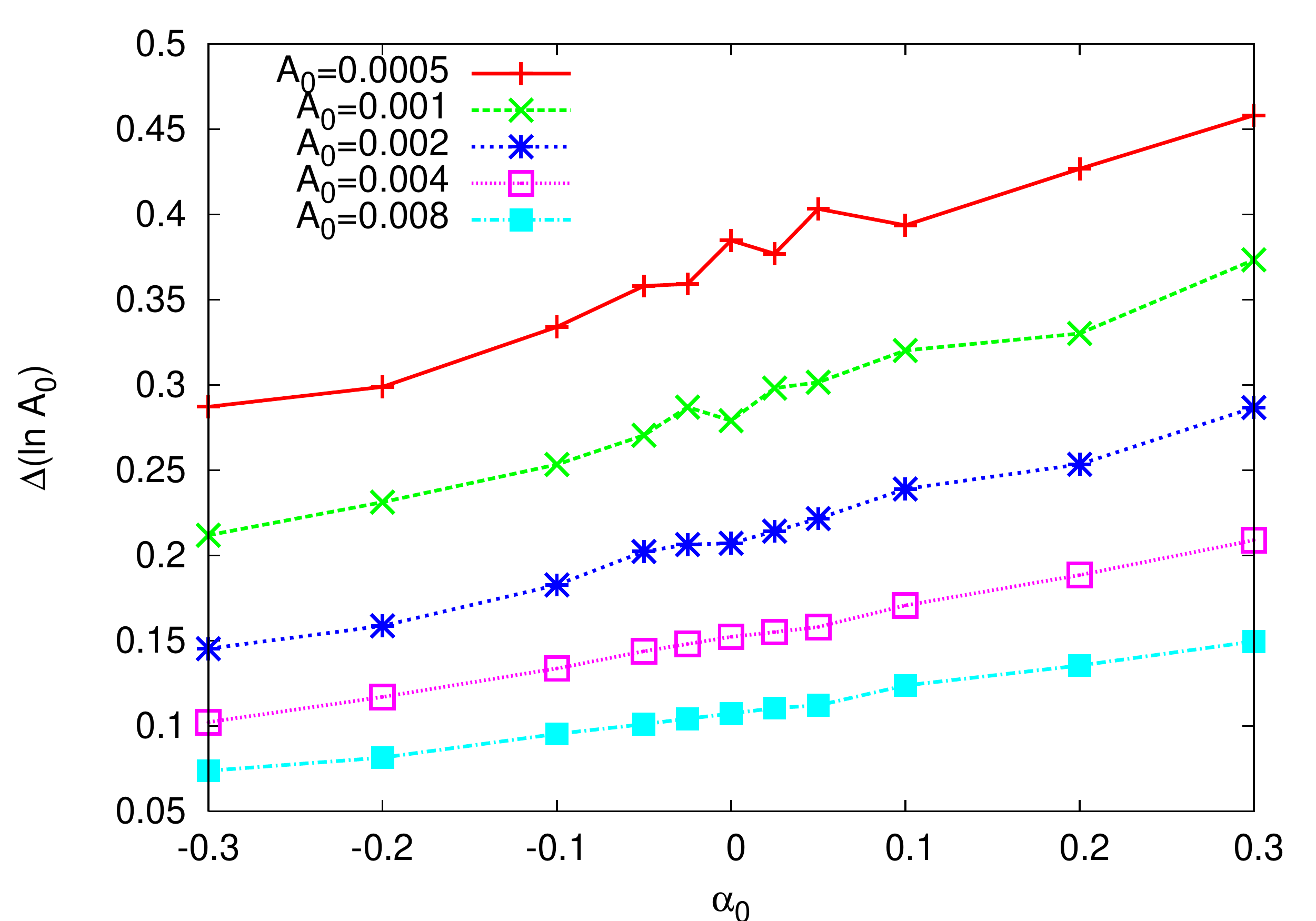}&
\includegraphics[width=0.5\textwidth,keepaspectratio=true]{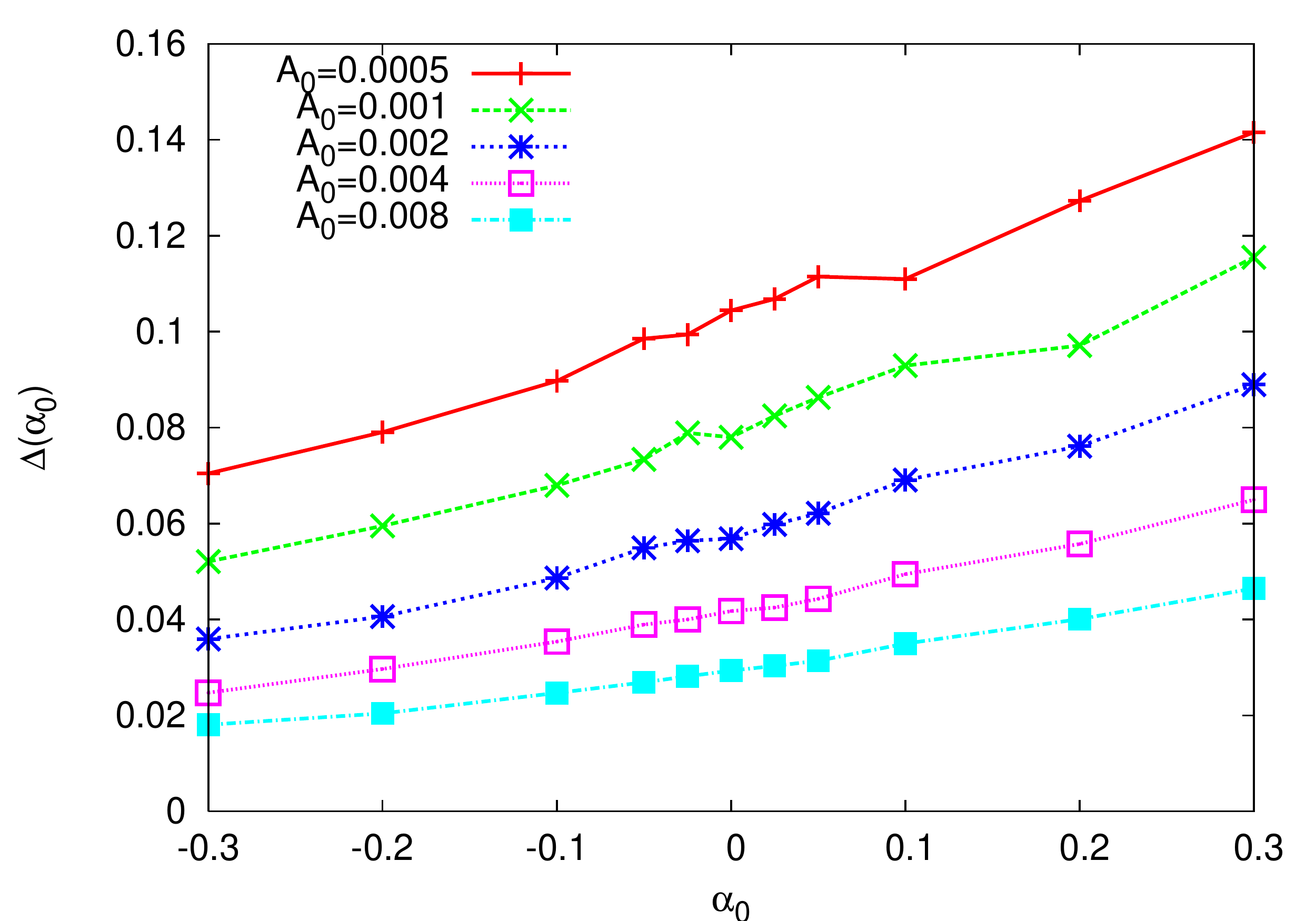}\\
\includegraphics[width=0.5\textwidth,keepaspectratio=true]{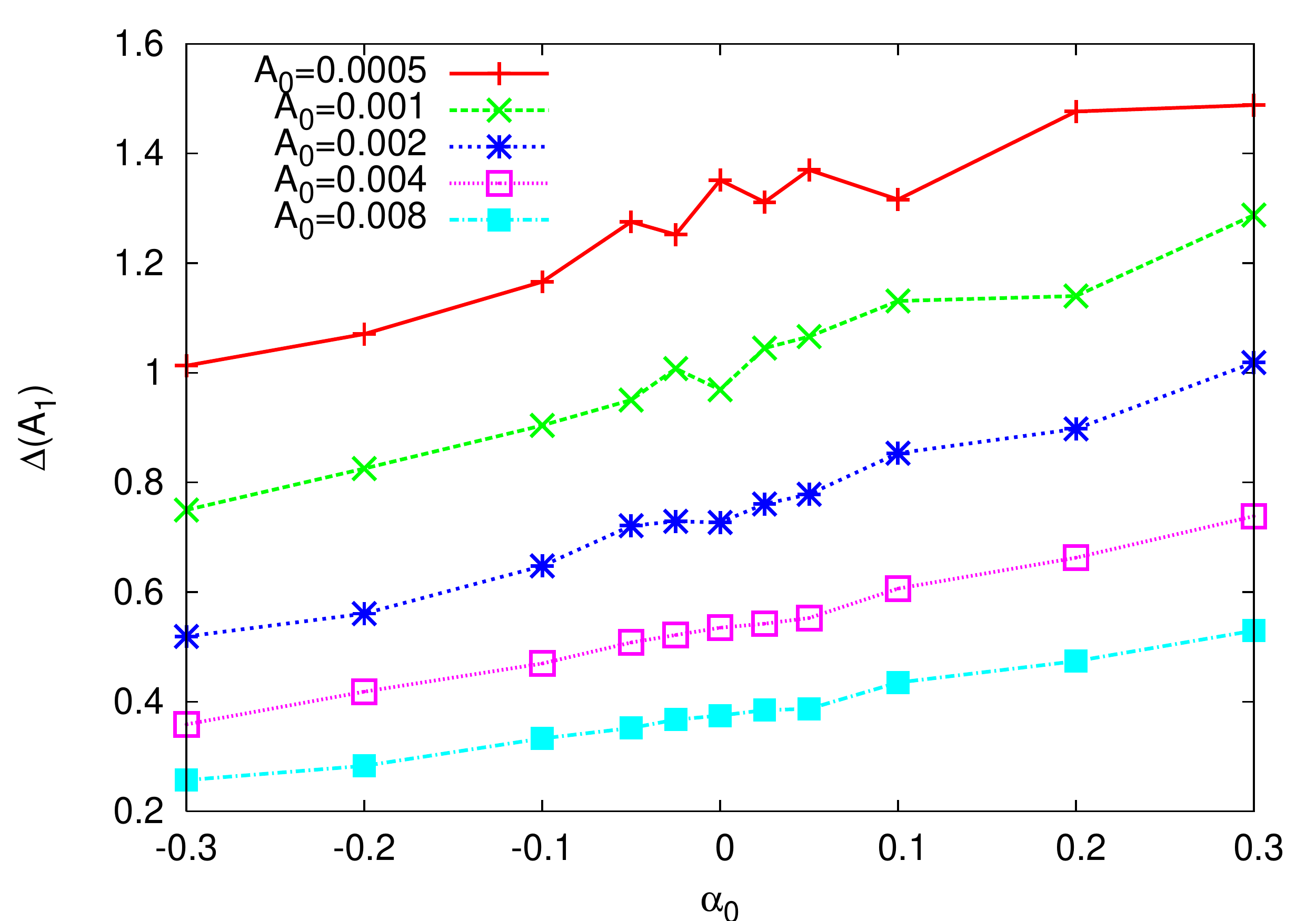}&
\includegraphics[width=0.5\textwidth,keepaspectratio=true]{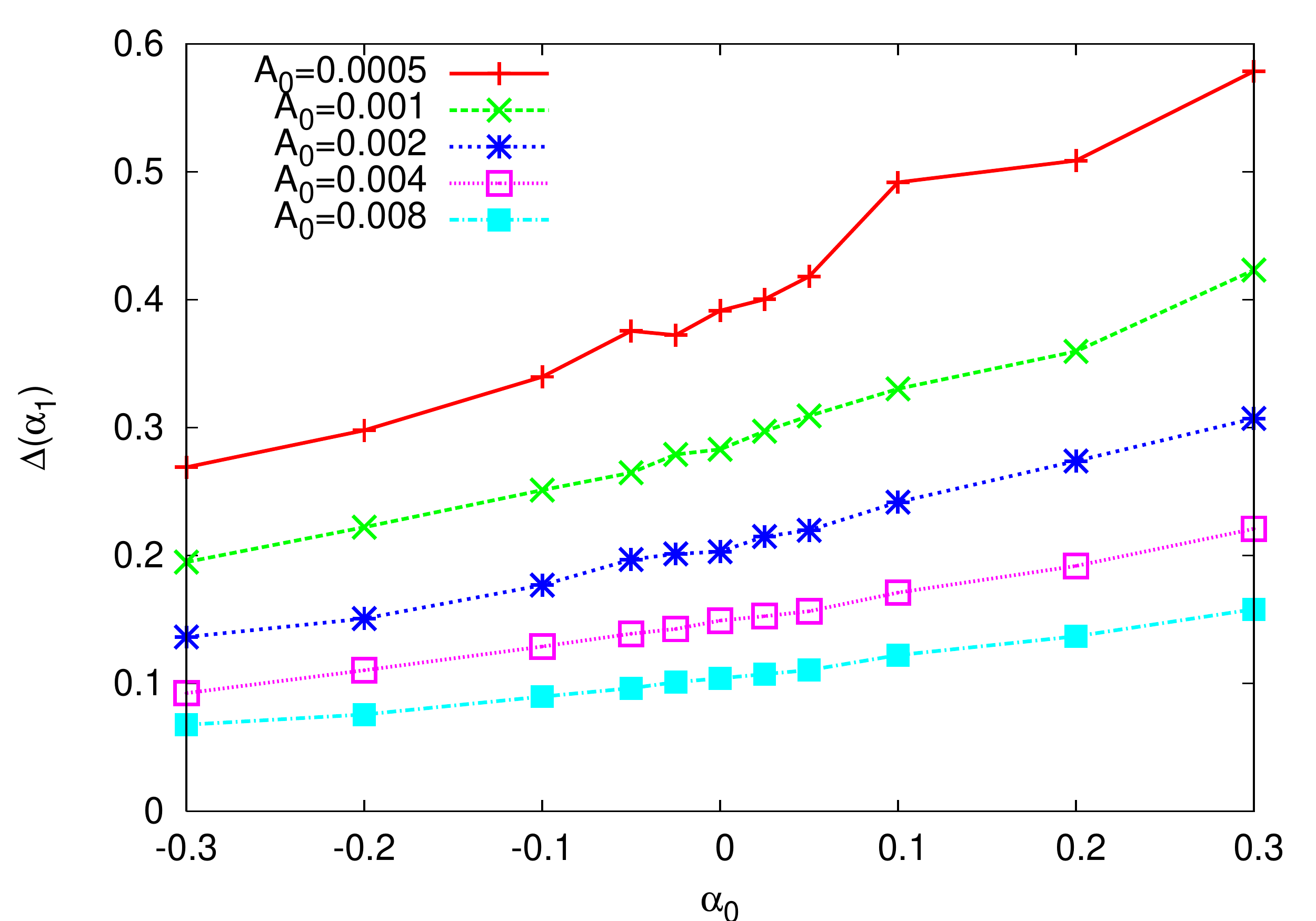}
\end{tabular}
\caption{Typical parameter estimation errors as a function of the parameters of the underlying distribution. We show the error in $\ln A_0$ (top left), $\alpha_0$ (top right), $A_1$ (bottom left) and $\alpha_1$ (bottom right) as a function of the values of $A_0$ and $\alpha_0$ of the underlying distribution. We keep $A_1=\alpha_1=0$ for all cases. The value of $\alpha_0$ is on the x-axis, while each line represents a different choice for $A_0$, as labelled in the key.}
\label{4parVarAalp}
\end{figure}

As in the redshift-independent case, it is informative to consider how the precision to which we can determine the parameters varies with the true parameters of the Universe. For this investigation, we chose to keep $A_1 = \alpha_1=0$ in all cases. This allows us to interpret the errors in $A_1$ and $\alpha_1$ as the amount of evolution that would have to exist for the LISA EMRI observations to be manifestly inconsistent with a non-evolving mass function. In Figure~\ref{4parVarAalp} we show how the typical errors in all four parameters vary as we change $A_0$ and $\alpha_0$. We see that the error in $A_1$ is typically $\sim 0.5$--$1$, and the error in $\alpha_1$ is typically $\sim 0.1$--$0.4$. There would therefore have to be a very significant change in the mass function between $z=0$ and $z=1$ for it to be measurable. These results lend further support to our previous conclusion that we will not be able to measure an evolution in the black hole mass function through EMRI observations alone.

Once again, much of the dependence of the errors on the underlying model can be explained by changes in the total number of observed events that the different models predict. All of the data shown in Figure~\ref{4parVarAalp} can be well approximated by $\Delta(\ln A_0) \approx 0.20 \sqrt{1000/N_{\rm obs}}$, $\Delta (\alpha_0) \approx 0.056 \sqrt{1000/N_{\rm obs}}$, $\Delta(A_1) \approx 0.72 \sqrt{1000/N_{\rm obs}}$ and $\Delta (\alpha_1) \approx 0.21 \sqrt{1000/N_{\rm obs}}$. As before, these results were found under the optimistic LISA assumption and assuming that all black holes have spin $a=0$. We now consider what happens when we relax these assumptions.

\subsubsection{Dependence on LISA performance and black hole spin}
\label{4pvaryLISA}
\begin{figure}[ht]
\begin{tabular}{cc}
\includegraphics[width=0.5\textwidth,keepaspectratio=true]{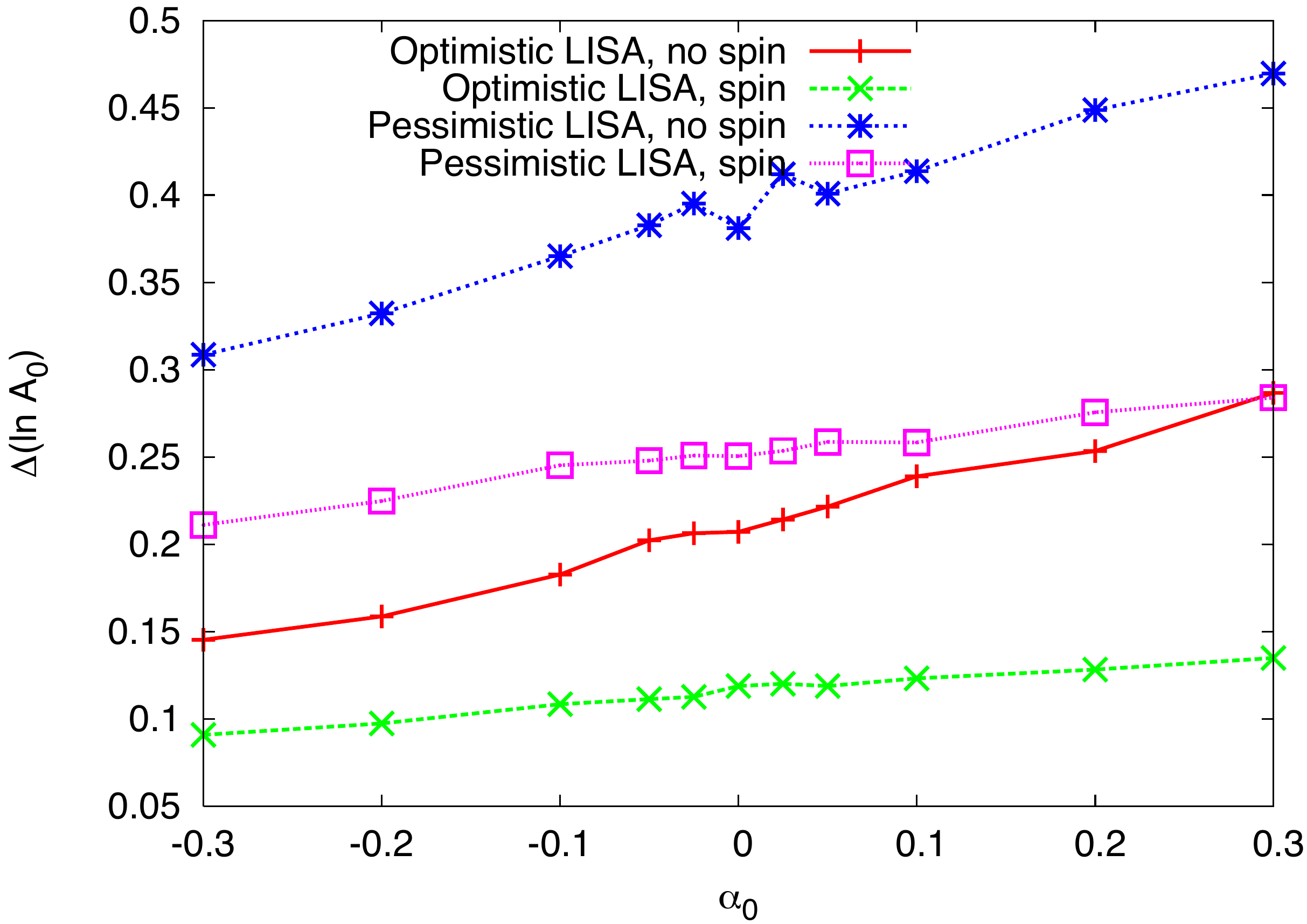}&
\includegraphics[width=0.5\textwidth,keepaspectratio=true]{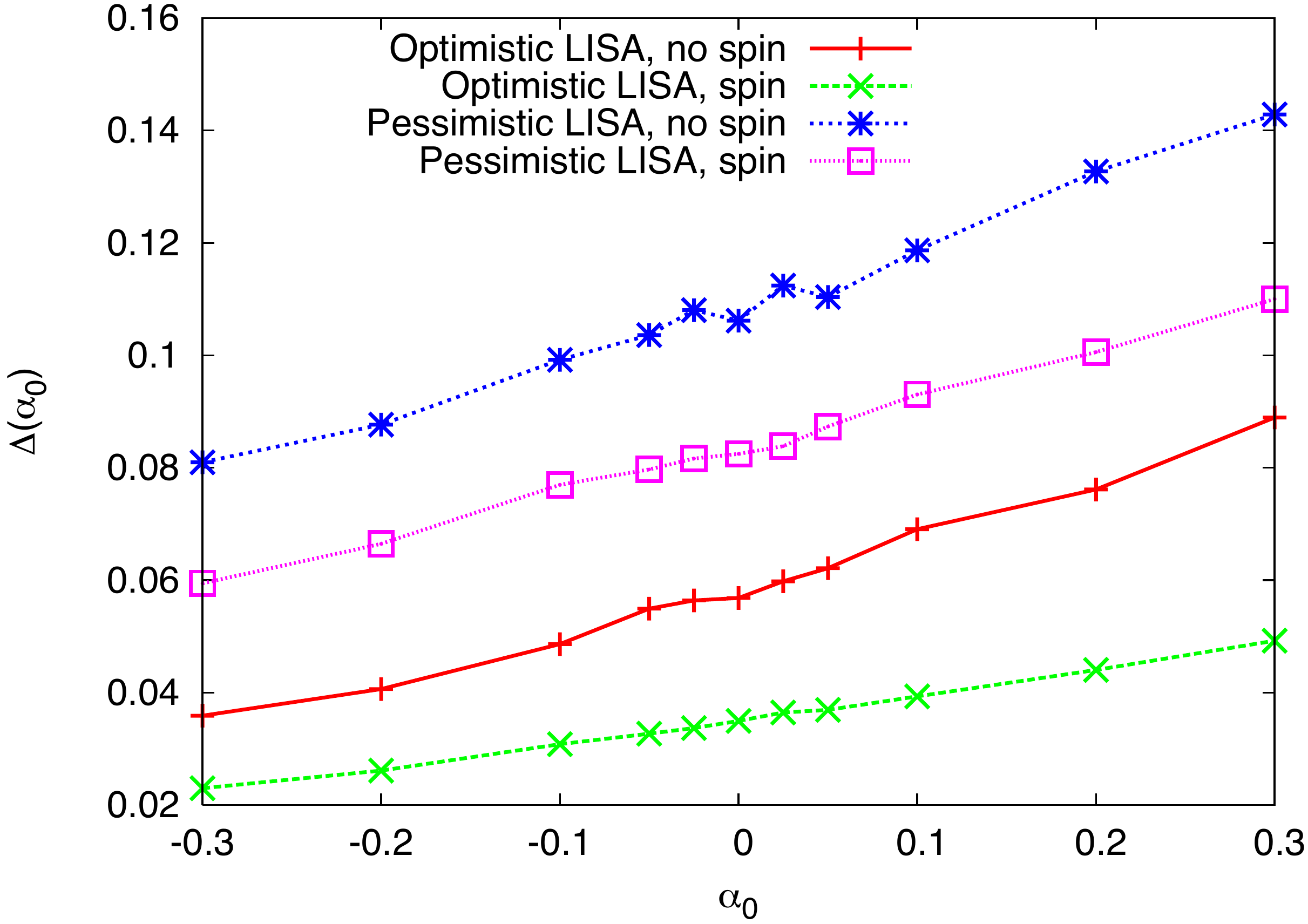}\\
\includegraphics[width=0.5\textwidth,keepaspectratio=true]{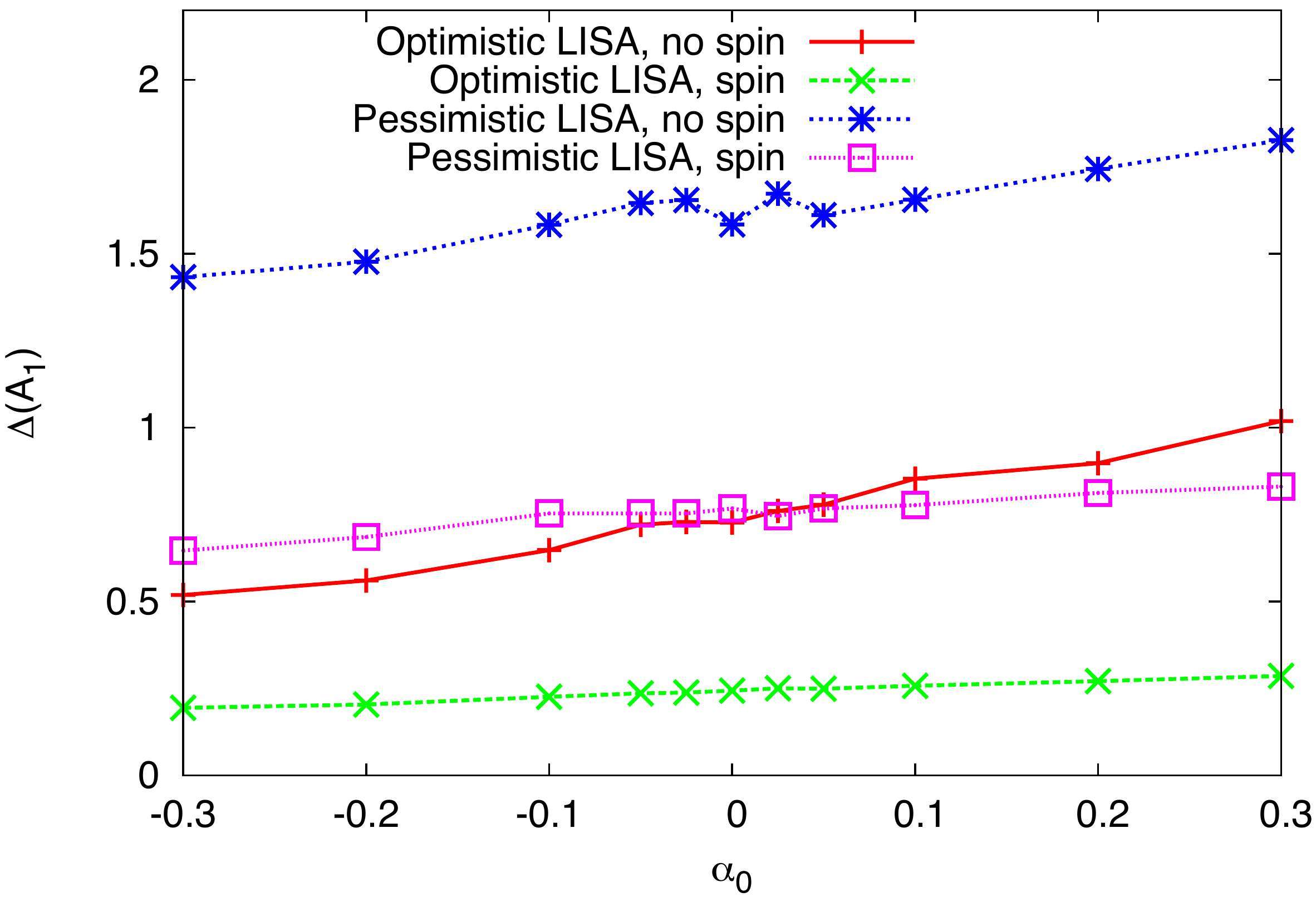}&
\includegraphics[width=0.5\textwidth,keepaspectratio=true]{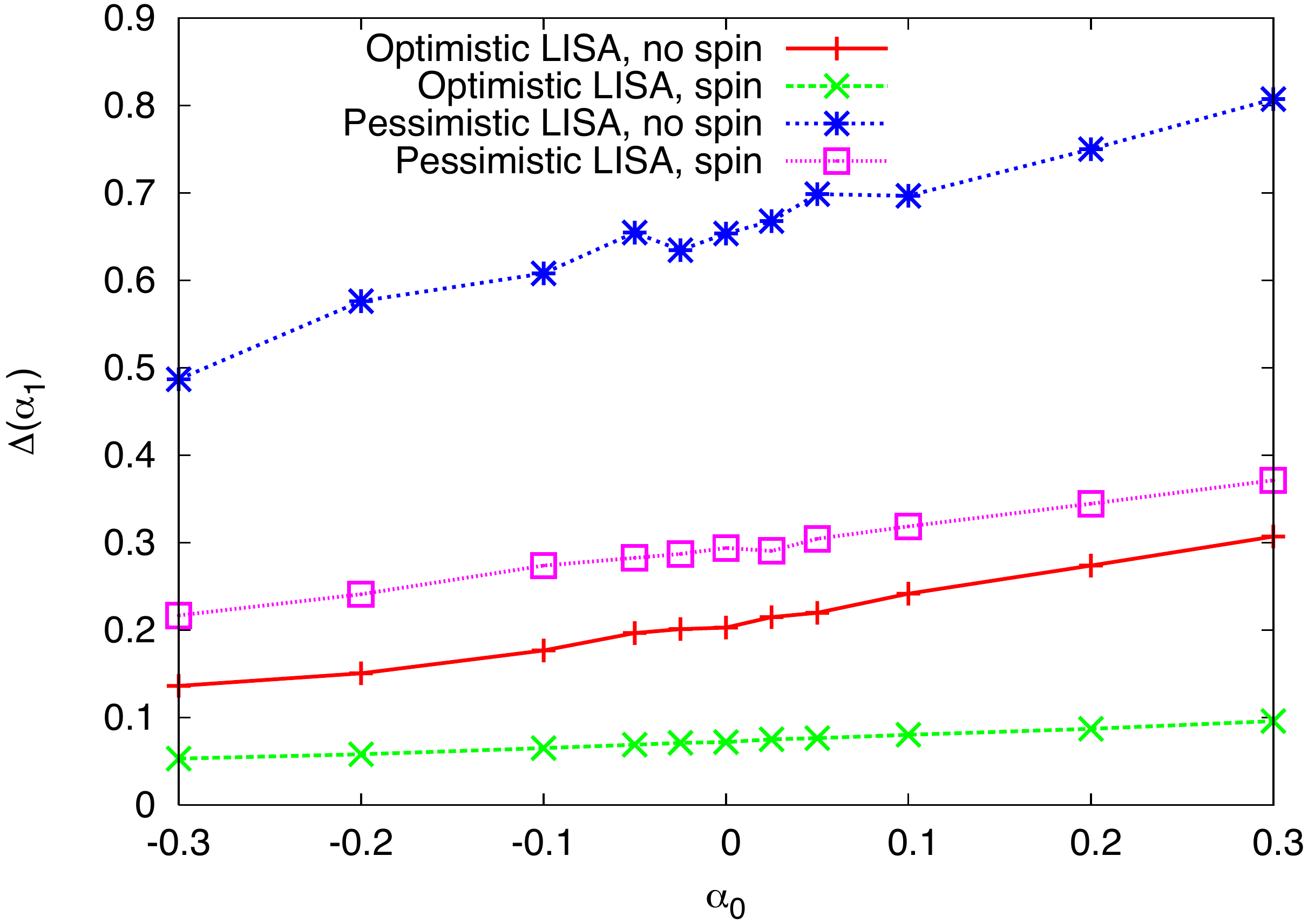}
\end{tabular}
\caption{As Figure~\ref{4parVarAalp}, but now showing results for various assumptions about LISA and about the spin of the central black holes of the EMRI sources. In all cases we have taken $A=0.002$Mpc$^{-3}$ and different curves correspond to different assumptions, as labelled in the key.}
\label{4presAllLISA}
\end{figure}

As before, we can explore how these results change when we vary the assumptions about LISA and about the black hole spins by modifying the observable lifetime function used in the analysis. We consider pessimistic LISA assumptions, and the case where all black holes have spin $a=0.9$, as we used in Section~\ref{varyLISA} for the redshift-independent case. In Figure~\ref{4presAllLISA} we show how the parameter errors vary under these various assumptions. The parameter errors are once again well fit by functions of the number of observed events of the form $\Delta(\ln A_0) = k_{A_0} \sqrt{1000/N_{\rm obs}}$,  $\Delta(\alpha_0) = k_{\alpha_0} \sqrt{1000/N_{\rm obs}}$,  $\Delta(A_1) = k_{A_1} \sqrt{1000/N_{\rm obs}}$ and $\Delta(\alpha_1) = k_{\alpha_1} \sqrt{1000/N_{\rm obs}}$. The various $k_X$ parameters are summarised in Table~\ref{fitz}.
As in the redshift-independent case, there is not much difference between optimistic and pessimistic LISA assumptions, other than in the number of events detected, but measurement precisions are somewhat improved if black holes are spinning.

\begin{table}
\begin{tabular}{|c|cc|cc|}
\hline
&\multicolumn{2}{c|}{$a=0$}&\multicolumn{2}{c|}{$a=0.9$}\\
&Optimistic LISA&Pessimistic LISA&Optimistic LISA&Pessimistic LISA\\\hline
$k_{A_0}$&0.2&0.15&0.15&0.13 \\
$k_{\alpha_0}$&0.056&0.043&0.044&0.041\\
$k_{A_1}$&0.72&0.64&0.30&0.38 \\
$k_{\alpha_1}$&0.21&0.24&0.092&0.15\\\hline
\end{tabular}
\caption{As Table~\ref{fitNoz}, but now for the redshift-dependent mass function. The dependence of the parameter uncertainty on the number of observed events is of the form $\Delta(X)=k_X \sqrt{1000/N_{\rm obs}}$, for the $k_X$ values tabulated here.}
\label{fitz}
\end{table}

\section{Conclusions and future work}
\label{sec:discuss}
We have explored the ability of the proposed space-based gravitational wave detector, LISA, to probe the properties of black holes through observations of extreme-mass-ratio inspiral events. We have presented a general framework for addressing such questions, and considered a particular special case in which we imagine that the events observed by LISA are divided into bins in mass and redshift. Assuming that the only model uncertainty is in the unknown number density of black holes in the LISA range, $10^4 M_{\odot} < M < 10^7 M_{\odot}$, and taking a simple, non-evolving, power-law model for the black hole mass function, ${\rm d}n/{\rm d}\ln M = A_0 (M/M_*)^{\alpha_0}$, we conclude that LISA will be able to measure the amplitude of the mass function to a precision $\Delta(\ln A_0) \sim 0.08$ and the slope to a precision $\Delta(\alpha_0) \sim 0.03$. These precisions scale with the number of observed events like $N_{\rm obs}^{-1/2}$ and have been normalised to a reference model with $N_{\rm obs} \approx1000$. The present uncertainty in the slope of the mass function in the LISA range is of the order of $\pm 0.3$, so LISA will beat this with as few as $10$ detected events.

If we allow for the mass function to evolve with redshift, using the ansatz ${\rm d}n/{\rm d}\ln M = A_0 (1+z)^{A_1} (M/M_*)^{\alpha_0-\alpha_1 z}$, but assume the Universe has a non-evolving mass function with $A_1=\alpha_1=0$, we find LISA will be able to measure the parameters of the distribution to a precision $\Delta(\ln A_0) \sim 0.2$, $\Delta(\alpha_0) \sim 0.06$, $\Delta(A_1) \sim 0.7$ and $\Delta(\alpha_1) \sim 0.2$. These again show a scaling like $N_{\rm obs}^{-1/2}$ and are normalised to a black hole distribution that predicts $1000$ events. The errors in the evolution parameters, $A_1$ and $\alpha_1$, are sufficiently large that we must conclude that EMRI observations alone will be unable to detect evolution in the black hole mass function.

The work presented here has made use of various simplifications, which we now discuss.

\subsection{EMRI parameter space}
We have considered measurements of the central black hole mass and the source redshift only, and have taken all EMRIs to be circular, equatorial inspirals into black holes with the same spin. We also used signal-to-noise ratios that were averaged over sky position and orientation when assessing the detectability of different sources. This allowed us to use the observable lifetime functions given in~\cite{gairEMRIastro}. It will be important to extend this work to generic EMRIs, which will require the extension of the observable lifetime calculation to generic orbits. In general, the addition of extra parameters to a model tends to decrease the precision to which the model parameters can be measured, as we saw when we allowed the parameters of the mass function to vary with redshift. However, in this case, we will not only be adding additional parameters, but additional measurements, since LISA EMRI observations will also measure the black hole spin, orbital eccentricity etc. to high precision~\cite{AK}. We would therefore not expect our conclusions about the precision to which we can measure the black hole mass function to change significantly. We have already seen that we can measure the slope of the mass function to higher precision if black holes tend to be more rapidly spinning, as the presence of spin increases the range in mass and redshift within which EMRIs can be detected. Eccentricity may have a similar effect, as the additional waveform harmonics that arise due to eccentricity will enter the LISA frequency band earlier, thus tending to enhance the signal-to-noise ratio. These effects must be quantified in the future.

The most important consequence of including new parameters will be the ability to ask additional astrophysics questions. EMRIs will have considerable power to probe the spin distribution of black holes in the appropriate mass range at low redshift. Measurements of the masses of the inspiraling objects and the orbital eccentricities will provide constraints on the processes occurring in stellar clusters in the centres of galaxies, such as mass segregation and the efficiency of the various channels that lead to EMRIs~\cite{astrogr}. It is important to understand what LISA will be able to tell us about these various processes and how correlations could affect our ability to address these questions independently.

\subsection{Data analysis technique}
There are various assumptions in the analysis technique described here that could be relaxed. We have ignored parameter errors, other than to verify they did not significantly affect our results. This was possible because we were using binning of the observations before analysing the data. We have described how errors can be included properly and how the data can be analysed in a continuum limit, but this requires having properly sampled posterior pdfs for the EMRI sources. While we do not anticipate that our results will change significantly under such an analysis, it might be possible to explore this using pdfs for EMRI sources constructed in the context of the Mock LISA Data Challenges~\cite{mldc4}. Additionally, the construction of the observable lifetime used here assumes that the LISA observation is $100\%$ complete for sources with SNR $>$ 30, and $0\%$ complete for sources with SNR$<30$. The exact LISA completeness function will depend on the algorithms employed to carry out data analysis, which are rather uncertain at present~\cite{BGPemri,Cemri}. Our present model is a reasonable approximation if an SNR cut is used for source selection for the follow-up. It is unlikely that LISA data analysis preparation will reach a point where we will have a better model for the completeness anytime soon, but it will be straightforward to recompute the effective observable lifetime once one is available. Finally, our analysis has used the same model, based on a Poisson probability distribution, to generate the data sets we have searched and for the posterior construction.This is likely to be a good model for the EMRIs occurring in a given galaxy, but there are various complicating factors, since it takes some time for galactic centres to become relaxed after galaxy mergers. Although the galaxies in which LISA can detect EMRIs are at low redshift and therefore are unlikely to have undergone recent mergers, the Poisson model could be checked by using numerical simulations, of the type described in~\cite{hopman09}. We are also assuming that all systems of a given type have the same EMRI rate. While there will certainly be an average rate for systems of particular type, there may be rate ``noise'' which we have ignored here. This could be included in the generation of the sets of events to search, but we would require a reasonable model for the amount of noise to include.

\subsection{EMRI rate uncertainties}
One significant uncertainty in these results is in the correct interpretation of what we are measuring. As discussed earlier, the rate of EMRIs of a particular type depends on both the number density of black holes, and on the intrinsic rate of inspirals per black hole. We assumed that the second of these was known, using results in~\cite{hopman09}, and therefore that the EMRI observations were telling us about the number density of black holes. In practice, we will be measuring the convolution of these two effects to the precisions quoted here. As discussed in Section~\ref{sec:probdist}, observations of low-mass galaxies and further theoretical work should better inform our understanding of the EMRI rate per galaxy and the level of our uncertainty in it before LISA flies. However, there will be some residual uncertainty in the model assumptions, and there may be correlations between the rate and the black hole mass, or even other parameters such as spin, which might not have been included in the models. LISA is unlikely to be able to tease apart these two effects directly, which must be borne in mind when the results are interpreted. The use of other observations, either of gravitational waves or in electromagnetic wavebands, might help with the final interpretation.

\subsection{Future applications}
The work described in this paper has illustrated the potential power of LISA observations for studying the astrophysics of black holes in the range $10^4M_{\odot} < M < 10^7 M_{\odot}$. We have focussed on EMRI observations as an illustration, but the same approach can be used for the interpretation of LISA observations of mergers between supermassive black holes (SMBHs). SMBH mergers and EMRIs probe different subsets of the same population of black holes --- SMBH events will probe the mergers between black holes up to the highest redshift and earliest cosmic times, while EMRIs will probe the whole population of black holes (not necessarily active or merging) at $z<1$ that are the end products of such SMBH mergers. Models that predict the rates of LISA SMBH mergers will therefore also make predictions for the number density of these low redshift black holes that play host to EMRIs. Thus, it should be possible to derive more powerful constraints on the models from combined observations. In particular, it might be possible to use SMBH mergers at lower redshift in conjunction with EMRI observations to place constraints on the evolution of the black hole mass function with redshift, which we have seen is not possible using EMRI observations alone. This should be explored in more detail in the future.

\begin{acknowledgments}
JG's work is supported by the Royal Society. CT's work was supported by a summer studentship, funded by the Royal Society. MV acknowledges support from NASA ATP Grant NNX07AH22G and a Rackham faculty grant. 
\end{acknowledgments}

\bibliography{EMRIsAsProbes}

\appendix
\section{Event distribution allowing for parameter errors}
\label{app:errdist}
Here we consider the effect of parameter errors on the distribution of observed events. For events in a given bin we assume that the error distribution,  $P(\vec{\lambda_{\rm o}}; \vec{\lambda_{\rm t}})$, is known, so that, by integrating this probability distribution over each bin, we can compute the probabilities, ${p_i}$, that, when observed by LISA, the event appears in bin $i$. If the process occurring in the original bin is a  Poisson process with mean $\beta$, then we claim that the observed data is a set of {\it independent} Poisson processes occurring in each bin, with mean $\beta p_i$ in bin $i$. This can be seen as follows. We denote the number of observed events in bin $i$ by $n_i$. The probability that we see a given set of observed events $\{n_i\}$ is equal to the probability, $P_1$, that the Poisson process generates a total of $N = \sum_i n_i$ events during the observation window, multiplied by the probability, $P_2$, that $N$ events would be distributed as $\{n_i\}$. The first of these probabilities is
\begin{equation}
P_1={\beta}^N {\rm e}^{-\beta}/N!, 
\end{equation}
while the second is
\begin{equation}
P_2 = \left(\prod_i p_i^{n_i}\right) \cdot \left[ \prod_j \left(\begin{array}{c} N-\sum_{k=1}^{j-1} n_k\\ n_j\end{array}\right) \right]
\end{equation}
in which $\left(\begin{array}{c} n\\ m\end{array}\right) = n!/((n-m)! \, m!)$ is the number of ways to choose $m$ objects from $n$. This probability is the probability that particular events fall into particular bins multiplied by the number of ways in which the events could be ordered into the bins to give the correct total number of events. Expanding out the second term we find that $P_2$ reduces to
\begin{equation}
P_2 = N! \prod_i \frac{p_i^{n_i}}{n_i!}
\end{equation}
and hence the total probability $P=P_1 \times P_2$ becomes
\begin{equation}
P = \beta^N {\rm e}^{-\beta} \prod_i \frac{p_i^{n_i}}{n_i!} = \prod_i \left( \frac{(\beta p_i)^{n_i} {\rm e}^{-\beta p_i}}{n_i!} \right)
\end{equation}
where we make use of the fact that $\sum_i p_i = 1$. This is just the product of the probabilities for Poisson processes with means $\beta p_i$ in each bin.

There will be many intrinsic bins that contribute to a given observed bin, but since the processes occurring in each galaxy are independent we can use the standard result that the distribution of the sum of two independent Poisson processes with means $\beta_1$ and $\beta_2$ is a Poisson process with mean $\beta_1+\beta_2$. We deduce that the number of observed events in each bin is drawn from an independent Poisson process, with mean given by the average over the intrinsic rates, weighted by the probability distribution of the errors in the measured parameters.

\end{document}